\documentclass[aip, amsmath,amssymb,
  preprint,%
]{revtex4-1}

\usepackage{xcolor,ulem}
\usepackage{graphicx,subfigure}% Include figure files
\usepackage{float} % Added by Ming

\usepackage{dcolumn}% Align table columns on decimal point
\usepackage{bm}% bold math
\usepackage[mathlines]{lineno}% Enable numbering of text and display math
\usepackage[utf8]{inputenc}
\usepackage[T1]{fontenc}
\usepackage{mathptmx}
\usepackage{etoolbox}
\usepackage{soul}
\usepackage{CJKutf8}
\usepackage{tabularx}
\usepackage{hyperref} % Allows clickable URLs and proper formatting

\usepackage{tikz}

\newcount\ndots
\def\drwln#1#2{\raise 2.5pt\vbox{\hrule width #1pt height #2pt}}
\def\spc#1{\hskip #1pt}
\def\solid{\drwln{24}{0.75}\ }

\def\dashed{\hbox {\drwln{6}{0.75}\spc{2}
                   \drwln{6}{0.75}\spc{2}\drwln{6}{0.75}}\nobreak\ }

\def\bdot{\hbox{\drwln{1}{.5}\spc{2}}}
\def\dotted{\hbox{\leaders\bdot\hskip 24pt}\nobreak\ }
\def\chndash{\hbox {\drwln{8.5}{0.75}\spc{2}
                    \drwln{3}{.75}\spc{2}\drwln{8.5}{.75}}\nobreak\ }

\def\square   {${\vcenter{\hrule height 0.4pt width 10pt
                 \hbox{\vrule width 0.4pt height 10pt \kern 9pt
                 \vrule width 0.4pt height 10pt}
                 \hrule height 0.4pt width 10pt}}$\nobreak\ }

\def\filsqr   {${\vcenter{\hrule height 2pt
                          \hbox{\vrule width 2.2pt height 0.2pt \kern 0.1pt
                                \vrule width 2.2pt}
                                \hrule height 2.2pt}}$\nobreak\ }

\newcommand\ming[1]{{\color{black}#1}}
\newcommand\vet[1]{{\color{black}#1}}
\newcommand\muela[1]{{\color{black}#1}}
\newcommand\jm[1]{{\color{black}#1}}

\makeatletter
\def\@email#1#2{%
 \endgroup
 \patchcmd{\titleblock@produce}
  {\frontmatter@RRAPformat}
  {\frontmatter@RRAPformat{\produce@RRAP{*#1\href{mailto:#2}{#2}}}\frontmatter@RRAPformat}
  {}{}
}%
\makeatother
\begin{document}
\preprint{AIP/123-QED}

\title[Atmospheric boundary layer over urban roughness: validation of large-eddy simulation]{Atmospheric boundary layer over urban roughness: validation of large-eddy simulation}
\author{
\begin{CJK*}{UTF8}{gbsn}
    Ming Teng (滕明)
\end{CJK*}
       }
\altaffiliation[Also at ]{ 
Dept. of Mechanical Engineering, University of Ottawa, Ottawa, Canada K1N 6N5
            }
\affiliation{
Barcelona Supercomputing Center (BSC), Barcelona, Spain
            }
\author{Josep M. Duró Diaz}
\affiliation{Universitat Politècnica de Catalunya (UPC), Terrassa, Spain
            }
\author{Ernest Mestres}
\affiliation{Universitat Politècnica de Catalunya (UPC), Terrassa, Spain
            }
\author{Jordi Muela Castro}
\affiliation{
Barcelona Supercomputing Center (BSC), Barcelona, Spain
            } 
\author{Oriol Lehmkuhl}
\affiliation{
Barcelona Supercomputing Center (BSC), Barcelona, Spain
            } 
\author{Ivette Rodriguez}
\affiliation{Universitat Politècnica de Catalunya (UPC), Terrassa, Spain
            }
\email{ivette.rodriguez@upc.edu}
\date{\today}% It is always \today, today,
             % but any date may be explicitly specified

\begin{abstract}
The study presents wall-modeled large-eddy simulations (LES)  characterizing the flow features of a neutral atmospheric boundary layer over two urban-like roughness geometries: an array of three-dimensional square prisms and the `Michel-Stadt' geometry model. The former is an arrangement of idealized building blocks and incorporates a $7 \times 7$ array of wall-mounted prisms with identical spacing ratios in both transversal and longitudinal 
directions. The latter mimics a typical central European urban geometry, which presents spatial inhomogeneity in all directions. In both cases, the incident wind angle is $0^\circ$. The Reynolds number for each case are $Re_H = 5.0 \times 10^6$ and 
$8.0 \times 10^6$, respectively ($Re_H = U_{ref} H/\nu$ with 
$U_{ref}$ and $H$ denoting the reference velocity and building height,  respectively, and 
$\nu$ the kinematic viscosity). The LES employs a high-order, low-dissipation numerical scheme with a spatial resolution of $0.75m$ within the urban canopy. An online precursor simulation ensures realistic turbulent inflow conditions, improving the accuracy of the results. The simulations performed successfully captures mean-velocity profiles, wake regions, and rooftop acceleration, with excellent agreement in the streamwise velocity component. While turbulent kinetic energy is well predicted at most locations, minor discrepancies are observed near the ground, partially due to insufficient near-wall resolution and measurement constraints. The analysis of scatter plots and validation metrics (FAC2 and hit rate) shows that LES predictions outperform the standard  criteria commonly used in urban flow simulations, while spectral analysis verifies that LES accurately resolves the turbulent energy cascade over approximately two frequency decades. The Kolmogorov $-2/3$ slope in the pre-multiplied spectra has been well reproduced below and above the urban canopy.  These findings reinforce the importance of spectral analysis in LES validation and highlight the potential of high-order methods for LES of urban flows.
\end{abstract}

\maketitle
\section{Introduction}
\subsection{Background}
Cities have long been regarded as centres of 
economic and demographic growth, resulting in a 
significant concentration of population in their 
cores. According to the United Nations, by 2030, two-thirds of the global population will reside in urban 
areas. Thus, the study and understanding of urban 
flows to improve forecasting and develop accurate 
prediction methods is at the focus of urban 
sustainability. Urban climate studies can be 
conducted at different scales, including the 
meteorological mesoscale, the meteorological 
microscale (up to about 2 km), the building scale 
(up to a few hundred meters), and the indoor 
environment (\(\approx 10~m\)), etc.

A large number of these studies have been conducted 
in generic or scaled-down geometries from two-dimensional (2D) street canyons to 3D generic 
geometries in order to understand the fundamental 
fluid dynamics associated with the urban flows. 
In these studies   
either Reynolds-averaged Navier-Stokes equations 
(RANS) or large-eddy simulations (LES) have been used.

\subsubsection{Simplified urban-like roughness}
\vet{Studies of flows over simplified 2D and 3D urban-like geometries have provided insight into the complex flow physics within street canyons due to their similarity to the topology of urban environments
\cite{meinders1999vortex, mathey1999large, 
xie2006and,Coceal2006,Blunn2022}.} \ming{Among these 
studies, the complex flow pattern, turbulence statistics, 
dispersion of passive scalars and wind loading are of 
particular interest. }

\citet{cheng2002near} considered a 
boundary-layer over a urban-like roughness in multiple 
configurations. Both staggered and in-line matrices of 
square prisms with identical height were investigated in the 
experiment. In addition, the study also included measurements of 
a staggered matrix comprising prisms of random heights. 
Prisms were equally-spaced 
throughout the investigations where 
Reynolds number 
considered ranged from $Re_H = 0.5 \times 10^4$ to $1.2 \times 10^4$ ($Re_H$ here is based upon \ming{the prism height, $H$, and} freestream 
velocity, $U_\infty$). The study focused on the 
characteristics of inertial and roughness sublayer. The roughness sublayer was found to have a depth of $1.80 - 
1.85H$ for prism matrices of identical $H$. Dispersive 
stresses arising from the inhomogeneity was negligibly 
small as compared to spatially averaged Reynolds shear 
stresses. On the other hand, the thickness of roughness 
sublayer increased substantially for the case with random 
heights. \citet{castro2006turbulence} extended the 
investigations and focused on the turbulence statistics
in the roughness layer for a staggered matrix of an 
identical height. A two-point correlation revealed  the 
dominant scales of turbulence was of the same order as the 
$H$ of the prisms. It also suggested the existence of a  
scale much smaller near the top of the canopy region. The 
results of stress anisotropy agreed with the 
previous findings in that the roughness increased the 
level of isotropy. This effect was most pronounced within 
the canopy layer where sweep events dominated the momentum 
transport.

\ming{\citet{xie2006and} performed LES and studied the turbulent flow over staggered wall-mounted cubes with both fixed and random heights. Reynolds numbers  $Re_H$ between $5 \times 10^3$ and $5 \times 10^6$ were considered. Numerical results were extensively validated against data from both direct numerical simulations and experiments, indicating the good predictive capability of LES for such flow scenarios. This is partially  flows over urban-like obstacles exhibit a wide inertial sub-range, suggesting that turbulence reaches a quasi-isotropic state at relatively lower frequency than non-vortex-shedding flows at similar Reynolds numbers. Moreover, steady RANS were also solved and proved inadequate especially in the canopy region due to  the inherent unsteadiness of the flow.}

\ming{\citet{inagaki2008turbulent} conducted field experiments using a scaled geometry to investigate the turbulence similarity over cubical obstacles. The roughness Reynolds number ranged between $10^4$ and $10^5$. The study examined turbulence similarity within the inertial sublayer using inner-layer scaling variables. It was concluded that inner-layer scaling effectively described the wall-normal velocity fluctuations and Reynolds stresses, regardless of surface roughness or outer-layer conditions, under near-neutral stratification in atmospheric flow. However, inner-layer scaling did not hold for horizontal velocity fluctuations due to the influence of outer-layer disturbances.}

While significant progress has been made using simplified geometries --providing insights into airflow and turbulence within cities, as well as improving the representation of drag forces and turbulent mixing-- certain limitations remain, specially when using scaled-down geometries. These simplifications often fail to capture the full complexity of real urban environments \cite{Chew2018}. In fact, scaled geometries may overlook the heterogeneity of building shapes, heights, and arrangements, leading to inaccuracies in representing wind flow patterns. Therefore, there is a need for improved parametrization of urban canopy turbulence that incorporates greater geometric detail \cite{Barlow2009}. 

\subsubsection{Full-scale urban roughness}
\vet{Studies of urban flows over semi-idealized or realistic urban geometries have been }mostly performed 
using either  RANS or unsteady RANS models. This preference is
mainly attributed to the complexity of the flow, 
which spans a wide range of spatial and temporal 
scales, combined with the large dimensions of the 
domain under investigation, making scale-resolving 
simulations computationally challenging. However, RANS models are inherently limited in their ability to capture the unsteady features of turbulent flows  within urban canopy layers, such as vortex shedding, separations, and reattachment. Consequently, the 
majority of such studies often suffer from reduced 
accuracy, particularly  in narrow street canyons and downwind regions of 
buildings, where flow accelerations are more pronounced
\cite{antoniou2019cfd, brozovsky2021validation}.

High-fidelity simulations have been proven to be more accurate in resolving the flow features in urban environments \ming{\cite{Stuck2021, li2010large, xie2009large}}.
\citet{xie2009large} studied the flow and 
dispersion characteristics in a 1:200 urban geometry 
corresponding to central London; they also concluded that 
a resolution  of $1~\text{m}$ at full-scale size would be appropriate to describe 
the mean turbulent flow behaviour. Other examples of this 
kind of studies on scaled-down urban areas include 
\citet{gousseau2011cfd} and \citet{hassan2022large}. While 
the former conducted RANS and LES 
simulations in a 1:200 geometric model to study pollutant
dispersion in a neighborhood of Montreal comparing their 
results against a wind tunnel prototype, the latter 
performed LES to examine traffic exhaust dispersion in a 
1:150 scaled design of a separated street canyon.
\vet{The critical role of LES to capture unsteady flow features, higher-order turbulence statistics, and complex flow patterns was highlighted in the study of \citet{tolias2018large}, who also emphasized the importance of conducting  a grid resolution study to achieve accurate results. }

\citet{kirkil2020large} conducted a LES over a 1.10 
$\times$ 0.76 km$^{2}$ region of downtown Oklahoma City to 
analyze the effect of the building height on the 
flow configurations. Similarities to flow fields around cylinders or inside cavities were observed. \citet{auvinen2020study} used LES to study the effect of 
several modeling parameters in the predictability of urban 
flows under realistic conditions, emphasizing the 
importance of mesh resolution at pedestrian level and the 
height of the domain when it comes to the simulation of 
the transition of a fully developed atmospheric boundary-layer into a \ming{urban boundary layer}.
More recently, the impacts of urban geometry 
heterogeneity on flow statistics 
were studied by \citet{cheng2023scaling}, who conducted simulations on
realistic urban geometries covering an area of 
about 1.0 $\times$ 0.5 km$^{2}$.  The study demonstrated the importance of turbulence statistics to improve the 
prediction of current urban canopy models.

\ming{\citet{OH2024105682} performed a LES using seasonal atmospheric data, with a 3 m resolution at the pedestrian level, to assess wind and thermal comfort in urban environments and evaluate the impacts of high-rise buildings. The study, which considered a canopy-layer-based Reynolds number of $5.4 \times 10^4$, focused on the thermal wind environment around the Yonsei University campus featuring two high-rise buildings. The results revealed that downdraft winds and large-scale circulations developed around realistic high-rise structures. These induced flows enhanced building-level wind speeds via the Venturi effect and reduced pedestrian-level temperatures by drawing cooler air downward.}

A recent study by \citet{Shaukat2024} has demonstrated that commonly used validation metrics for evaluating simulation quality should not be the sole criteria for assessment. Similarly, research by \citet{Tian2024} has once again highlighted the necessity of using full-scale geometries to effectively analyze and capture the complexities of urban flows. Together, these findings emphasize the importance of conducting high-accuracy simulations on real-world urban geometries.

\subsection{Scope}
In a review of the published literature, \citet{toparlar2017review} reported that less than 3\% of 
investigations employed LES for modeling the urban microclimate in full-scale scenarios. With the 
increase in the computational capacity, LES is increasingly used to simulate 
various phenomena such as wind patterns, pollutant dispersion (e.g., \citet{moon2014large, 
chew2018buoyant}) and thermal effects (e.g., \citet{chew2018buoyant}) in full-scale urban 
environments in the presence of the \ming{atmospheric boundary layer (ABL)}.

In the present study, wall-modeled LES of the \ming{ABL} over an urban roughness is implemented. 
Two geometries are considered, namely, a wall-mounted \ming{array} of 
\ming{cubic} prisms and the “Michel-stadt” model. Simulations of both cases 
play an important role in validation and 
understanding of the micro-climate in full-scale urban areas for 
their resemblance to flow over complex terrain and in the 
vicinity of full-scale buildings. \ming{In most of the aforementioned 
LES studies, the inflow boundary conditions implemented fall into 
one of three categories: periodic boundary condition in the 
streamwise direction \cite{xie2006and, li2010large, 
moon2014large}, 
prescribed velocity profile at inflow \cite{chew2018buoyant, 
kirkil2020large, Stuck2021, hassan2022large}, or artificially 
generated turbulent inflow \cite{xie2009large, gousseau2011cfd, 
tolias2018large, OH2024105682}. In the current study, a realistic 
turbulent ABL is implemented at the inlet via an online precursor 
simulation. This approach ensures a fully developed ABL with 
stochastic fluctuations that accurately represent the  turbulent approaching flow.} 

\vet{The main objectives of the present work are to assess the performance of the methodology used to model the turbulent flow in complex urban scenarios, as well as to demonstrate the potential of this type of simulations for accurately modeling urban flows.
A thorough validation, including direct comparison of turbulence statistics with wind tunnel data, point-to-point analysis of various quantities, and evaluation via figures of merit, as well as energy spectra, gives credibility to the current setup. In addition, }
visualization of instantaneous flow 
structures is presented providing insight into the flow evolution. In the following, we first present the definition of the study cases in Section \ref{sec:cases} and the numerical methodology in Section \ref{sec: method}. The results are compared with wind tunnel data in Section \ref{sec: results}. The main findings and the accuracy of the numerical setup are discussed in Section \ref{sec: Discu}, and conclusions are provided in Section \ref{sec: conclusions}.

\section{Cases definition}\label{sec:cases}

\vet{ In order to demonstrate the potential of using high-fidelity simulations to study turbulent flow in full-scale urban geometries, two different cases with available wind tunnel measurements have been considered. The selected cases are: a 3D cubic prism array \cite{BRO01,brown2001comparison} \ming{--- an arrangement of idealized building blocks}, and the 'Michel-Stadt' BL3-3 case, which represents urban areas in central European cities \cite{leitl2024}.}

\textbf{Case Study 1. 3D square prism array}

The first case studied is the wind tunnel experiment conducted by \citet{brown2001comparison}, which also serves to validate the numerical framework. This experiment investigated a neutral \ming{ABL} approaching a $1:200$ scaled array of 3D buildings arranged in seven rows. \ming{Each building has} dimensions $L = W = H = 0.15$ m, where $L$ and $W$ represent the building's length and width. The streamwise and spanwise distances between blocks were set to $H$.

To replicate the ABL in the wind tunnel experiments, spires and floor roughness elements were used to create an equivalent ABL, characterized by a reference velocity $U_{ref} = 3 \mathrm{m/s}$ at the reference height, $H_{ref}$ ($H_{ref}=H$) and a friction velocity $u^* = 0.24 \mathrm{m/s}$. In the experiments, vertical profiles of streamwise and spanwise velocities, together with turbulent kinetic energy ($TKE$), were measured at various streamwise positions along the centerline of the array.
  
\textbf{Case Study 2. “Michel-Stadt” BL3-3 }

The second case considered is part of the CEDVAL-LES database -- a collection of datasets designed for validating \ming{LES} \cite{leitl2024}. The experiments were conducted at the wind tunnel facility of the Meteorological Institute at the University of Hamburg. The specific urban geometry under study, referred to as the “Michel-Stadt” case (reference BL3-3), represents a semi-idealized urban layout typical of residential areas in Central European cities. The scaled model, at $1:225$, consists of 60 flat-roof building blocks with courtyards and roof heights of $0.625H  (15~\text{m})$, $0.75H (18~\text{m})$, and $H  (24~\text{m})$, covering a total area of $1320~\text{m} \times 830~\text{m}$ at full scale.

In the experiments, the approaching \ming{ABL} was modeled using roughness elements to replicate a highly rough flow environment. This setup was characterized by a surface roughness length $z_0 = 1.53~\text{m}$ and a friction velocity $u^* = 0.596 ~\text{m/s}$. The reference velocity was defined as $U_{ref} = 6.1~\text{m/s}$ at a reference height of $H_{ref}=100~\text{m}$.

Laser Doppler Anemometry (LDA) measurements were used to obtain time-series data for two velocity components (streamwise and lateral) across 40 vertical profiles and 5 horizontal levels within the central part of the city. These levels correspond to heights of $2~\text{m}$, $9~\text{m}$, $18~\text{m}$, $27~\text{m}$, and $30~\text{m}$, as described in \citet{Hertwig2012}. Additionally, first- and second-order flow statistics are available for comparison. For the remainder of the paper, the aforementioned cases  will be referred to as Case 1 and Case 2, respectively.

\section{Methodology}\label{sec: method}
\subsection{Governing Equations}
The isothermal and incompressible flow in an urban environment is governed by the equations of conservation of mass and momentum. By spatially filtering these equations, the LES formulation can be obtained,

\begin{equation}
    \frac{\partial \tilde{u}_i}{\partial x_i} = 0,
    \label{eq: conserv}
\end{equation}
\begin{equation}
    \frac{\partial \tilde{u}_i}{\partial t} + \frac{\partial 
    }{\partial x_j}\tilde{u}_i \tilde{u}_j = -\frac{1} {\rho}\frac{\partial \tilde{p}}{\partial x_i} + \nu \nabla^2 
    \tilde{u}_i -\frac{\partial 
    \tau_{ij}}{\partial x_j}.
    \label{eq: mov_filtred}
\end{equation}

In Eqs. \eqref{eq: conserv} and \eqref{eq: mov_filtred},  
\(\tilde{(\cdot)}\) represents the filtered variables. \(x_i\) (or 
\(x\), \(y\), and \(z\), denoting streamwise, spanwise, and wall-normal directions) are the spatial coordinates and \(u_i\) 
(or \(u\), \(v\), and \(w\)) are the corresponding 
velocity components. \(p\), \(\nu\) and \(\rho\) represent the 
hydrodynamic pressure, kinematic viscosity and the density of the 
fluid, respectively. The last term on the right-hand side of Eq. \eqref{eq: mov_filtred} \vet{ represents the non-resolved scales and has to be modeled. Its deviatoric part is modeled as, }

\begin{equation} 
    \label{sgss}
   {\tau}_{ij}- \frac{1}{3}{\tau}_{kk} \delta_{ij} =
    -2\nu_{sgs} \tilde{{\cal S}}_{ij}.
 \end{equation}
 
\vet{ where $\tilde{{\cal S}}_{ij} = 1/2 (\partial \tilde{u}_i/\partial x_j + \partial \tilde{u}_j/\partial x_i)$ is the rate-of-strain tensor of the resolved velocity field, and $\delta_{ij}$ is the Kronecker delta. }The subgrid-scale viscosity, $\nu_{sgs}$, in the present study is modeled by the \citet{vreman2004} model.

The filtered incompressible Navier-Stokes equations, Eqs. 
\eqref{eq: conserv} and \eqref{eq: mov_filtred},  are solved 
using SOD2D (Spectral high-Order coDe 2 solve partial 
Differential equations) \cite{gasparino2024}. 
\muela{It is an open-source code featuring high-fidelity and low-dissipation, based upon a high-order Continuous Galerkin spectral element method (CG-SEM) \cite{sod2drepo}. It employs the Gauss-Lobatto-Legendre (GLL) quadrature, thus the nodes are non-equispaced, avoiding the Runge effect on high-order interpolations. The quadrature points coincide with the element nodes, resulting in a closed rule integration. Since this can lead to aliasing effects due to the reduced integration order, the skew-symmetric convective operator split detailed by \citet{kennedy2008} is employed, which counters these undesired aliasing effects. For the temporal discretization, the BDF-EXT3 high-order operator splitting approach proposed by \citet{Karniadakis1991} is used to solve the velocity-pressure coupling, with the pressure Poisson equation solved by means of a conjugate-gradient solver. Regarding spatial discretization, in the present study the mesh is based on a $4^{th}-$order hexahedral elements. 
}

\vet{ SOD2D has been developed to efficiently utilize the growing computational power available globally, with a particular emphasis on leveraging GPUs, which play a central role in modern high-performance computing (HPC). To ensure versatility, the code is designed to run on both GPU and CPU architectures. Specifically, SOD2D is written in Fortran and employs MPI for coarse-grained parallelism and OpenACC for fine-grained parallelism, enabling efficient execution on heterogeneous systems. Additionally, the code uses HDF5 for input/output operations, a robust and widely validated library for HPC applications. For further details, readers are referred to \ming{\citet{folk1999hdf5}}.}

\subsection{Computational domain and mesh}

\begin{figure}
    \subfigure[]{
\includegraphics[width=0.48\linewidth]
    {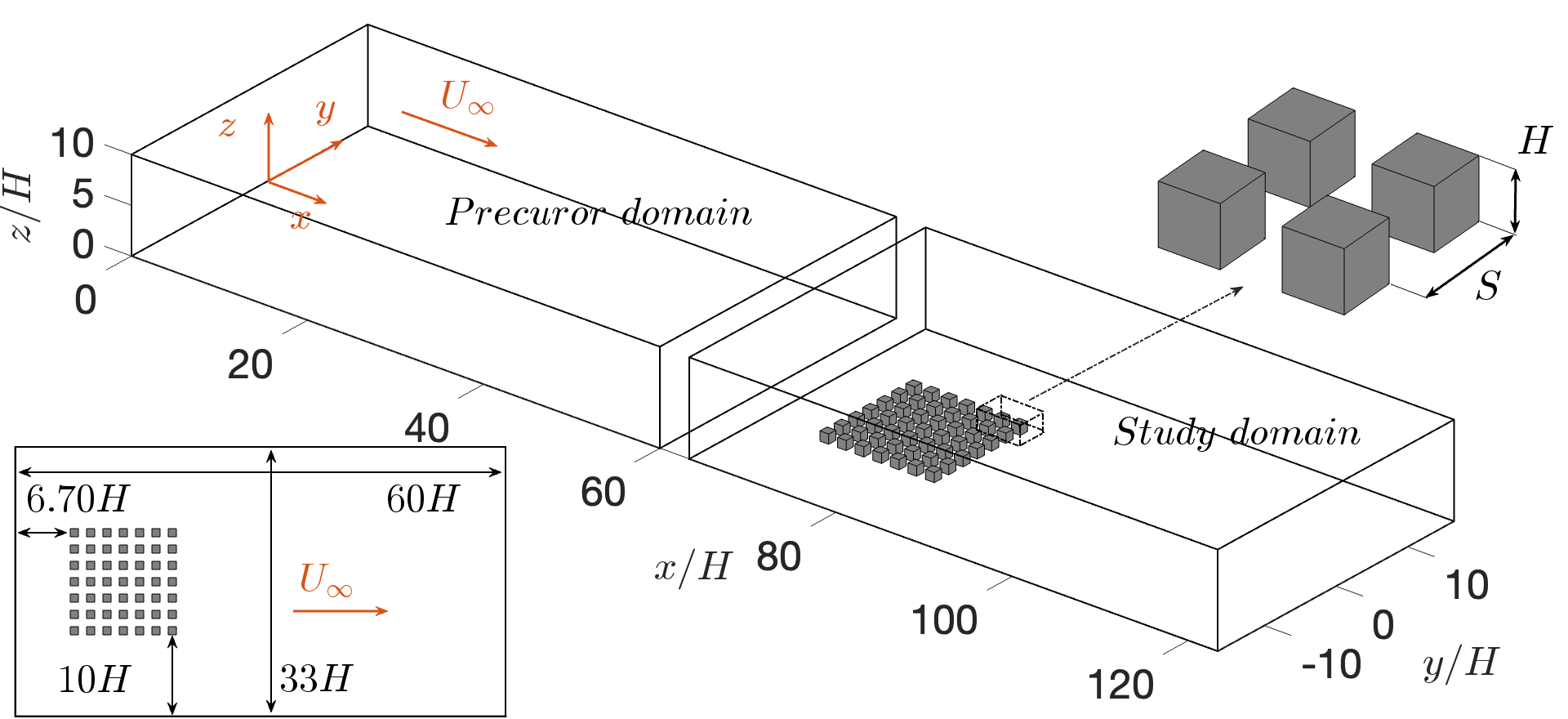}}
   \subfigure[]{
\includegraphics[width=0.48\linewidth]
 {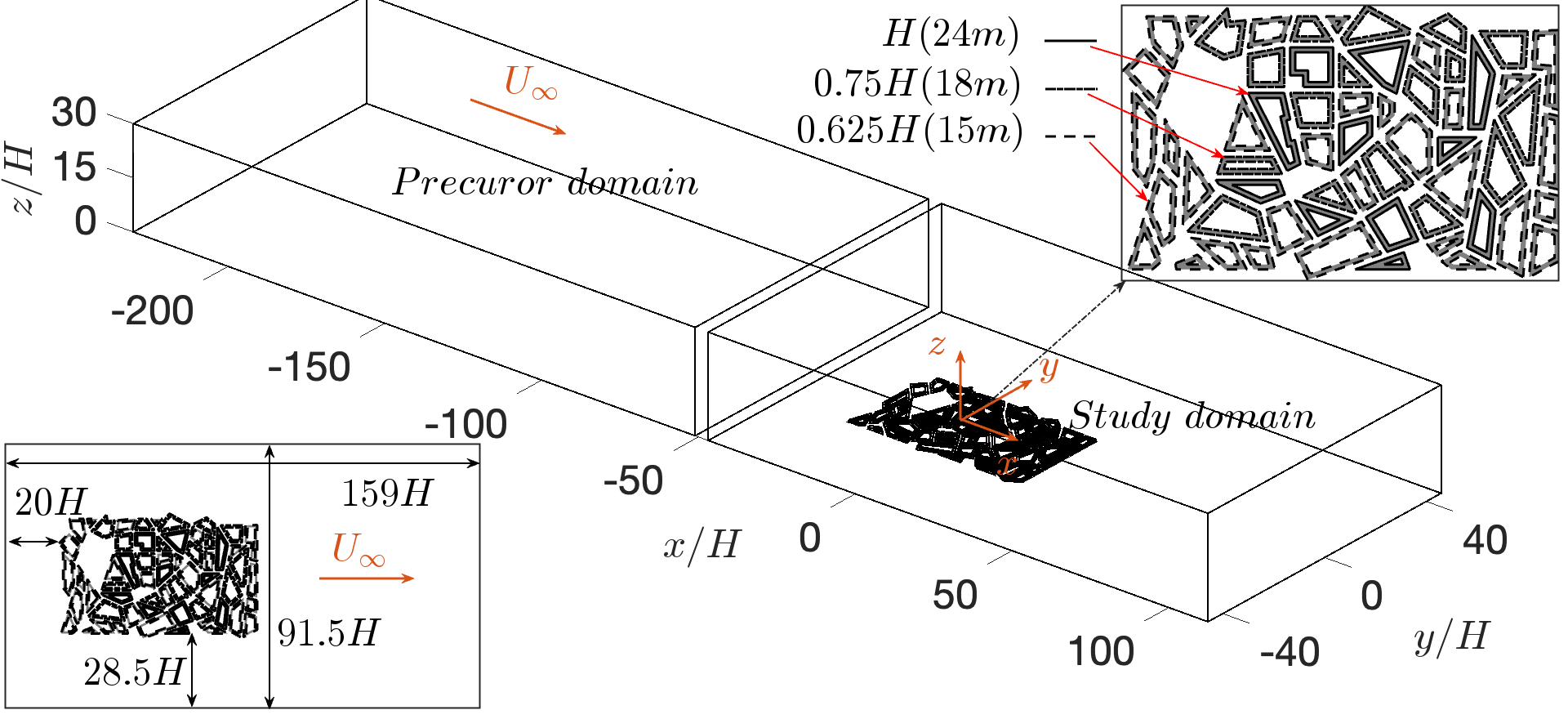}}
    \caption{Sketches of the computational domains used for the 
    calculations of an atmospheric boundary-layer over $(a)$ a 3D square prism array and $(b)$ a “Michel-Stadt” BL3-3 geometry 
    model, respectively.} 
    \label{fig:Geometry}
\end{figure}

\begin{figure}
    \centering   \subfigure[]{
    \includegraphics[width=0.48\linewidth]{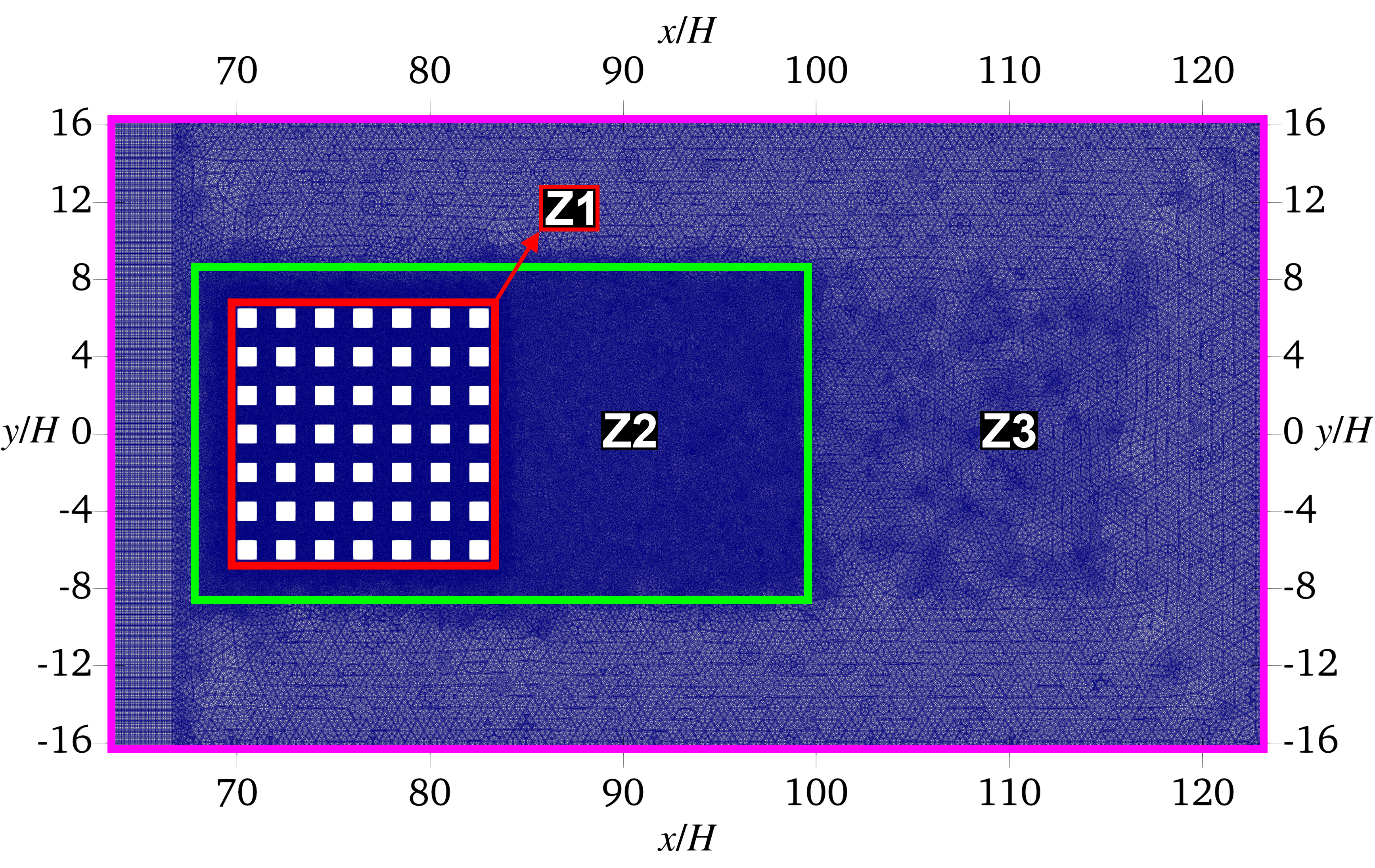}}
       \subfigure[]{
    \includegraphics[width=0.48\linewidth]{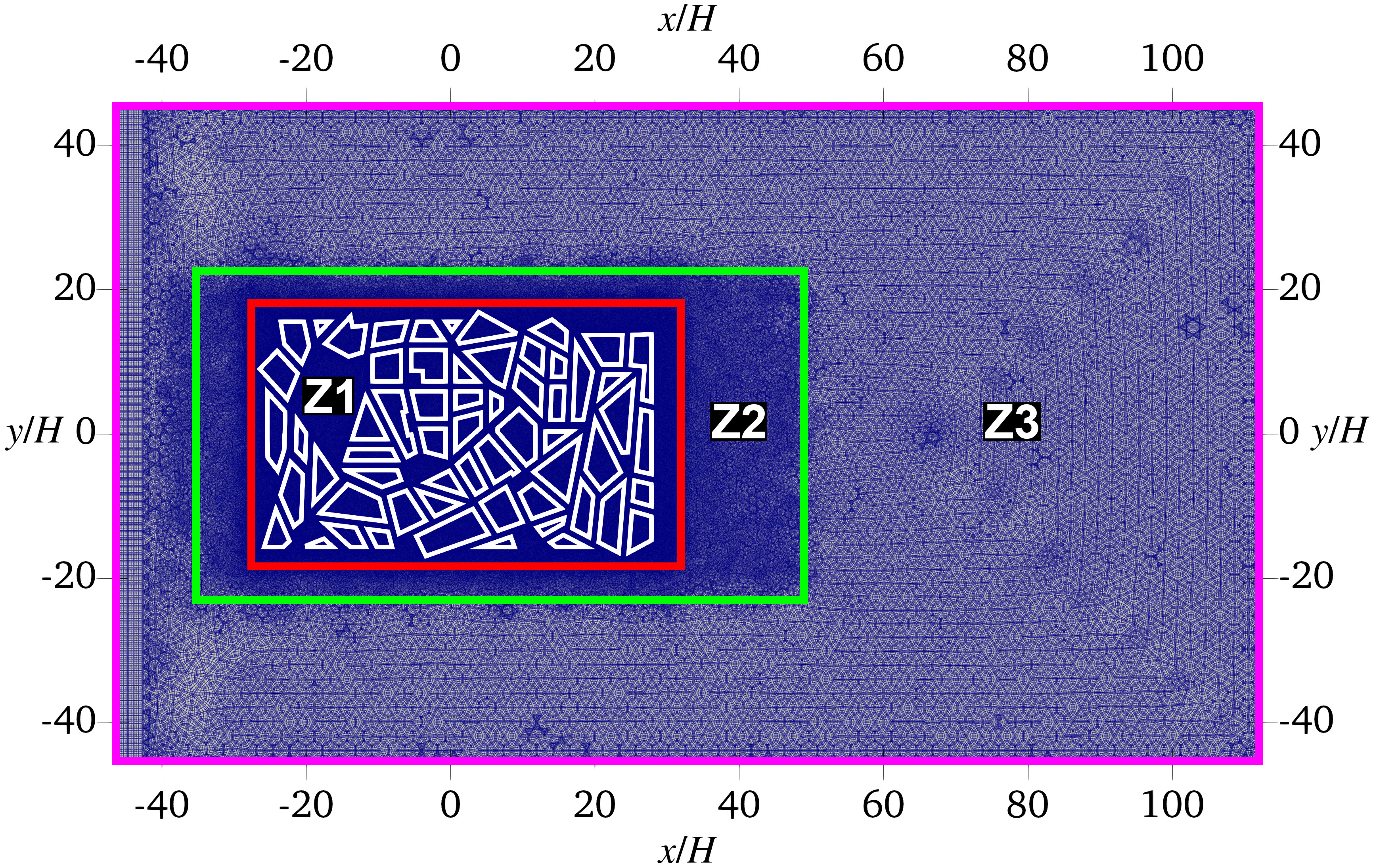}}

    \caption{Grid resolution used for the study domains shown in Fig. \ref{fig:Geometry}: $(a)$ a 3D square prism array and $(b)$ a “Michel-Stadt” BL3-3 geometry. The specifications of three zones, $Z1 - Z3$, are given in Table \ref{table:meshes}. } 
    \label{fig:meshResolution}
\end{figure}

\vet{Figure \ref{fig:Geometry} shows the numerical setup and computational domains for both cases considered. For \ming{Case 1}, shown in Fig. \ref{fig:Geometry}$(a)$, 
the computational domain has a size of $x \times y \times z = $ \ming{$60H\times 33H \times 10H$} (where $H = 30~\text{m}$ denotes the height of the square prism). 
\ming{In accordance with best practice guidelines for simulating urban flows \cite{franke2007best}, the 3D block array is placed $6.7H$ from the inlet and $10H$ from the lateral boundaries of the domain. }\jm{The outlet of the domain is placed $40.3H$ downstream the \ming{array}.}

For \ming{Case 2}, as seen in Fig. \ref{fig:Geometry}$(b)$, the domain has a 
size of $x \times y \times z =$ \jm{$159.4H 
\times 91.7H \times 30H$} ($H$ denotes maximum building height $H=24~\text{m}$) incorporating a 
heterogeneous geometry representative of a semi-idealised neighborhood.} \jm{In the study domain, the geometry is placed along 
the centerline of the horizontal plane, $20H$ 
downstream of the inlet. The lateral boundaries of the 
domain are placed at a distance of $28.5H$ from the 
city, while the outlet of the domain is located at 
$84.4H$ downstream of the city.}

\vet{In order to simulate the flow around the complex urban geometries, unstructured hexahedron-element meshes are constructed. Within each element, the solution is represented by tensor products of $4^{\text{th}}$-order polynomials in each direction.

For both cases, three levels of refinement have been considered. Details about the grid resolutions used for both cases are given in Fig. \ref{fig:meshResolution} and Table \ref{table:meshes}. The computational meshes are designed to resolve an important part of the inertial subrange in the region of interest. To achieve this,
\ming{three zones} of refinement with progressively increasing mesh resolution are implemented. \ming{Within the city, the finest meshes are specified 
with resolutions below  $1\ \text{m}$} (see \ming{Z1 in} Fig. \ref{fig:meshResolution}). This results in meshes of $22.8 \times 10^6$, $62.2 \times 10^6$, and \jm{$207.07 \times 10^6$ nodes} for Case 1 \ming{, and $22.6 \times 10^6$, $78.11 \times 10^6$, and $217.84 \times 10^6$ nodes for Case 2.} According to observations of \citet{xie2009large}, a 
resolution of about $1\ \text{m}$ is sufficient to 
achieve reasonable accuracy in the predictions of 
turbulence statistics in full-scale urban geometries. 
Thus, in the present work, within the zone of interest—
i.e., the urban geometry identified as $Z1$—grid mesh with 
resolution below $1\ \text{m}$ is employed.} 

\begin{table}
\caption{Grid meshes evaluated in the simulations of both cases considered. $N_{DoF}$ denotes the degrees of freedom. $\Delta_{Z1}$, $\Delta_{Z2}$ and $\Delta_{Z3}$ represent the corresponding grid resolutions in the $Z1-$, $Z2-$ and $Z3-$ zones shown in Fig. \ref{fig:meshResolution}}\label{table:meshes}
\begin{tabular}{l l l  l l l l l l }
\hline
Grid resolutions && $N_{DoF} (\times 10^6)$ && $\Delta_{Z1}[m]$ && $\Delta_{Z2}[m]$ && $\Delta_{Z3}[m]$ \\\hline
\multicolumn{9}{c}{Case 1}\\
Coarse && \jm{$22.8$} && 1.875 && 3 && 6 \\
Medium && \jm{$62.2$} && 1.25 && 2 && 4 \\
Fine && \jm{$207.1$} && 0.75 && 1.25 && 2.5 \\\hline
\multicolumn{9}{c}{Case 2}\\
Coarse && \jm{$22.6$} && 3 && 6.25 && 15 \\
Medium && \jm{$78.1$} && 1.5 && 5 && 11.5 \\
Fine && $217.8$ && 0.75 && 3.75 && 6.25 \\\hline

\end{tabular}
\end{table}

\subsection{Boundary conditions and computational details}\label{comp_detail}

\vet{To impose a realistic turbulent \ming{ABL} inlet condition, a precursor simulation is implemented simultaneously \ming{with the downstream simulation of the study domain} (online precursor). Typically, two methodologies are commonly used to impose inflow conditions: synthetic data generation (see, for instance \citet{Xie2008}) or \ming{database storing the spatial and temporal variation of the inflow conditions from a precursor simulation} (e.g. \citet{Shaukat2024}). While the former may not fully replicate the turbulent characteristics of a real ABL, the latter requires storing a substantial amount of high-frequency data which should match the time-step and the integration time of the main simulation.
In this study, we utilize two domains, i.e., the precursor and the main domain, which are solved simultaneously. This approach eliminates the need to store inflow data and seamlessly adapts to the spatial dimensions of the domain and the time-step variations of the city simulation.
To appropriately set up the inflow conditions, periodic boundary conditions are applied in 
the $x-$ and $y-$directions within the precursor domain (see Fig. 
\ref{fig:Geometry}) where the turbulent ABL is driven by an imposed pressure gradient in 
the $x-$direction. The outlet of the precursor simulation serves to “drive” the 
inlet of the study domain: \ming{the inflow of the study domain} is 
updated at each time step with data from the outlet of the precursor simulation.
 
To ensure that the inflow aligns with the conditions documented in the experiments, Fig. \ref{fig:precursor} compares the numerical results for the fine 
mesh in both cases with the measurements described in Section \ref{sec:cases}. 
Both profiles of streamwise mean velocity and $TKE$
are plotted. As shown in Fig. \ref{fig:precursor}, the precursor simulation accurately reproduces these statistics in both cases.
\begin{figure}
    \centering   
    \subfigure[]{
    \includegraphics[width=0.28\linewidth]{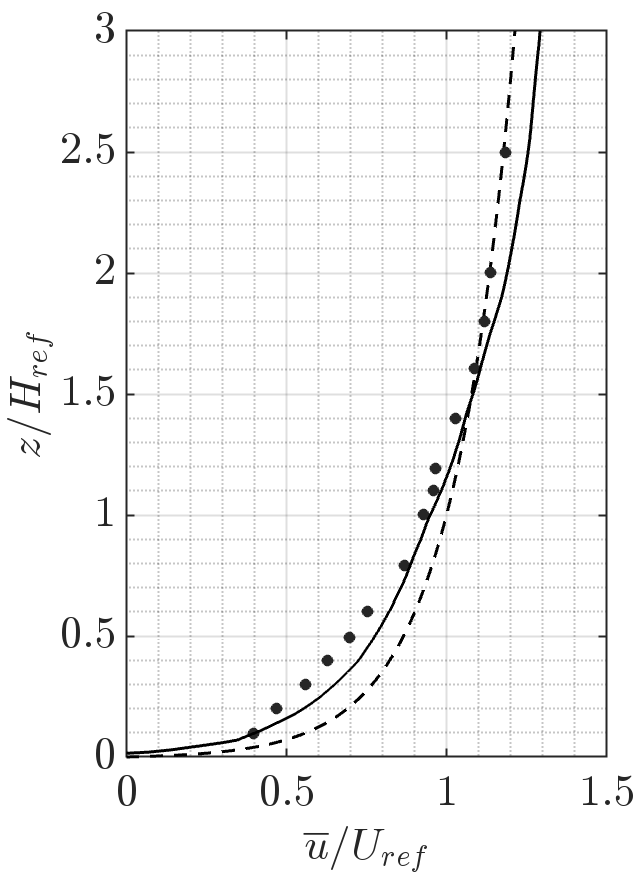}}
    \subfigure[]{    
    \includegraphics[width=0.50\linewidth]{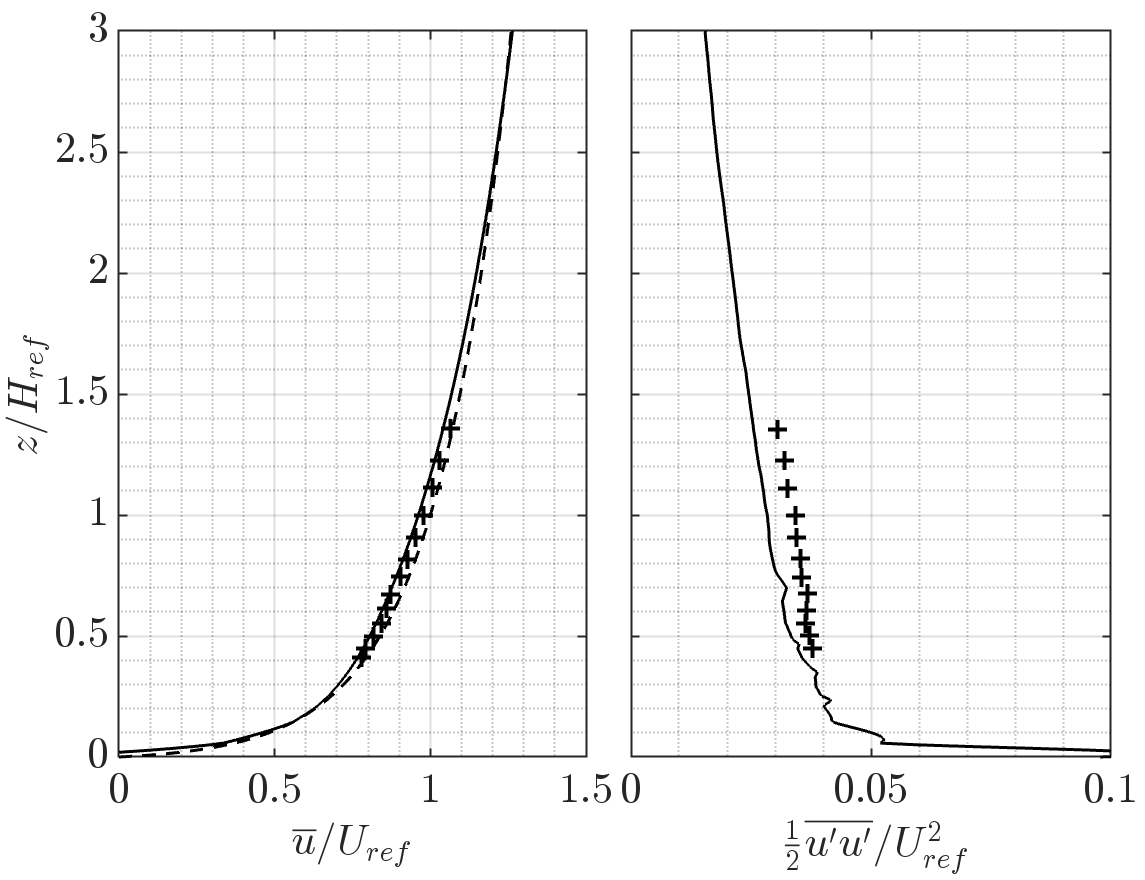}}
    \caption{Profiles of streamwise mean-velocity, $\overline{u}/U_{ref}$, and $TKE$, $\frac{1}{2}\overline{u'_iu'_i}/U^2_{ref}$, at inflow in comparison with experimental measurements. $(a)$ Case 1 and $(b)$ Case 2. \solid~Numerical data (fine mesh); \dashed~logarithmic law; $\bullet$~\citet{brown2001comparison}; {\textnormal{+}}~\citet{leitl2024}.}
    \label{fig:precursor}
\end{figure}
    
In addition, the top and lateral boundaries of the computational 
domain are modeled as free-slip walls, assuming zero normal gradients for all variables. At 
the outlet, zero static pressure is imposed. Within urban canyons, non-equilibrium 3D flows 
are expected around  buildings. Consequently, applying an equilibrium wall model on the wall 
of the buildings (e.g., a log-law near the wall) is inadequate. Therefore, no-slip 
conditions are applied at the walls, while for the ground  a wall model is employed. }

The flow field takes approximately  10 flow-throughs ($FT$) to pass the transient phase and reach the quasi-steady state. One $FT$ is defined based upon the length of the city and the velocity at the edge of the atmospheric boundary layer: $FT = L_{city}/U_{edge}$. This corresponds to 48$T$ and 125$T$, respectively, \ming{for each case considered ($T$ denotes one eddy turnover time, $T = H/u^{*}$)}. Following the initial transient phase, flow statistics are collected over approximately $40 FTs$ for both cases. This corresponds to $192T$ and $500T$ for Case 1 and 2, respectively. For further details regarding the temporal convergence of the flow, please refer to the discussion in Section\ref{sec: tmpConv}.

\vet{Finally, it is important to highlight the computational effort it takes for the present simulations. Both cases were executed on Marenostrum V, utilizing 8 nodes from the accelerated partition, each equipped with 4 H100 GPUs.
For Case 1, the setup achieved a performance of $0.128$ seconds per iteration, corresponding to $0.62$ ns/dof. Statistical data collection on the finest grid (207 million grid points) over 40 $FT$s required a wall time of 0.4 days, resulting in a computational cost of 404.9 GPU-hours. Similarly, for Case 2, simulations on the finest grid required a wall time of 6.7 days using 32 GPUs, with a total computational cost of 4744 GPU-hours.
These results emphasize the efficiency of the current setup in performing accurate and computationally efficient simulations for complex urban environments, achieving a balance between precision and resource utilization.}

\section{Results}\label{sec: results}

Extensive comparison of the results have been performed for both cases. Both profiles of streamwise mean velocity and $TKE$ have been compared to experimental measurements obtained from wind tunnel. Moreover, for Case 2, validation metrics have been used, along with energy spectra obtained at different stations.

\subsection{Grid-convergence study and validation}\label{sec: gridconvergence}

\begin{figure}
    \centering
    \includegraphics[width=0.495\linewidth]{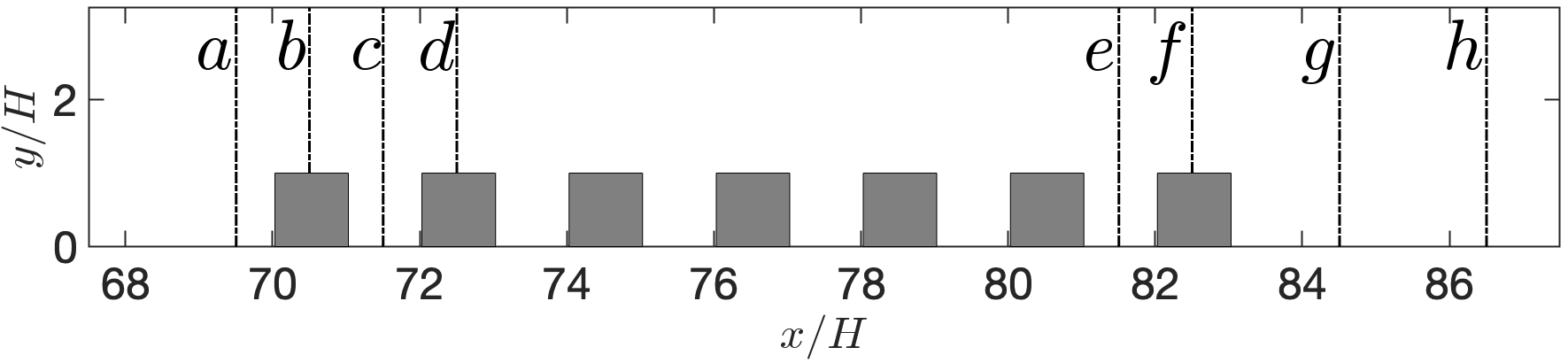}\\
    \vspace{0.10in}
    \includegraphics[width=0.495\linewidth]
    {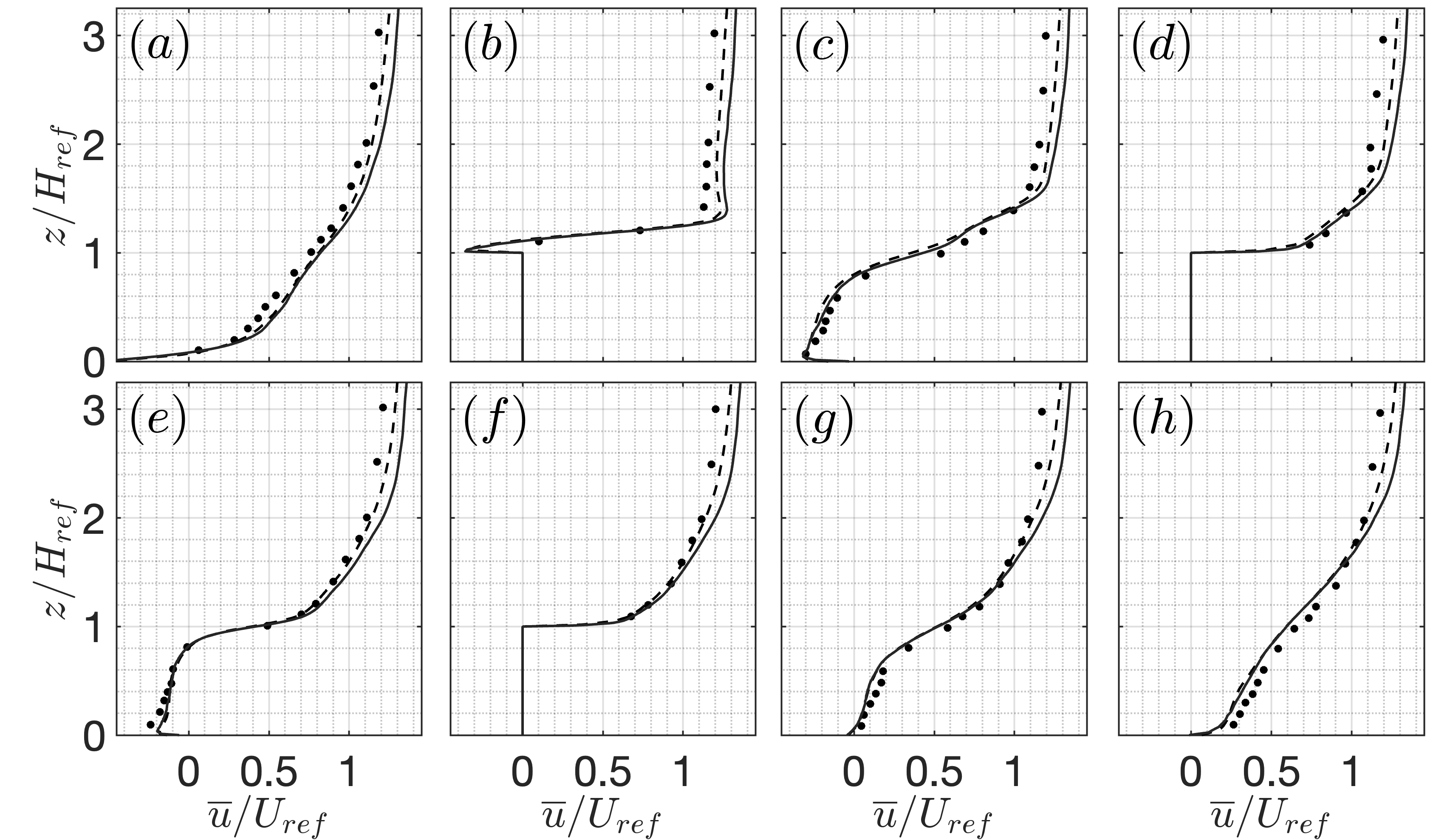}\\
    \caption{Case 1: comparison of streamwise mean-velocity profiles, $\overline{u}/U_{ref}$, with those of experimental results at 
    selected locations along $x-z$ plane at $y = 0$ as shown in top figure. \color{black}\dashed~Coarse mesh; 
    \color{black}\solid~fine mesh; $\bullet$~ 
    \citet{brown2001comparison}.} 
    \label{fig:gridCongergence_cubes_u}
\end{figure}

\begin{figure}
    \centering
    \includegraphics[width=0.495\linewidth]{Figures/Geo2D.png}\\
    \vspace{0.10in}
    \includegraphics[width=0.495\linewidth]{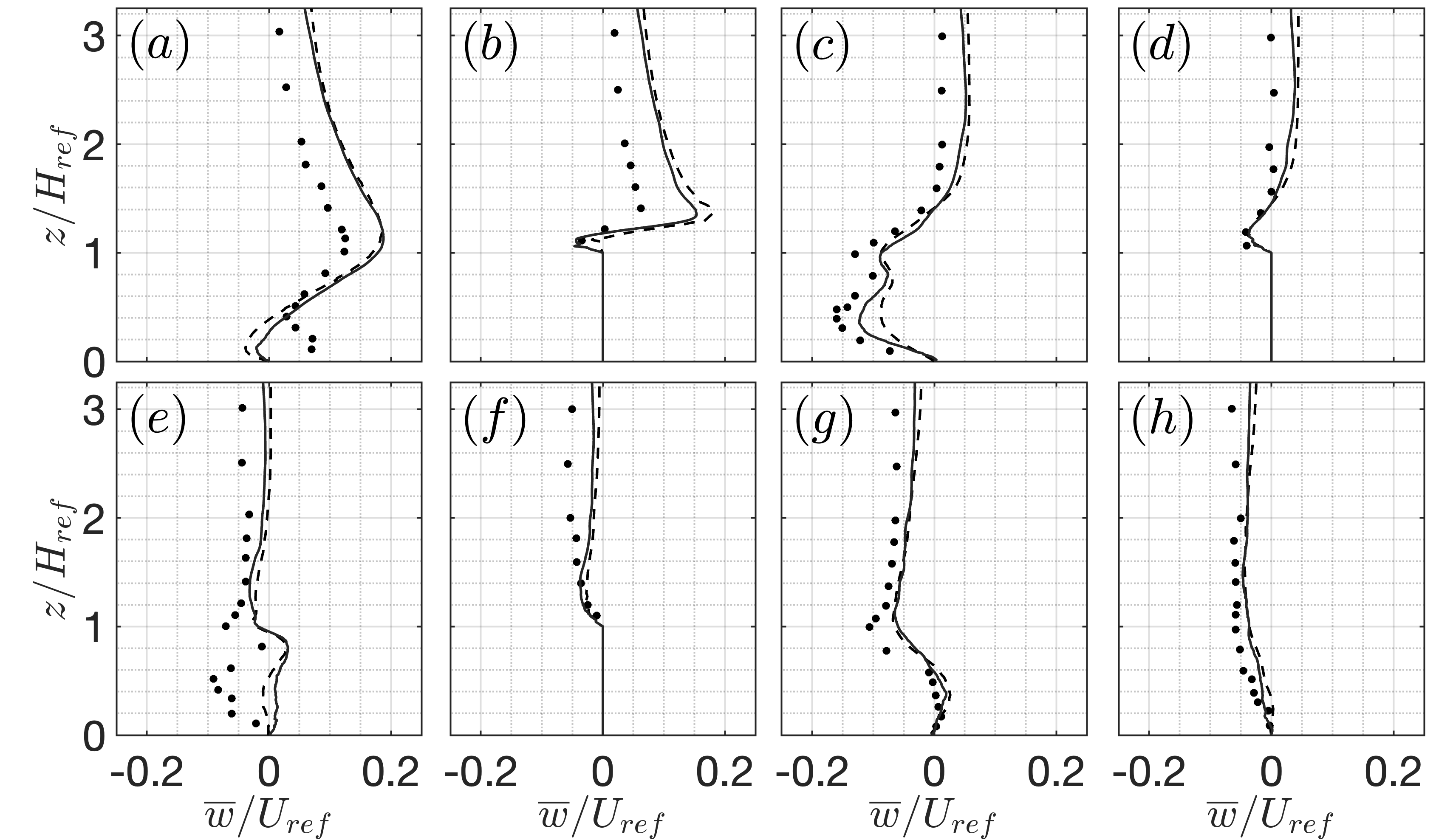}\\
      \caption{Case 1: comparison of wall-normal mean-velocity profiles, $\overline{w}/U_{ref}$, with those of experimental results at 
    selected locations along $x-z$ plane at $y = 0$ as shown in top figure.
    \color{black}\dashed~Coarse mesh; 
    \color{black}\solid~fine mesh; $\bullet$~ 
    \citet{brown2001comparison}.} 
    \label{fig:gridCongergence_cubes_w}
\end{figure}

\begin{figure}
    \centering
    \includegraphics[width=0.495\linewidth]{Figures/Geo2D.png}\\
    \vspace{0.10in}
     \includegraphics[width=0.495\linewidth]{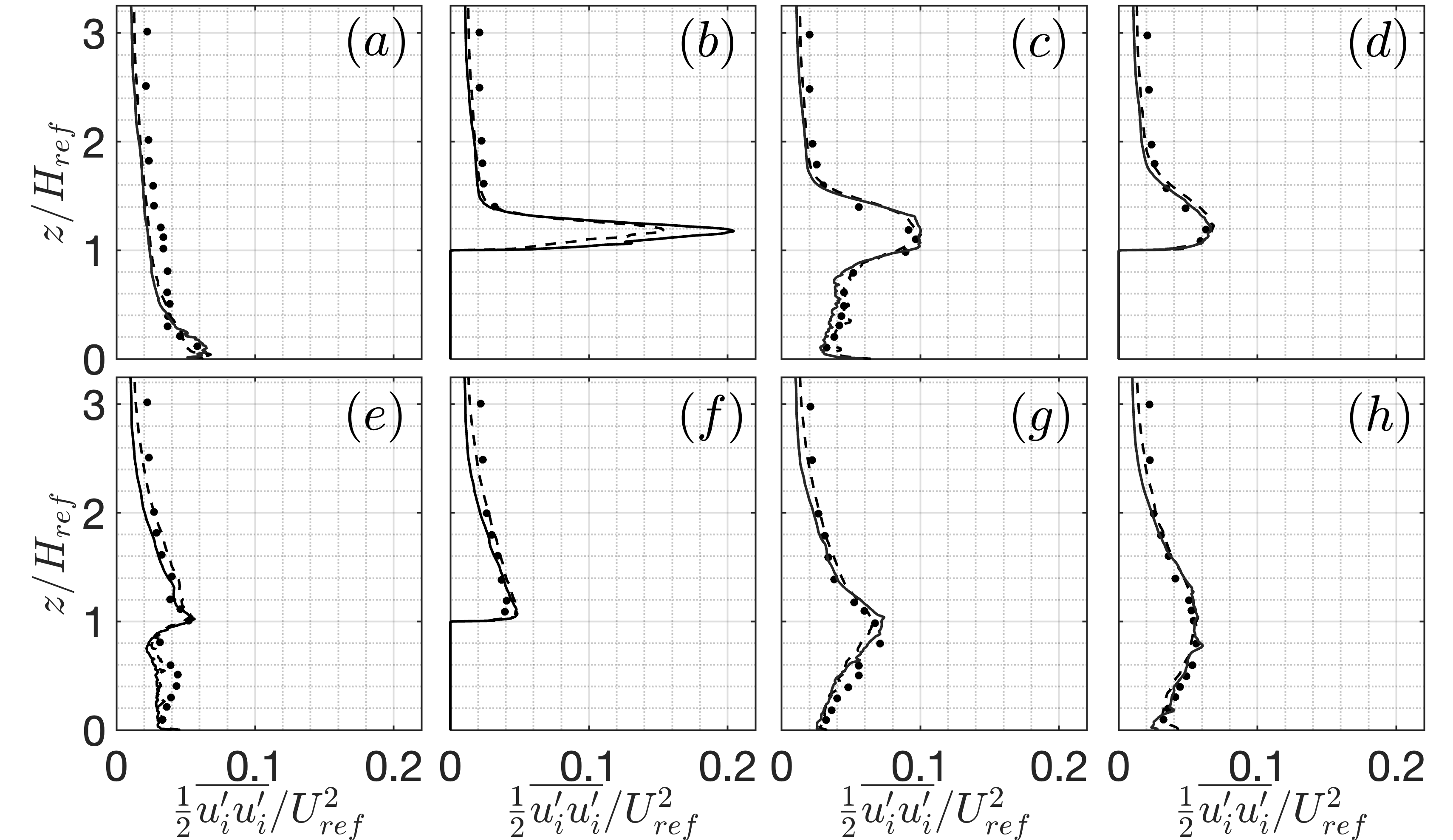}    
    \caption{Case 1: comparison of $TKE$, $\frac{1}{2}\overline{u_i'u_i'}/U_{ref}^2$, with those of experimental results at 
    selected locations along $x-z$ plane at $y = 0$ as shown in top figure. 
    \color{black}\dashed~Coarse mesh; 
    \color{black}\solid~fine mesh; $\bullet$~ 
    \citet{brown2001comparison}.}
    \label{fig:gridCongergence_cubes_tke}
\end{figure}

Validation of the numerical results begins with grid-convergence studies. For each case, three grid resolutions are employed within the study domain to assess the influence of grid refinement on the numerical results. For Case 1, Figs. \ref{fig:gridCongergence_cubes_u} to \ref{fig:gridCongergence_cubes_tke} compare profiles of streamwise mean velocity, $\overline{u}/U_{ref}$ (where $\overline{\cdot}$ represents temporal averaging, and $U_{ref}$ denotes the reference velocity), wall-normal mean velocity, $\overline{w}/U_{ref}$, and $TKE$, $\frac{1}{2} \overline{u_i'u_i'}/U_{ref}^2$ ($u_i' = \overline{u_i} - u_i$ represents the fluctuating components), with experimental results at selected locations along the plane $y = 0$. For clarity, results for the medium grid resolution are omitted. In general, discrepancies between the results obtained using the coarse and fine meshes are minor for both first- and second-order statistics, indicating that the fine mesh resolution is sufficient to achieve good agreement with experimental data.

The $\overline{u}/U_{ref}$ profiles (see Fig. \ref{fig:gridCongergence_cubes_u}) demonstrate good agreement  across all stations. Notably, the reverse flow in the recirculation region behind the blocks is reasonably predicted (see Figs. \ref{fig:gridCongergence_cubes_u}$(c,~e)$). However, a slight under-prediction of $TKE$ is observed in these regions (Fig. \ref{fig:gridCongergence_cubes_tke}), which can likely be attributed to an insufficient resolution close to the ground. \ming{Meanwhile,} a wall-function is not advisable \ming{in this region} due to the lack of local stress equilibrium in the areas around buildings where the flow exhibits significant separation. The simulations capture the $TKE$ peak above building heights well, corresponding to the shear layer development along rooftops.

Despite the overall good agreement in $\overline{u}/U_{ref}$ profiles, discrepancies are observed in $\overline{w}/U_{ref}$ profiles across all stations (see Fig. \ref{fig:gridCongergence_cubes_w}). They persist even with grid refinement, suggesting that they may stem from experimental uncertainties rather than insufficient resolution. It should be pointed out that reproducing an ABL in a wind tunnel is a challenging task,
as it depends not only on the turbulence generation method but also on factors such as Reynolds number mismatch and wall effects. \ming{These factors} can influence turbulence structures and introduce discrepancies between experimental results and full-scale atmospheric simulations.

\begin{figure}
    \centering
    \includegraphics[width=0.495\linewidth]{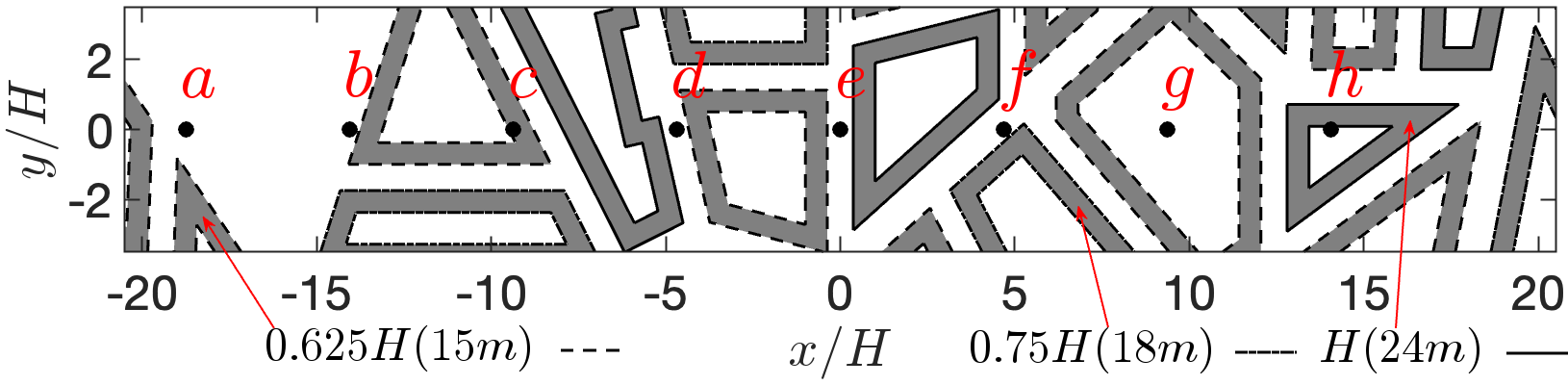}\\
    \vspace{0.10in}
    \includegraphics[width=0.495\linewidth]{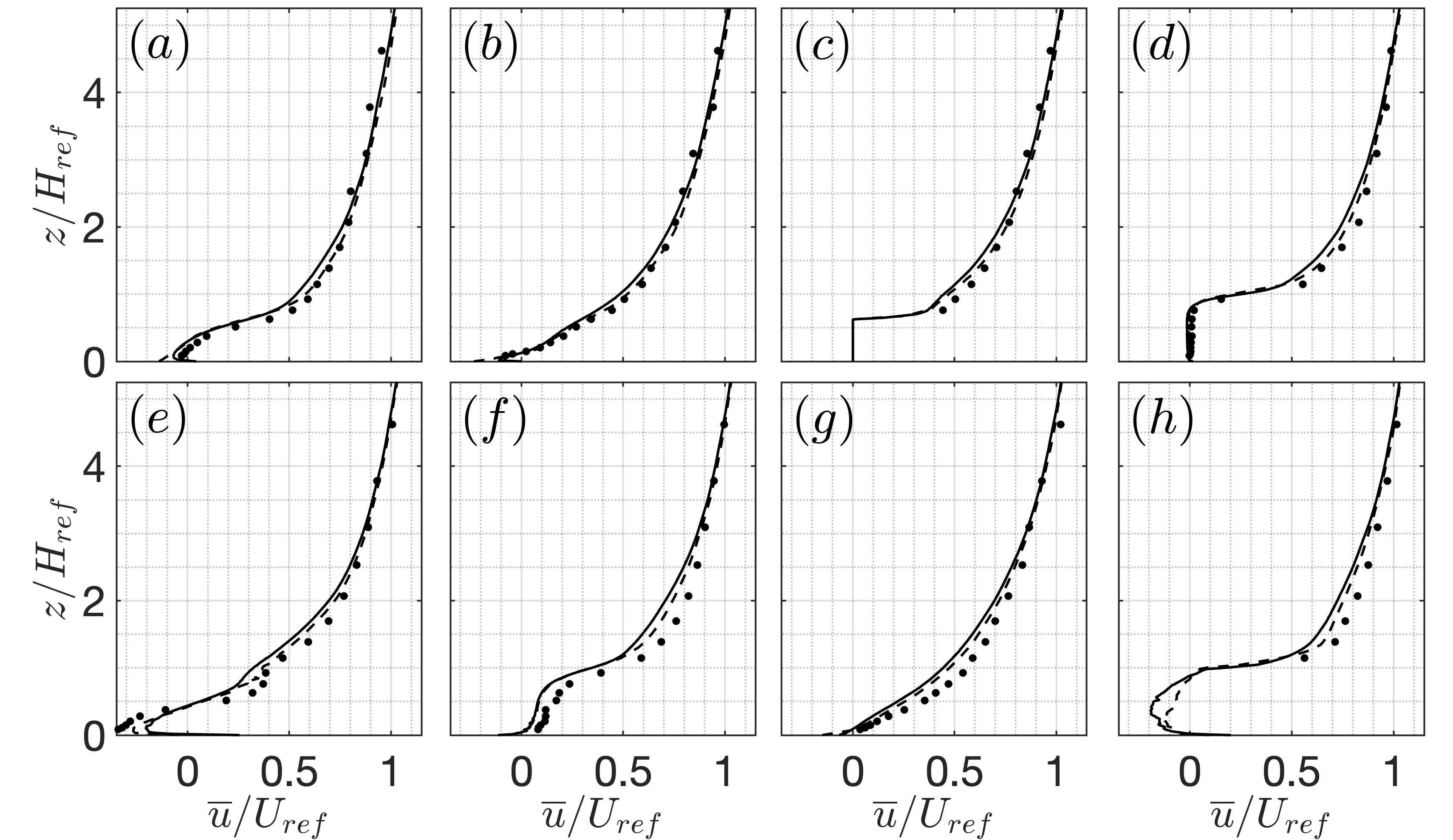}  
    \caption{Case 2: comparison of streamwise mean-velocity profiles, $\overline{u}/U_{ref}$, with those of 
    experimental results at selected locations along $x-z$ plane at $y = 0$ 
    as shown in the top figure. 
    \color{black}\dashed~Coarse mesh; 
    \color{black}\solid~fine mesh; $\bullet$~ 
    \citet{leitl2024}.}
    \label{fig:gridCongergence_MichelStadt1_u}
\end{figure}
\begin{figure}
    \centering
    \includegraphics[width=0.495\linewidth]{Figures/Geo2D_MichelStat_1.png}\\
    \vspace{0.10in}
    \includegraphics[width=0.495\linewidth]{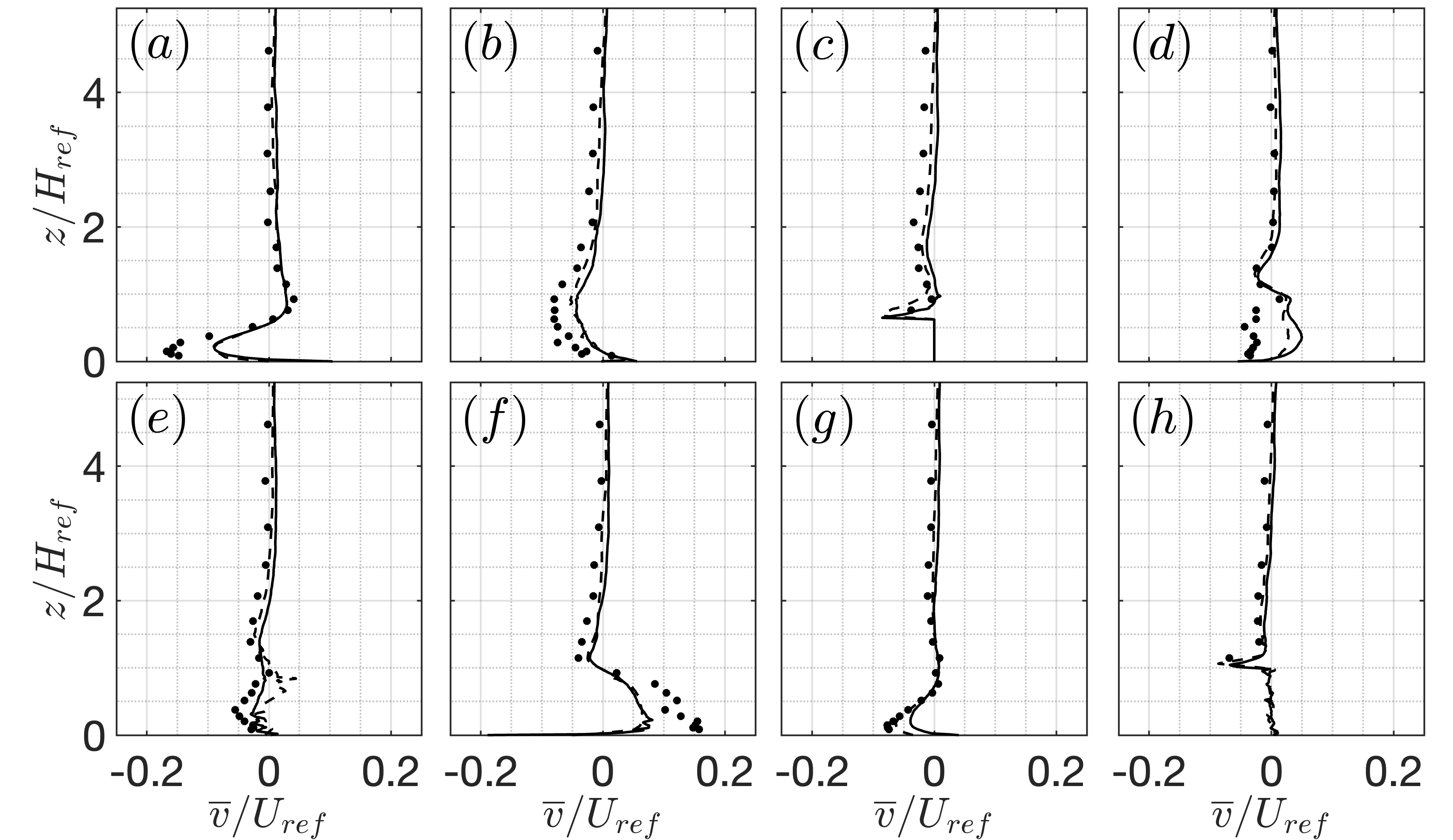}
     \caption{Case 2: comparison of spanwise mean-velocity profiles, $\overline{v}/U_{ref}$, with those of 
    experimental results at selected locations along $x-z$ plane at $y = 0$
    as shown in the top figure. 
    \color{black}\dashed~Coarse mesh; 
    \color{black}\solid~fine mesh; $\bullet$~ 
    \citet{leitl2024}.}
    \label{fig:gridCongergence_MichelStadt1_v}
\end{figure}
\begin{figure}
    \centering
    \includegraphics[width=0.495\linewidth]{Figures/Geo2D_MichelStat_1.png}\\
    \vspace{0.10in}
    \includegraphics[width=0.495\linewidth]{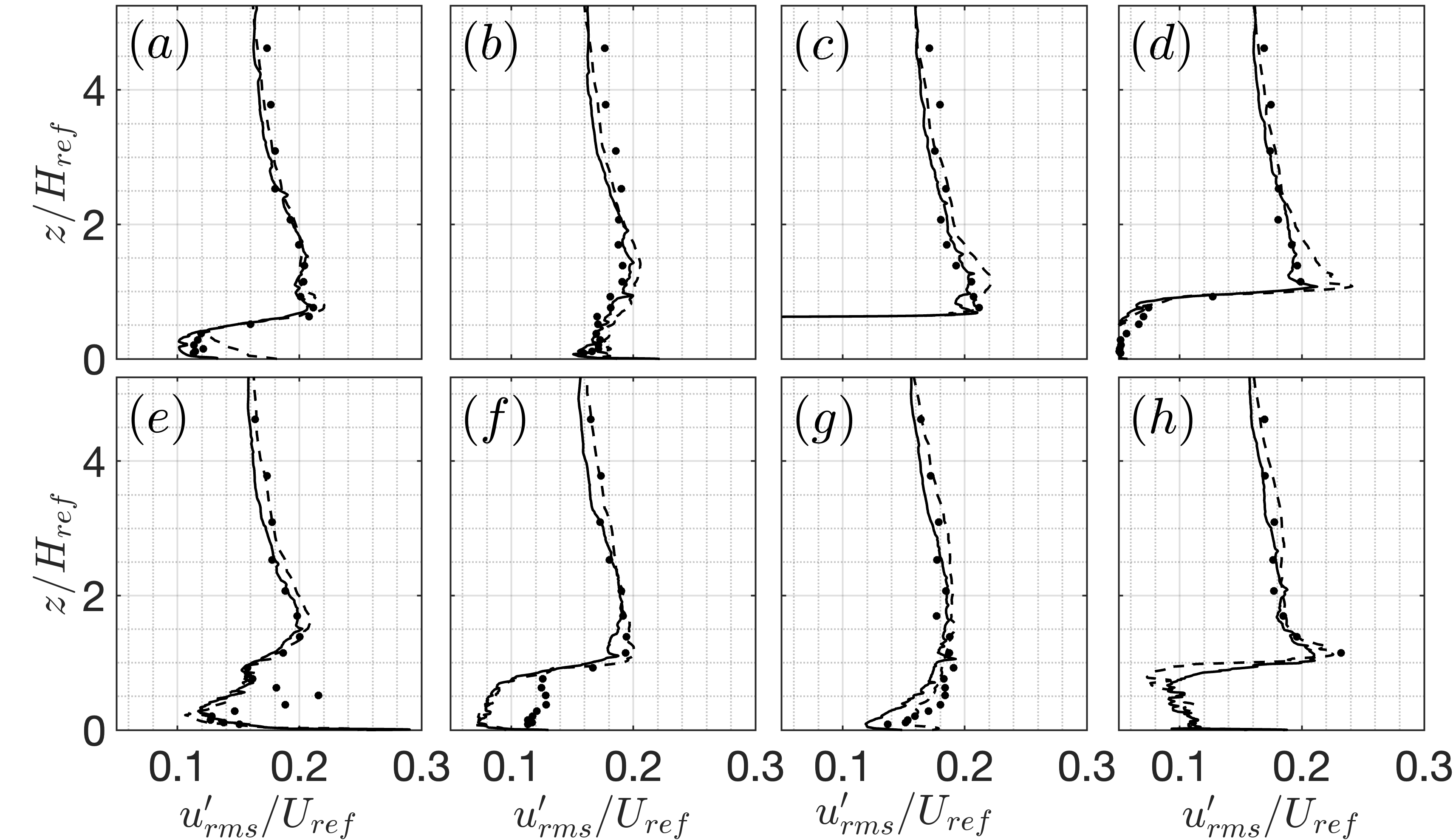}    
    \caption{Case 2: comparison of \textit{rms} of streamwise velocity fluctuations, $u_{rms}'/U_{ref}$, with those of 
    experimental results at selected locations along $x-z$ plane at $y = 0$
    as shown in the top figure. 
    \color{black}\dashed~Coarse mesh; 
    \color{black}\solid~fine mesh; $\bullet$~ 
    \citet{leitl2024}.}
    \label{fig:gridCongergence_MichelStadt1_RMS}
\end{figure}

For Case 2, experimental data include 
$\overline{u}/U_{ref}$, $\overline{v}/U_{ref}$, 
and 
% streamwise velocity fluctuations at various locations 
root-mean-square ($rms$) of the streamwise fluctuating 
component, $u'_{rms}$, at various locations. \ming{For clarity, only the comparison of these statistics in the $x-z$ plane 
along $y = 0$ is presented} in Figs. 
\ref{fig:gridCongergence_MichelStadt1_u} to 
\ref{fig:gridCongergence_MichelStadt1_RMS}, \ming{while the remaining data is provided in the Appendix \ref{sec: appedix}}. Similar to Case 1, good agreement is observed 
between the results of different mesh resolutions. 
$\overline{u}/U_{ref}$ profiles are accurately predicted at all 
locations, discrepancies are evident in $\overline{v}/U_{ref}$ 
profiles though. This is particularly the case near building corners and cross-road areas (e.g., Figs. \ref{fig:gridCongergence_MichelStadt1_v}$(a,~d,~f)$). Above rooftops, the flow becomes nearly one-dimensional, and the agreement with experimental data is excellent, reflecting the accuracy of the inflow conditions generated by the upstream precursor simulation.

As in Case 1, the simulations accurately capture the $u'_{rms}/U_{ref}$ peaks above building heights \ming{(see Fig. \ref{fig:gridCongergence_MichelStadt1_RMS})}. These peaks are indicative of shear layers formed near the rooftop level of the buildings. However, numerical predictions fail to fully resolve the fluctuations within the reverse flow regions between buildings (e.g., Figs. \ref{fig:gridCongergence_MichelStadt1_RMS}$(e,~f)$. This under-prediction may be attributed to the inherent challenges in modeling separated flows. Overall, the grid-convergence study confirms that the fine meshes with resolution down to $0.75m$ are capable of producing reliable results.

\subsection{Instantaneous flow field }\label{sec: instan}

\begin{figure}
    \centering
    \subfigure[]{
        \includegraphics[width=0.70\linewidth]{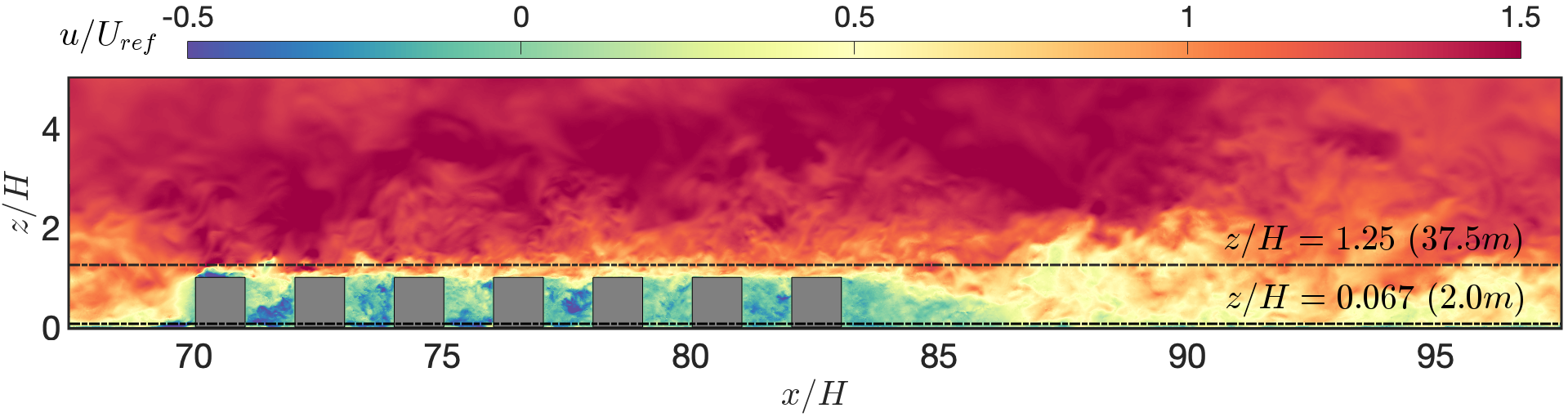}
        }
    \subfigure[]{
    \includegraphics[width=0.45\linewidth]{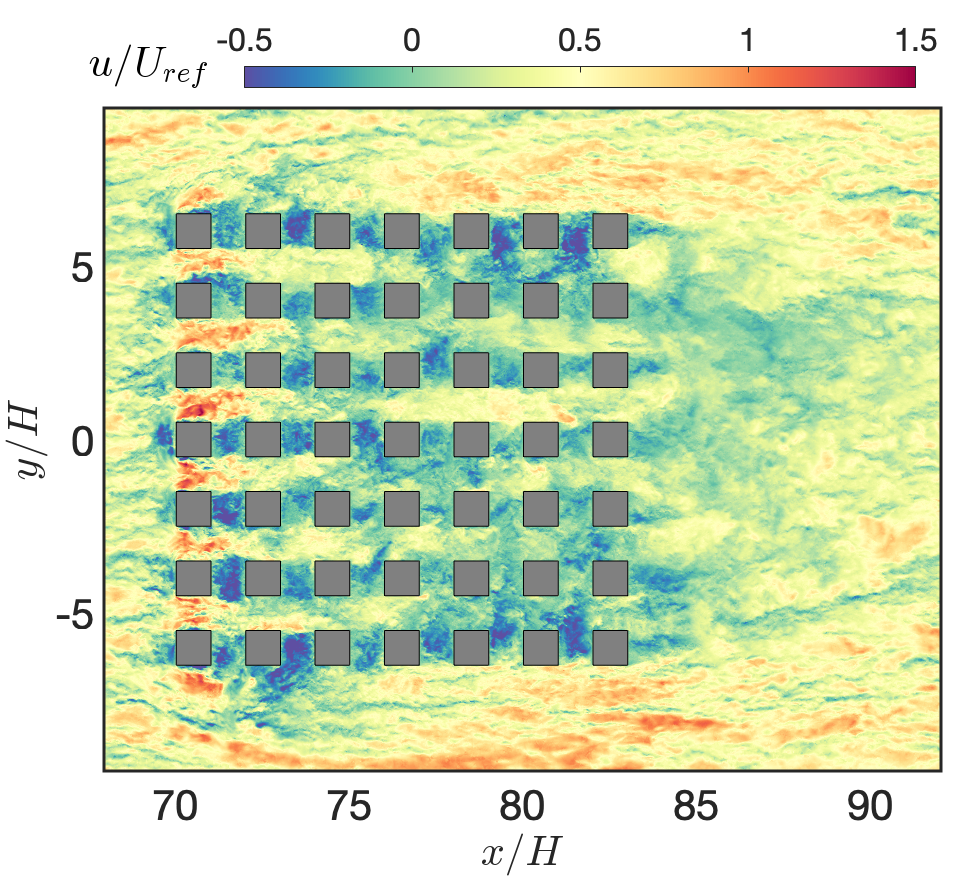}}
    \subfigure[]{    
    \includegraphics[width=0.45\linewidth]{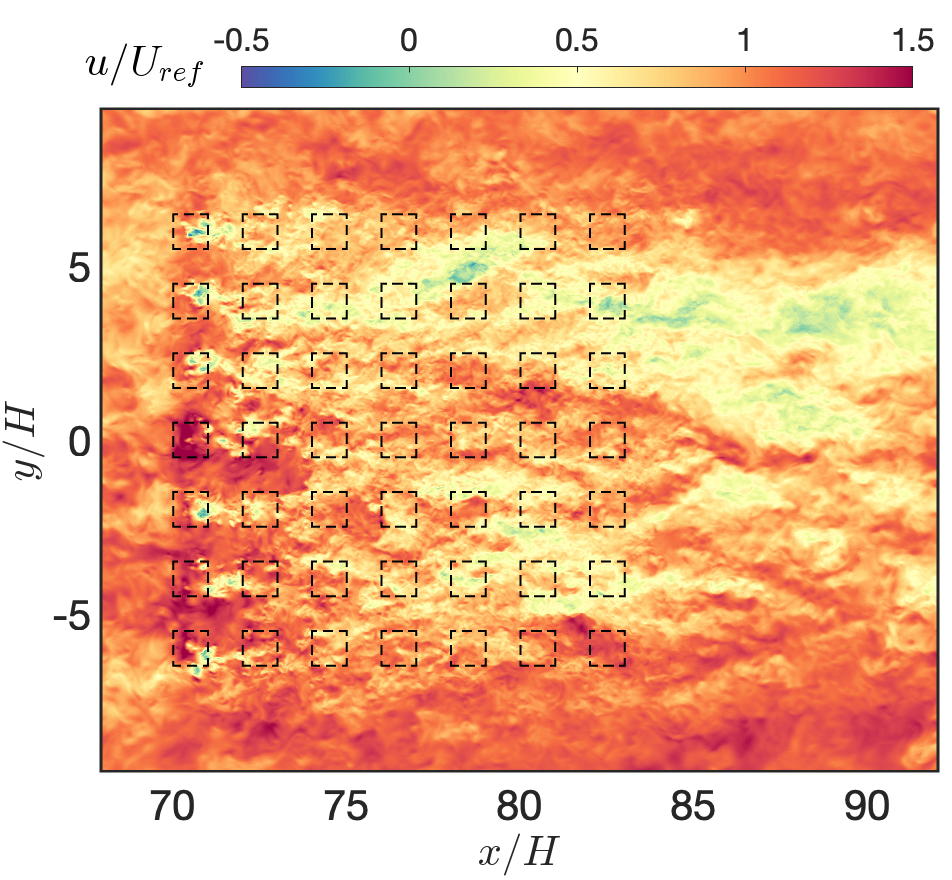}}
    \subfigure[]{    
    \includegraphics[width=0.45\linewidth]{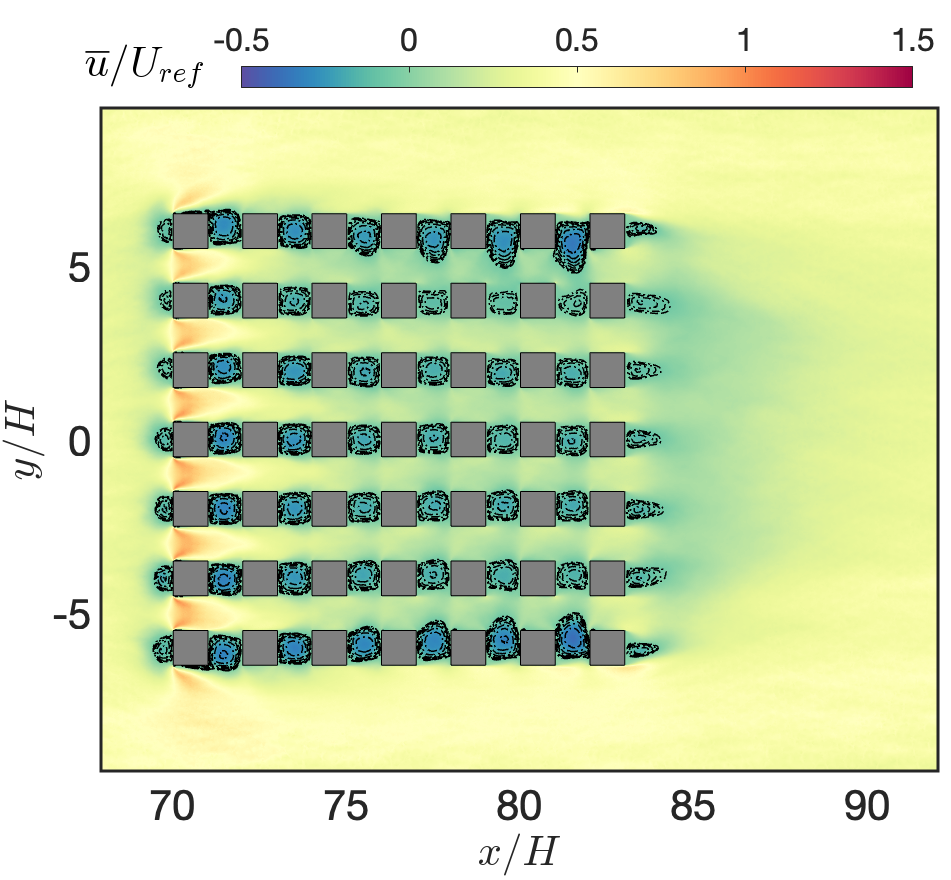}}
    \subfigure[]{    
    \includegraphics[width=0.45\linewidth]{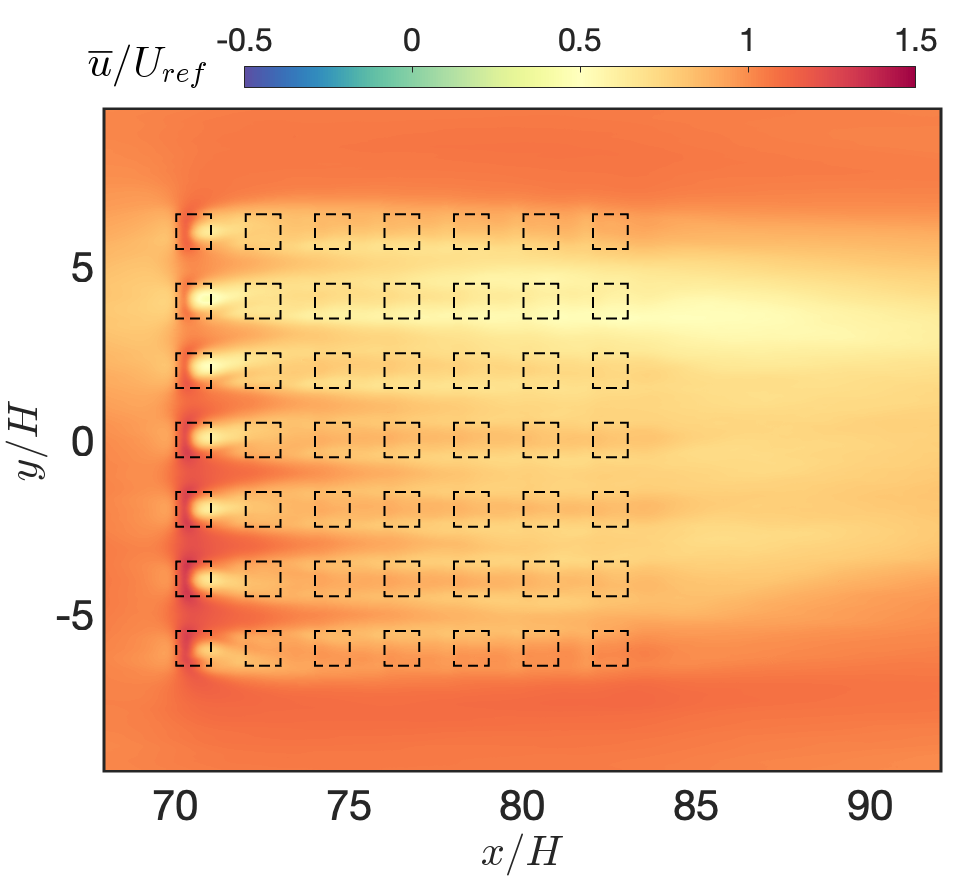}}
    \caption{Case 1: instantaneous streamwise velocity field, 
    $u/U_{ref}$, in the $(a)$ $x-z$ mid-plane ($y = 0$) and 
    $x-y$ plane at a distance of $(b)$ $z/H = 0.067~(2.0~\text{m})$ 
    and $(c)$ $z/H = 1.25~(37.5~\text{m})$ away from the ground. 
    Streamwise mean-velocity field, $\overline{u}/U_{ref}$, 
    at a distance of $(d)$ $z/H = 0.067~(2.0~\text{m})$ and $(e)$ 
    $z/H = 1.25~(37.5~\text{m})$ away from the ground. 
    \color{black}\chndash~ 10 log-spaced contours between $\overline{u}/U_{ref} = -0.50$ and $-0.05$;     \color{black}\dashed~locations of the square prisms.}
    \label{fig:cubicArray}            
\end{figure}

\begin{figure}
    \centering
    \subfigure[]{
    \includegraphics[width=0.98\linewidth]{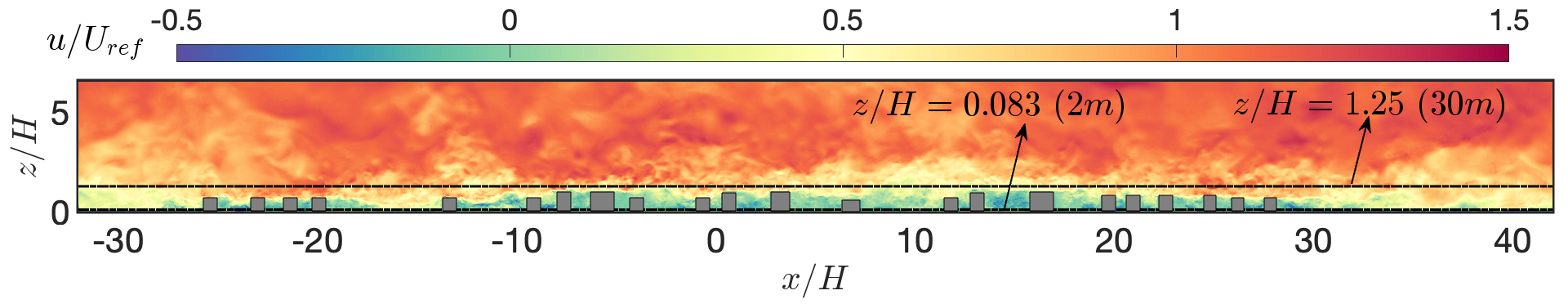}}
    \subfigure[]{
    \includegraphics[width=0.49\linewidth]{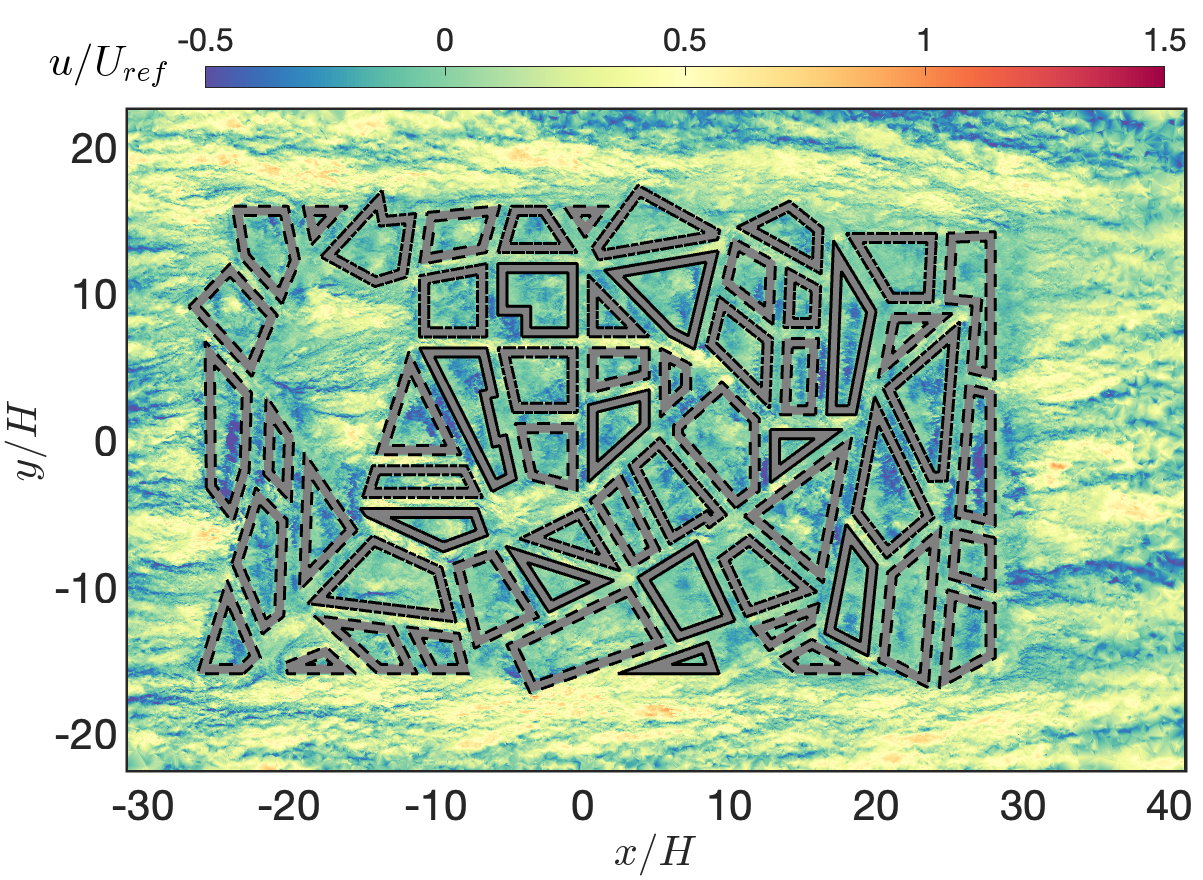}}
    \subfigure[]{    
    \includegraphics[width=0.49\linewidth]{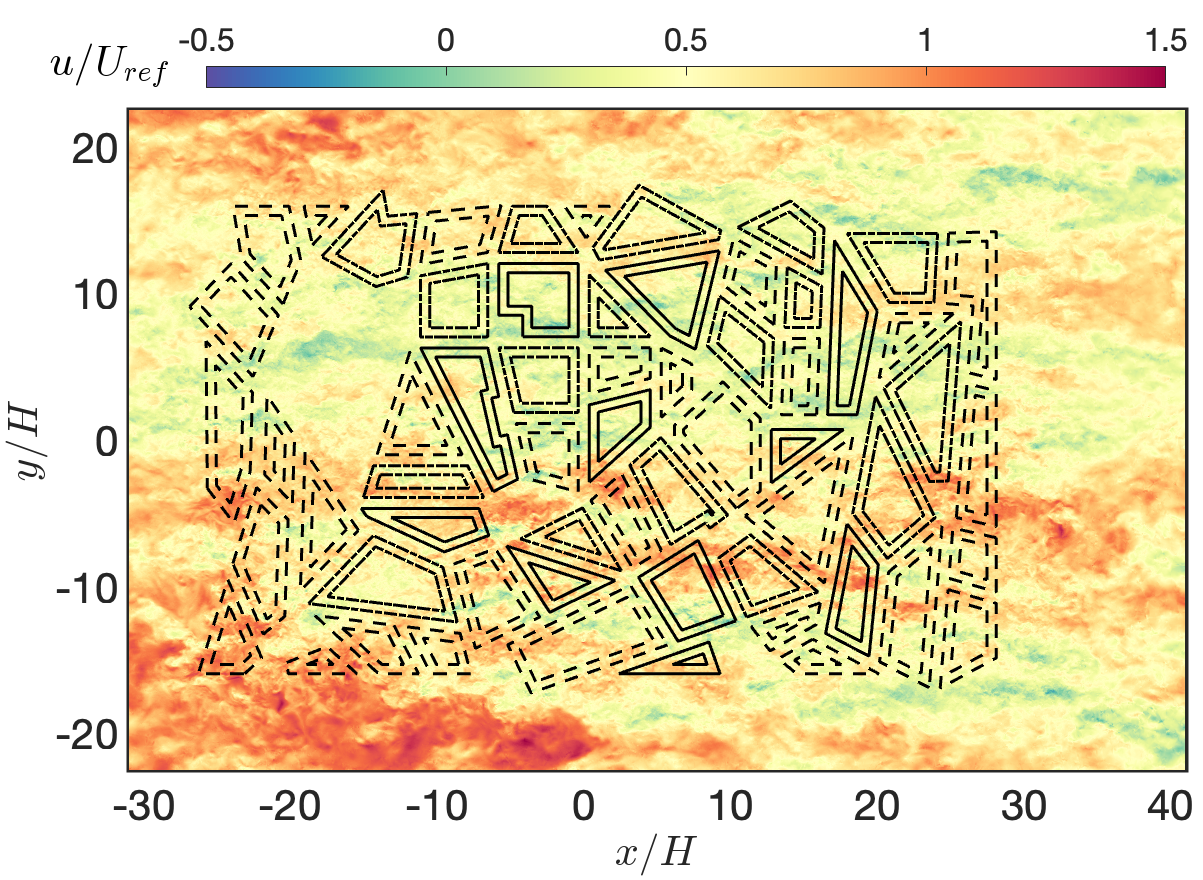}}
    \subfigure[]{
    \includegraphics[width=0.49\linewidth]{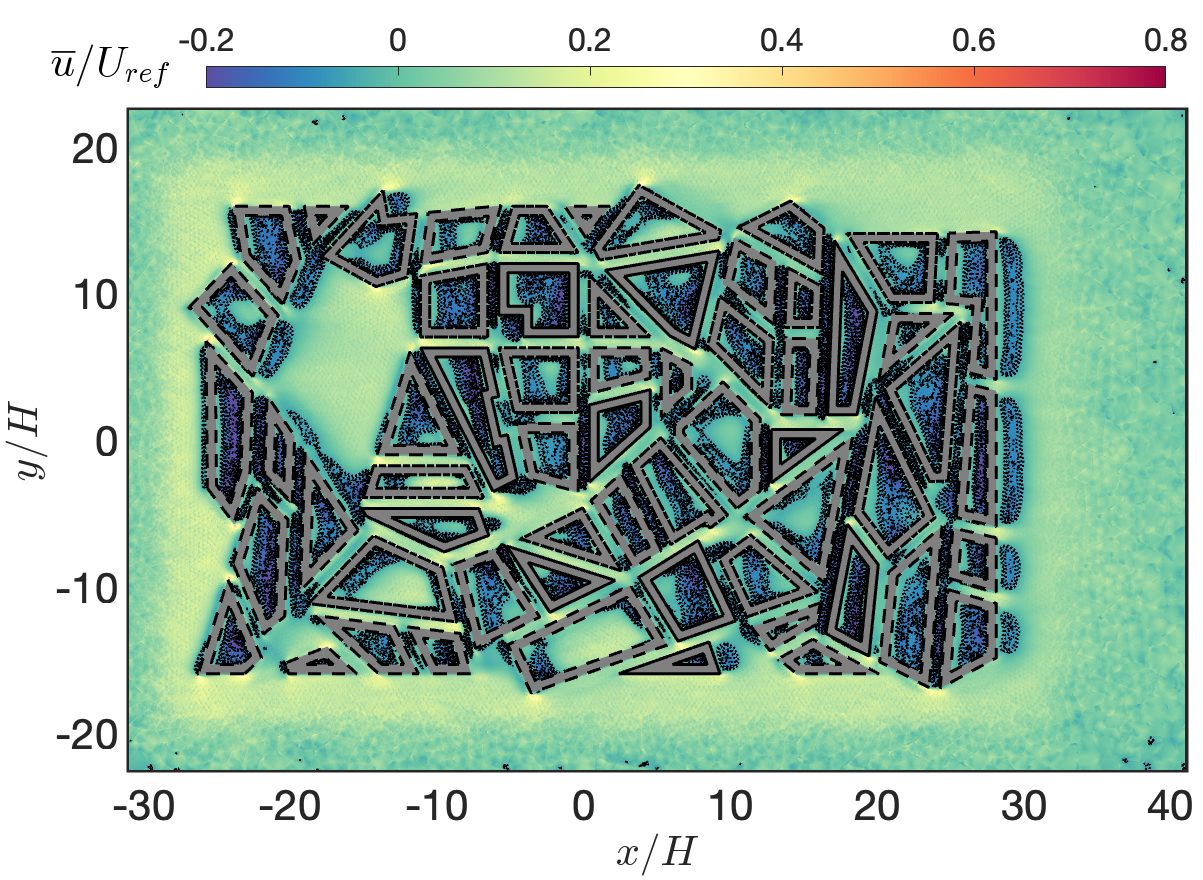}}
    \subfigure[]{    
    \includegraphics[width=0.49\linewidth]{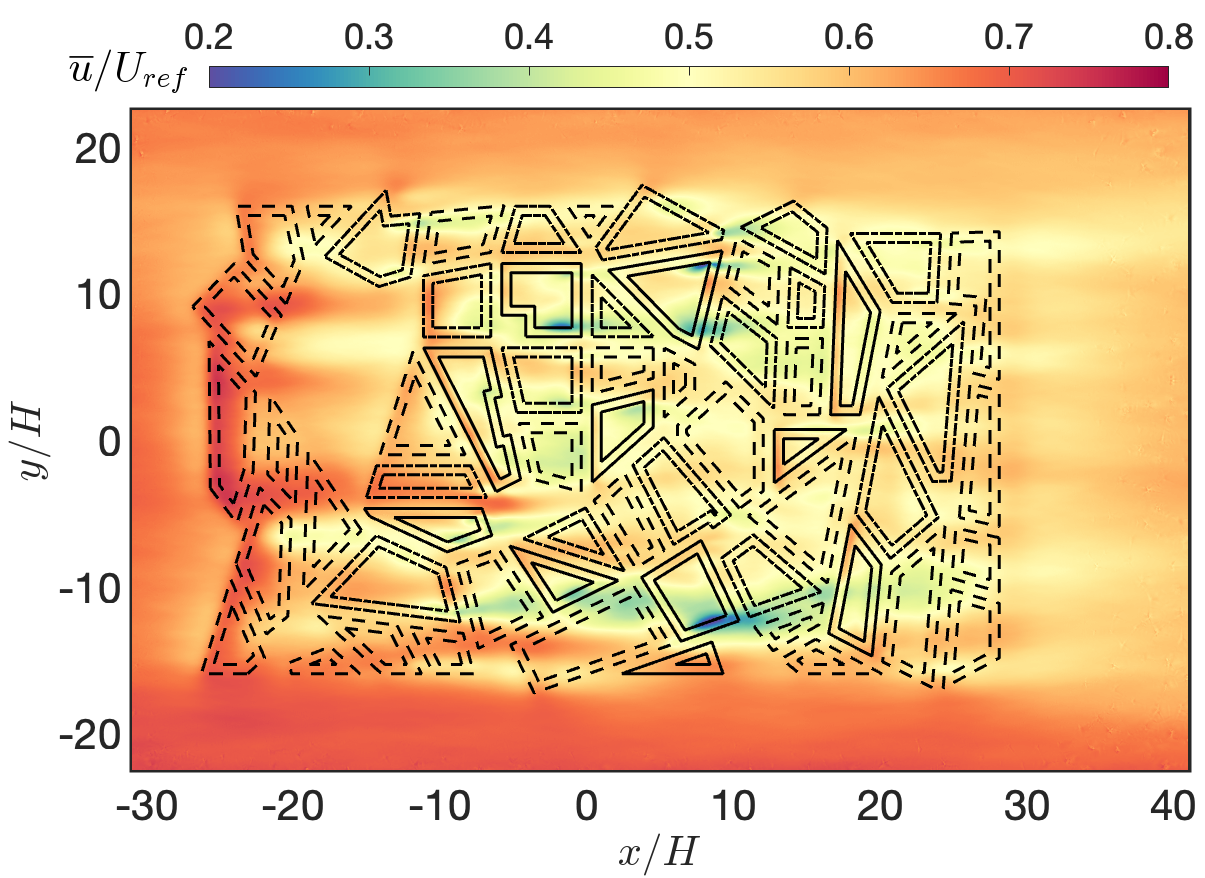}}
    \caption{Case 2: instantaneous streamwise velocity 
    field, $u/U_{ref}$, in the $(a)$ $x-z$ mid-plane ($y = 0$) and $x-y$ plane at a distance of 
    $(b)$ $z/H = 0.083~(2~\text{m})$ and $(c)$ $z/H = 1.25~(30~\text{m})$ away from the ground. Streamwise mean-velocity field, $\overline{u}/U_{ref}$, 
    at a distance of $(d)$ $z/H = 0.083~(2.0~\text{m})$ and $(e)$ $z/H = 1.25~(37.5~\text{m})$ away from the ground. 
    \color{black}\dotted~10 log-spaced contours between $\overline{u}/U_{ref} = -0.50$ and $-0.05$;
    \color{black}\dashed~$15~\text{m}-$buildings; 
    \color{black}\chndash~$18~\text{m}-$buildings; 
    \color{black}\solid~$24~\text{m}-$buildings.}
    \label{fig:MichelStadt}              
\end{figure}

Figures \ref{fig:cubicArray} and 
\ref{fig:MichelStadt} visualize the instantaneous streamwise velocity fields, $u/U_{ref}$, for both cases, 
providing an insight into the flow 
features at different planes. For Case 1, see Fig. 
\ref{fig:cubicArray}$(a)$, a large-scale vortex dominates the spaces between the square prisms, while an extensive wake region develops downstream of the trailing prism. For the current aspect ratio \vet{$S/H = 2$ \ming{(i.e, a ratio of the face-to-face distance between the prisms, $S$, to the height of a cube)}, according to \citet{oke1988street} a skimming flow configuration is expected. That is, the flow above the rooftop bypasses the canyon and a stable vortex is formed  within the canyon. }
At the pedestrian height $(z/H = 0.067~(2~\text{m}))$, see Fig. \ref{fig:cubicArray}$(b)$, the velocity field is characteristic of the reverse flow in the wake of each square prism. 
Due to the conservation of mass, the flow accelerates 
through the aisles as it enters the matrix. \vet{This results in the formation of a stable recirculation within the canyon between two consecutive buildings (not shown here) as predicted by \citet{oke1988street}}. This phenomenon is well pronounced in the mean-flow field as shown in Fig. \ref{fig:cubicArray}$(d)$ where a spanwise symmetry about the center plane ($y = 0$) and a quasi-periodicity in the $x-$direction within the array of prisms is also observed.
As the distance increases beyond the prism height $(z/H = 1.25~(37.5~\text{m}))$, see Fig. 
\ref{fig:cubicArray}$(c,~e)$, the impact of the square prisms on 
the flow field is primarily observed in the reduction of velocity magnitude within the matrix, where the reverse flow in the wake region is no 
longer perceptible. 

Figure \ref{fig:MichelStadt} presents the 
instantaneous flow structures of Case 2. One distinctive feature of a semi-idealized urban configuration is its geometrical complexity, shown in Fig. \ref{fig:MichelStadt}$(a)$. A large-scale vortex occupies compact inter-building spaces, while an extensive wake region forms in larger open areas. Close to the 
ground, at $z/H = 0.083~(2.0~\text{m})$ (see Fig. \ref{fig:MichelStadt}$(b)$), flow field within the enclosed building block is mostly dominated by the reverse flow. The wake region downstream of the trailing column of the buildings, see Fig. \ref{fig:MichelStadt}$(d)$, does not appear 
particularly extensive. This is likely a result of spatial inhomogeneity upstream. As the distance increases beyond the buildings and reaches $z/H = 1.25~(30~\text{m})$, 
the reverse flow is barely observed within and downstream of the high-rise buildings $(z = 24~\text{m})$. The impact of the low- and high-rise buildings on the flow field reduces as 
distance increases away from the ground as seen in Fig. \ref{fig:MichelStadt}$(c,~e)$. The specific distance at which the urban geometry considered
no longer influences the flow field can be evaluated by identifying the roughness sublayer\cite{cheng2002near}. This question will be addressed in detail via mean statistics  in further studies.

\section{Discussion}\label{sec: Discu}

\subsection{On the accuracy of the simulations}

In the previous section, we provided a detailed comparison of the results obtained at different locations with those measured in wind tunnel experiments, based on first- and second-order statistics. Although the analysis of these statistics provides critical insights by highlighting areas where predicted values differ \ming{from experimental measurements}, such as recirculating flows specially close to the ground, these discrepancies are inherently challenging to quantify.

In this sense, validation metrics can serve as a complementary quantitative assessment of the high-fidelity simulation results. These metrics, proposed by \citet{franke2007best}, are frequently used in existing studies (for instance, \citet{Vranckx2015, Yu2017,tolias2018large,Ding2022} etc.) and here they are applied to Case 2, "Michel-Stadt" BL3-3, as extensive measurements at more than 1000 locations are available in the database.

The validation metrics employed are the hit rate ($HR$) and the fraction of predictions within a factor of two of the observations ($FAC2$). The hit rate quantifies the proportion of predictions that deviate from the observed values by no more than a specified range. This range is determined either by a relative error threshold, denoted as $D$, or by an absolute difference limit, 
$W$. In essence, $HR$ evaluates the accuracy of predictions based on predefined tolerances for acceptable deviations. On the other hand, $FAC2$ measures the fraction of numerical predictions that lie within a factor of two of the corresponding experimental measurements. These metrics offer a straightforward and intuitive way to assess the agreement between simulated and observed data, emphasizing the capability of the model to stay within acceptable bounds of variation relative to experimental measurements.

\begin{equation}
    HR=\frac{1}{N}\sum\limits_i N_i, \hskip 3ex N_i=\begin{cases}
    1,& \text{if }  \frac{|P_i-O_i|}{O_i}\leq D \hskip 2ex \text{or} \hskip 2ex |P_i-O_i| \leq W  \\
    0,              & \text{otherwise}
\end{cases}
\end{equation}

\begin{equation}
    FAC2=\frac{1}{N}\sum\limits_i N_i, \hskip 3ex N_i=\begin{cases}
    1,& \text{if } 0.5\leq \frac{P_i}{O_i}\leq 2 \hskip 2ex \text{or} \hskip 2ex (O_i\leq W  \& P_i\leq W)\\
    0,              & \text{otherwise}
\end{cases}
\end{equation}

In this study, $P_i$ and $O_i$ denote the normalized numerical predictions and experimental values, respectively. The parameter $D$ represents the allowable relative deviation, which accounts for the reproducibility of wind tunnel measurements in built environments and the permissible inaccuracies in simulation results. Based on the recommendations of \citet{Schatzmann2010}, $D$ is set to $0.25$.  
The parameter $W$ defines the allowable absolute deviation, reflecting uncertainties caused by interpolation and measurement repeatability. This value is case-dependent, with reported ranges in the literature from $0.01$ to $0.1$ \cite{Franke2012,Kakosimos2013,Yu2017,tolias2018large}. In this study, $W$ is assigned values of $0.0165$ for streamwise velocity and $0.0288$ for spanwise velocity, following the work of \citet{tolias2018large}. These values align with the experimental measurements conducted for Case 2.  The results for these parameters are summarized in Table \ref{table:qualitative}.

\begin{table}
\caption{Summary of validation metrics. Quality criteria used are: $FAC2\geq 0.3$ and $HR \geq 0.66$.}\label{table:qualitative}
\begin{tabular}{l c c c c c c}
\hline
height[m]& &\multicolumn{2}{c}{$\overline{u}/U_{ref}$}& &\multicolumn{2}{c}{$\overline{v}/U_{ref}$}\\
&& FAC2 & HR && FAC2 & HR \\\hline
  2   && 0.73&0.41&&0.83&0.38 \\
  9 && 0.86&0.46&&0.88&0.68\\
  18 &&0.91&0.69&&0.94&0.81\\
  27 && 1.00&0.99&&0.98&0.96\\
  30 && 1.00&1.00&&0.99&0.97\\
  Total& & 0.86&0.64&&0.91&0.70\\
\hline
\end{tabular}
\end{table}

\begin{figure}
    \centering  
    \subfigure[]{    \includegraphics[width=0.48\linewidth]{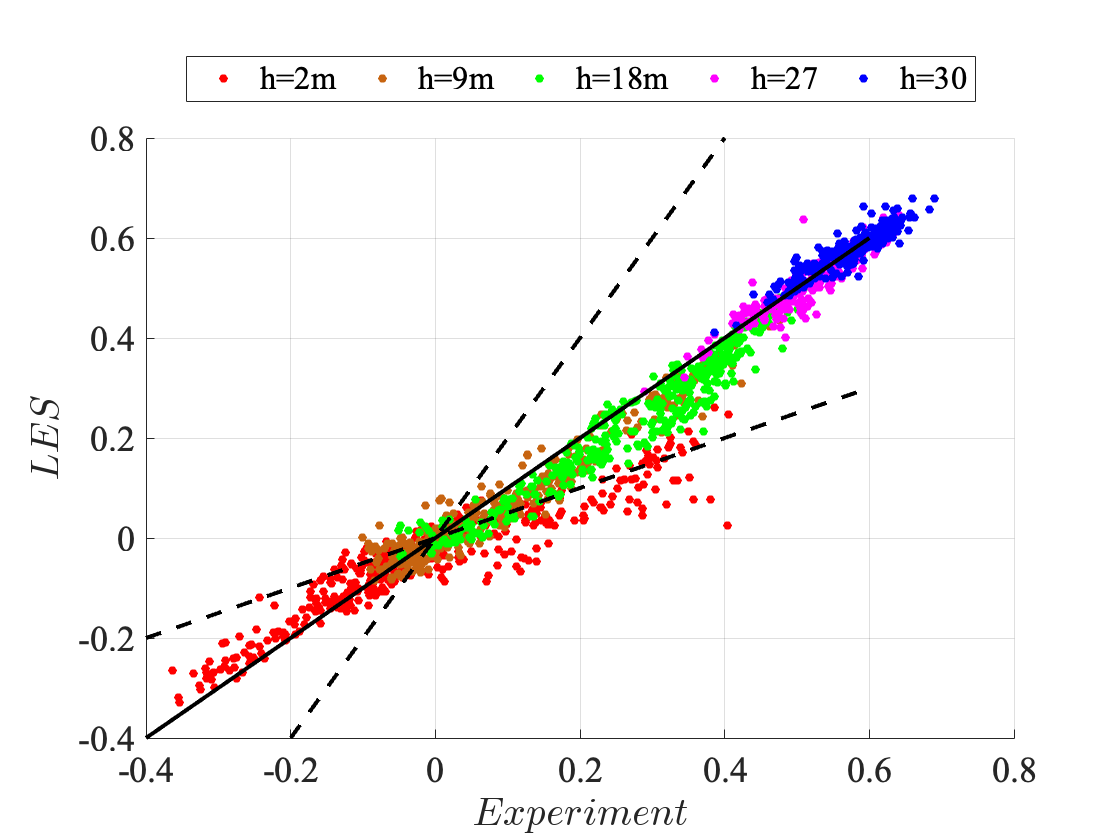}}
     \subfigure[]{    \includegraphics[width=0.48\linewidth]{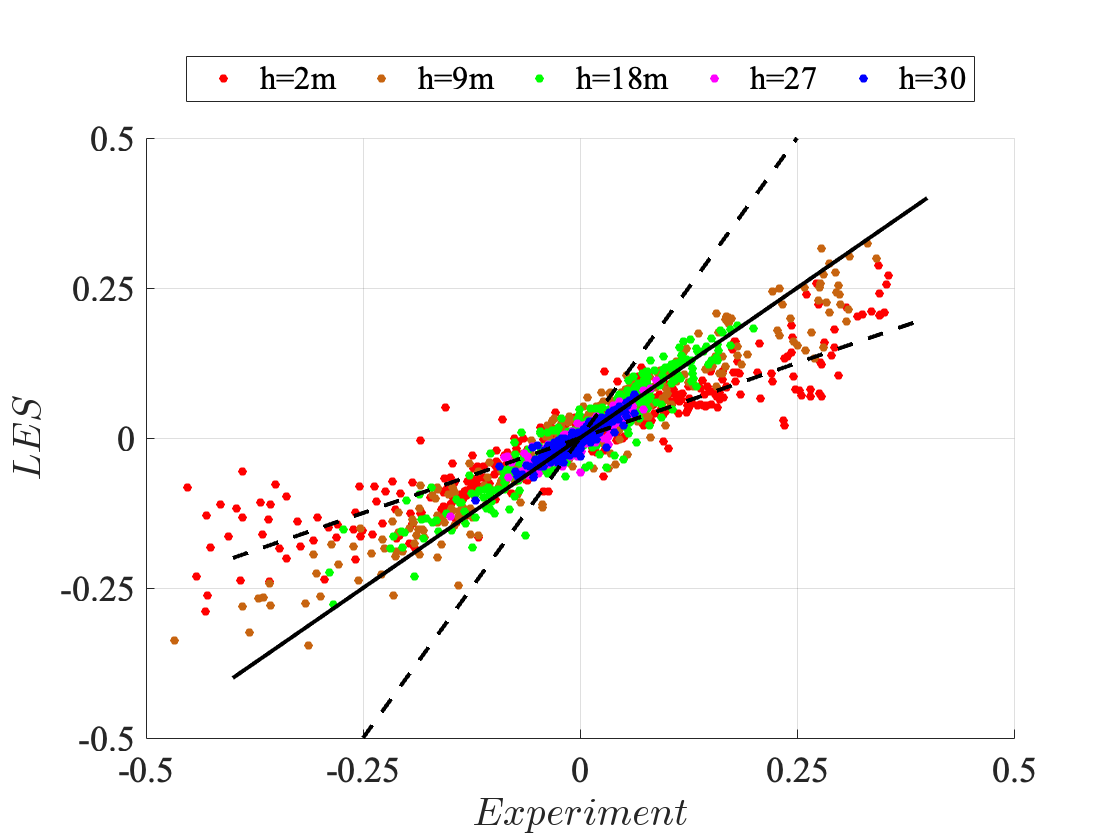}}
      \subfigure[]{    \includegraphics[width=0.48\linewidth]{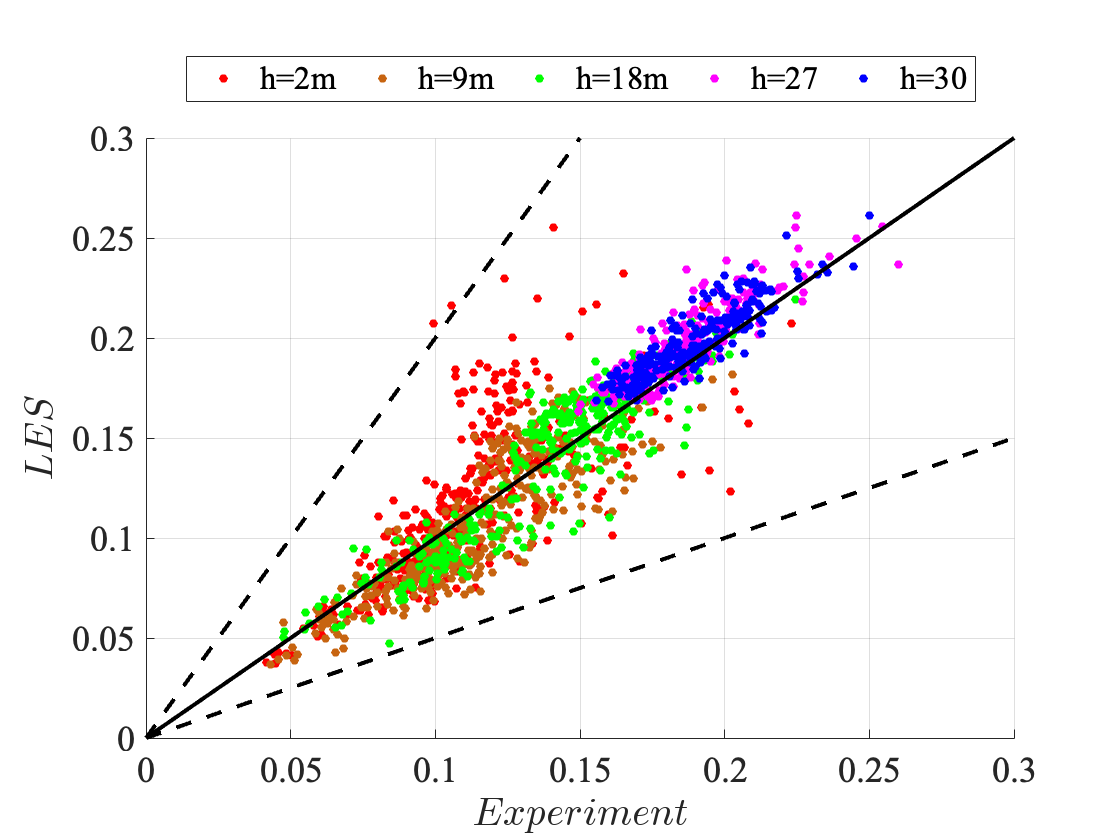}}
       \subfigure[]{    \includegraphics[width=0.48\linewidth]{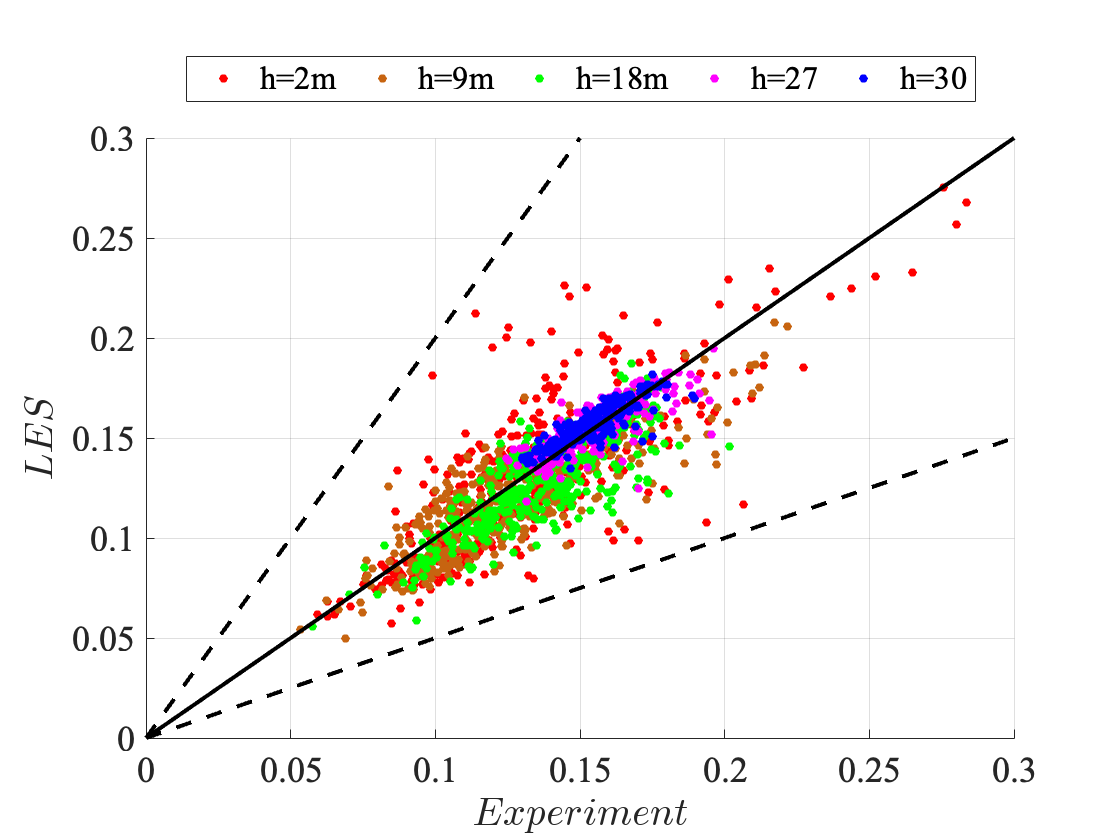}}
    \caption{Scatter plots comparing LES and experimental measurements for velocity components and their $rms$ values at different heights. $(a)$ Streamwise velocity component, $u/U_{ref}$, $(b)$ spanwise velocity component, $v/U_{ref}$, $(c)$ $rms$ of streamwise velocity, and $(d)$ $rms$ of spanwise velocity. \color{black}\solid~ the 1:1 agreement; \color{black}\dashed~ a factor of two difference.}
    \label{scatter}
   \end{figure} 
 
Table \ref{table:qualitative} presents the validation metrics for data at different heights, along with the overall values for both streamwise and spanwise velocity components. 
Scatter plots comparing high-fidelity simulation results with experimental data, focusing on $u/U_{ref}$ and $v/U_{ref}$ and their fluctuations, can also provide a valuable metric for assessing the confidence level of the numerical measurements. These plots, shown in Fig. \ref{scatter}, include the 1-to-1 line along with the 1-to-2 and 2-to-1 reference lines to better contextualize the level of agreement.

The comparison between LES and wind tunnel experiments shows a generally good agreement in $u/U_{ref}$, with higher accuracy at greater heights. The $FAC2$ and $HR$ values show that LES performs well, particularly above $18 m$, where both metrics exceed 0.9. At lower heights, discrepancies become more evident, though $FAC2$ remains above 0.7, surpassing the $0.3$ threshold  considered acceptable for urban flow simulations, as suggested by \citet{Hanna2012}. Similarly, $HR$ values are mostly above 0.66, the admissible limit for this quantity\cite{VDI2000}. These results suggest that LES provides a higher level of accuracy than typically required for urban flow modeling, particularly in capturing the mean flow field.

The scatter plots further illustrate this agreement (see Fig. \ref{scatter}$(a,~b)$). The LES predictions align well with the experimental data, with the majority of points clustering around the identity line (solid black line), especially for $u/U_{ref}$ (Fig. \ref{scatter}$(a)$). However, for $v/U_{ref}$, the scatter is noticeably larger (Fig. \ref{scatter}$(b)$), indicating a higher degree of variability between LES and experimental results. This discrepancy is consistent with the lower $HR$ values for $\overline{v}/U_{ref}$, suggesting that LES predicts larger lateral velocities than observed in the wind tunnel experiments. Given that LES employs a low-dissipation numerical scheme and achieves a high resolution of 0.75m, it is unlikely that numerical errors are the solely responsible for this difference. The wind tunnel measurements  may also underestimate lateral velocity fluctuations due to physical constraints, such as probe alignment issues, wind tunnel sidewall effects, or Reynolds number mismatches. These effects can also be added to the fact that wind tunnel turbulence generation using spires and roughness elements may not fully replicate full-scale turbulence structures, particularly in the lateral direction. Consequently, the summation of numerical uncertainties along with those in wind tunnel might be the reason of the larger deviations observed in this velocity component, compared to those in streamwise direction.

The $rms$ velocity comparisons (Fig. \ref{scatter}$(c,~d)$) provide additional insights into the representation of turbulence intensity. The LES captures the fluctuations in the streamwise component  relatively well,  especially at mid and upper heights (green, blue, magenta points). However, at lower heights (red points, $h = 2~\text{m}$), there is a tendency for LES to slightly deviate form the $rms$ values of the experiments. The velocity fluctuations  for the lateral component tend to have a larger scattering along the identity line, with a slight underestimation compared to the experiments, especially at low and mid-heights. As discussed before, deviations at lower heights might in general be attributed to insufficient grid resolution necessary to represent small-scale turbulence near walls, but also to some limitations in the experimental setup.  Given that urban flow turbulence is highly anisotropic, with complex wake interactions and shear layers, the Reynolds number mismatch between LES (full-scale) and the wind tunnel (1:200 scale) may lead to differences in how turbulence develops and is measured. 
While $FAC2$ and $HR$ are useful for assessing LES accuracy, these metrics have limitations and should not be the sole basis for validation. FAC2 does not account for systematic biases; it simply evaluates whether LES predictions fall within a factor of two of experimental values. HR is more restrictive, but it still does not fully capture whether the LES reproduces the correct turbulence structures, spectral content, or flow dynamics. Given these limitations, energy spectra analysis is essential to assess whether LES captures the correct distribution of $TKE$ across scales. This is particularly important in urban environments, where multi-scale turbulence, anisotropy, and wake interactions play a crucial role in flow development.

\begin{table}
  \begin{center}
   \caption{Locations of the numerical probes for the calculations of pre-multiplied energy spectra of the streamwise fluctuating velocity component.}
  \label{tab:spectra}
\def~{\hphantom{0}}
  \begin{tabular}{ c c c c c c c }
      \hline
      Probes & $x(m)$ & $y(m)$ & $z(m)$ & $x/H$ & $y/H$ & $z/H$\\[3pt]
      \hline
P1	& -57.17& -31.99 & 2.0 & -2.38 & -1.33 & 0.083\\
P2	& 3.85	& -88.34 & 2.0 & 0.16  & -3.68 & 0.083\\
P3	&-3.19  & 102.76 & 2.0 & -0.13 &  4.28 & 0.083\\
P4	&-108.56& -22.48 & 2.0 & -4.52 & -0.94 & 0.083\\
      \hline
P5	& -112.5& -135 & 30.13 & -4.69 & -5.63 & 1.26\\
P6	& -157.5& -45  & 30.13 & -6.56 & -1.88 & 1.26\\
P7	& -135  &  45  & 30.13 & -5.63 &  1.88 & 1.26\\
P8	& -67.5 &  90  & 30.13 & -2.81 &  3.75 & 1.26\\
      \hline
  \end{tabular} 
  \end{center}
\end{table}

\begin{figure}
    \centering
    \includegraphics[width=0.60\linewidth]{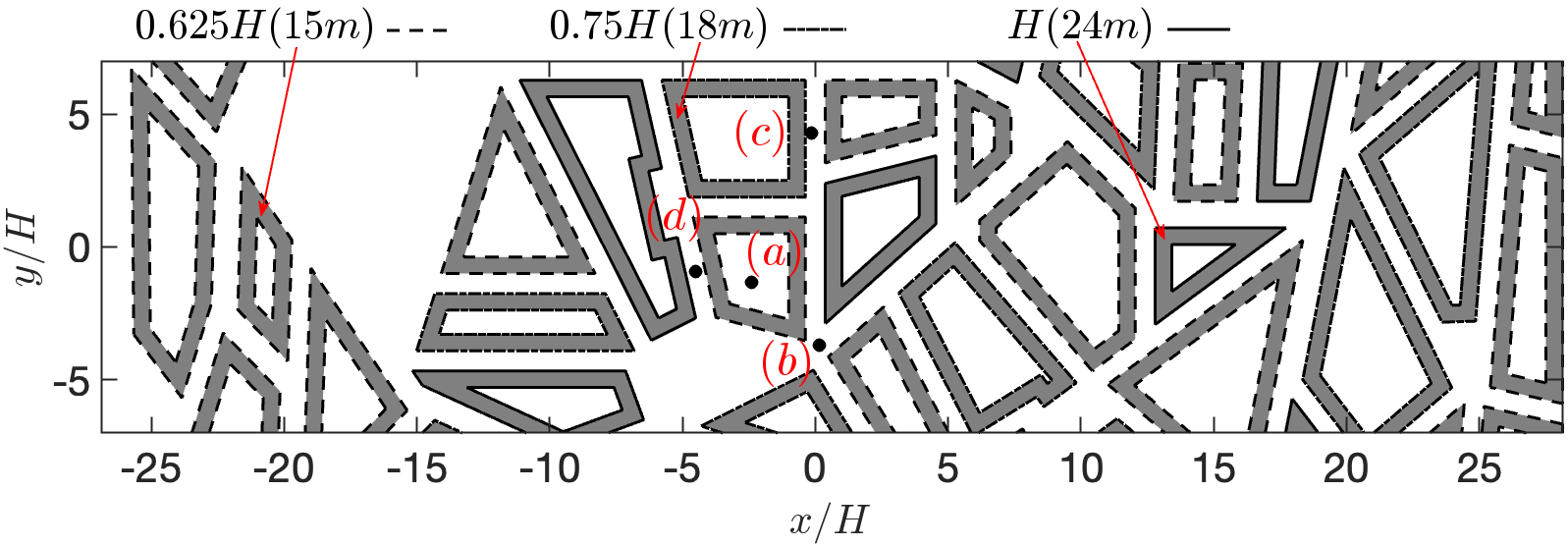}\\    
    \subfigure[]{
    \includegraphics[width=0.48\linewidth]{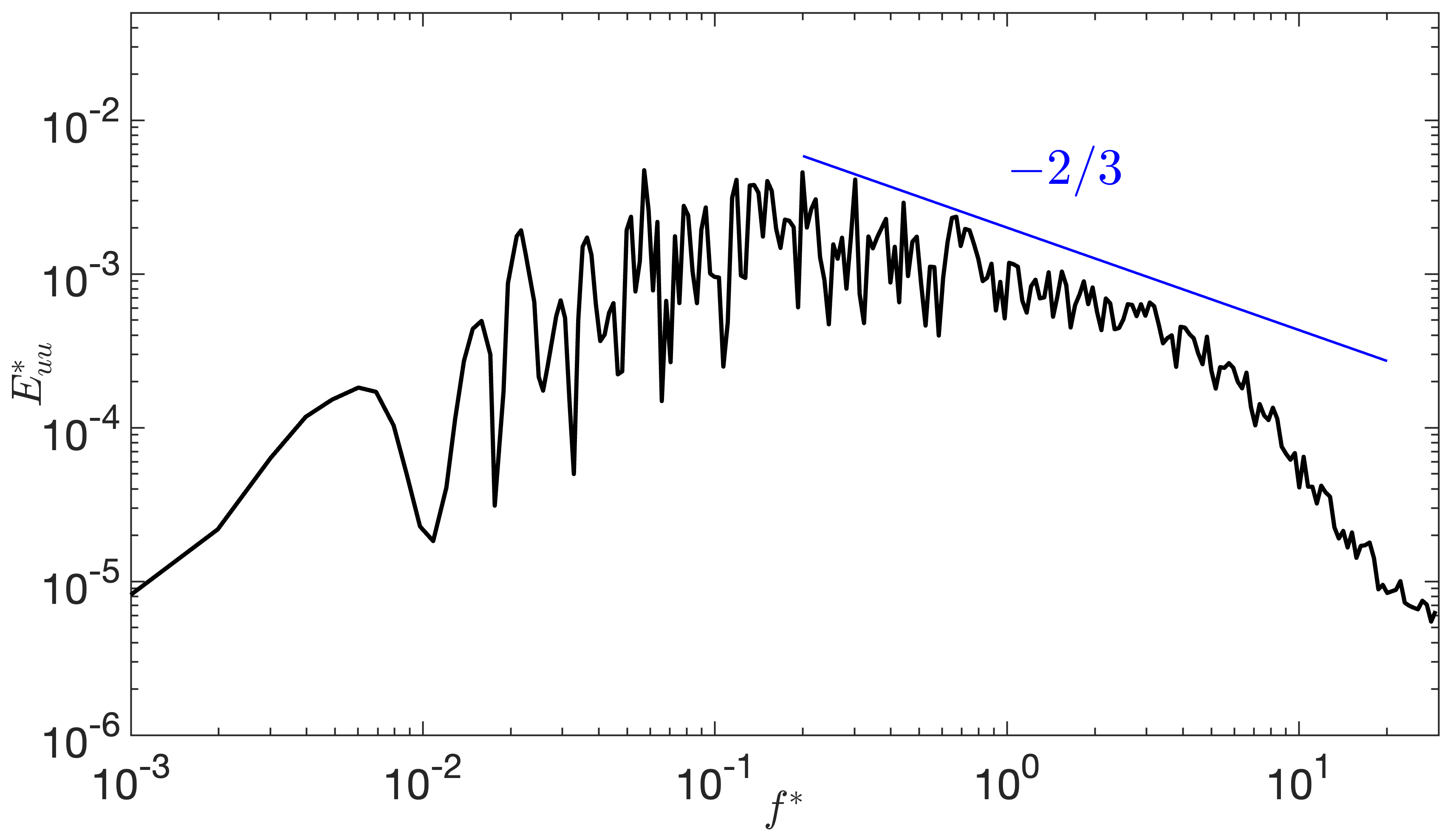}
    }
    \subfigure[]{
    \includegraphics[width=0.48\linewidth]{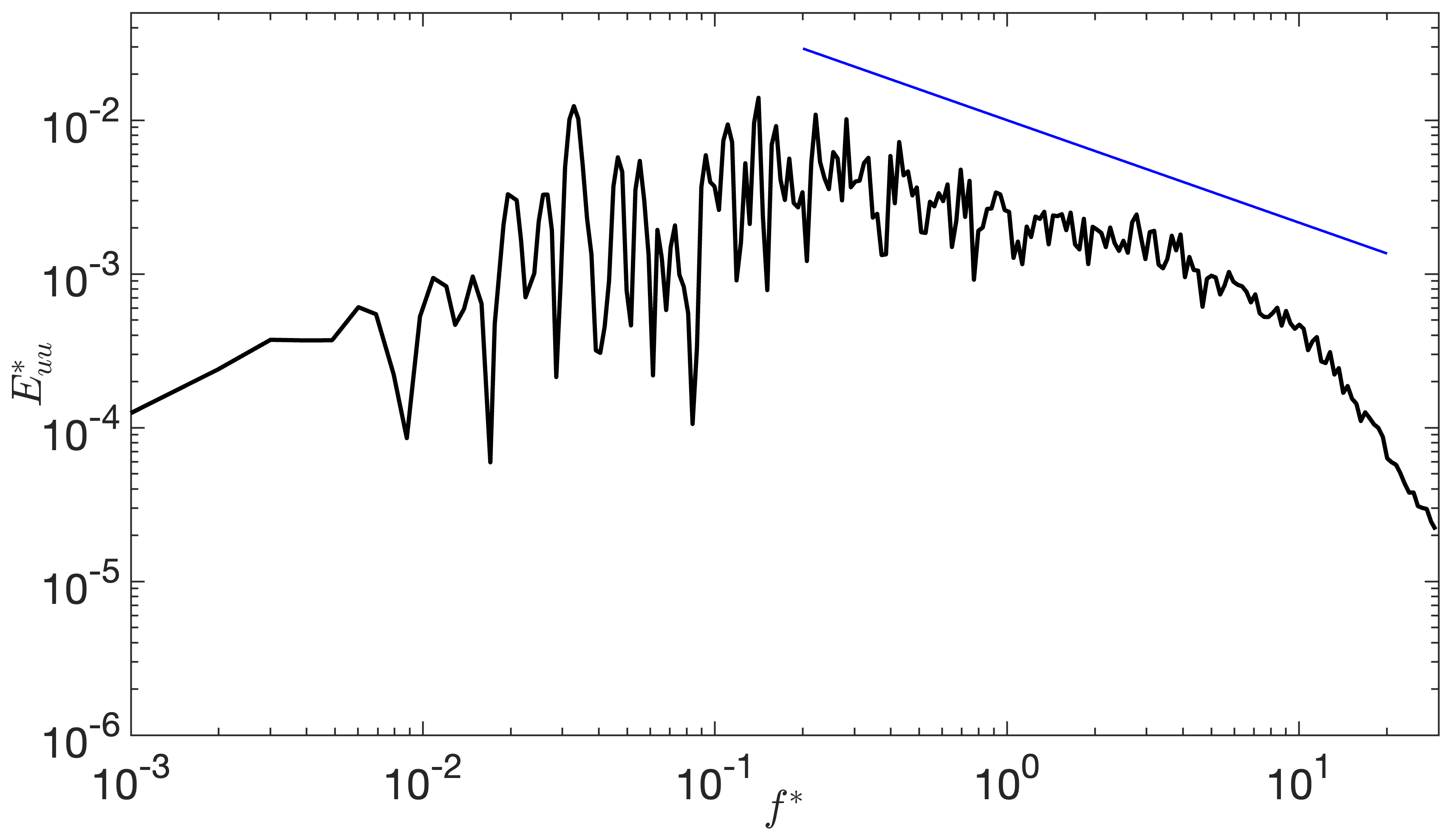}
    }\\
    \subfigure[]{
    \includegraphics[width=0.48\linewidth]{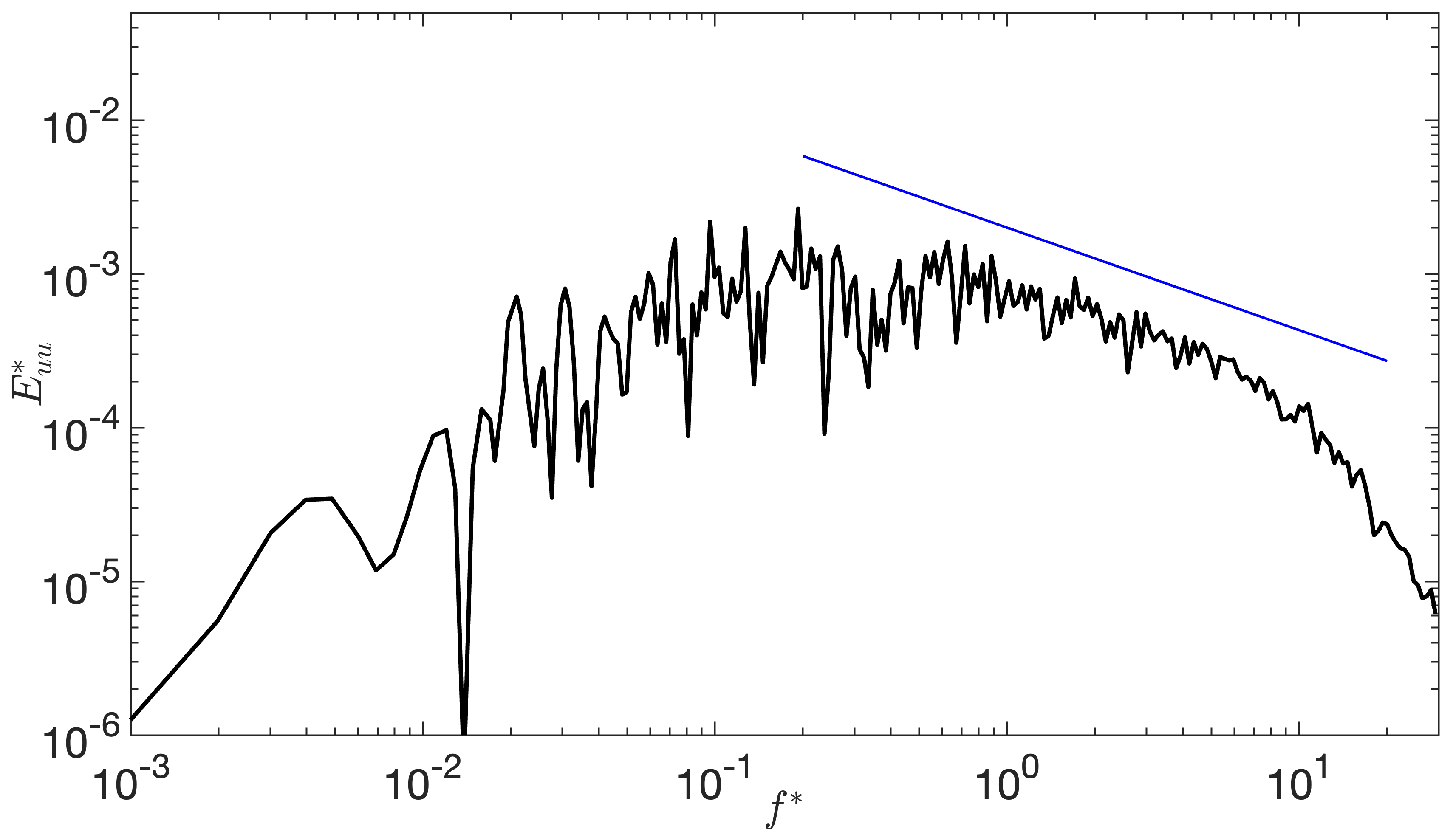}
    }
    \subfigure[]{
    \includegraphics[width=0.48\linewidth]{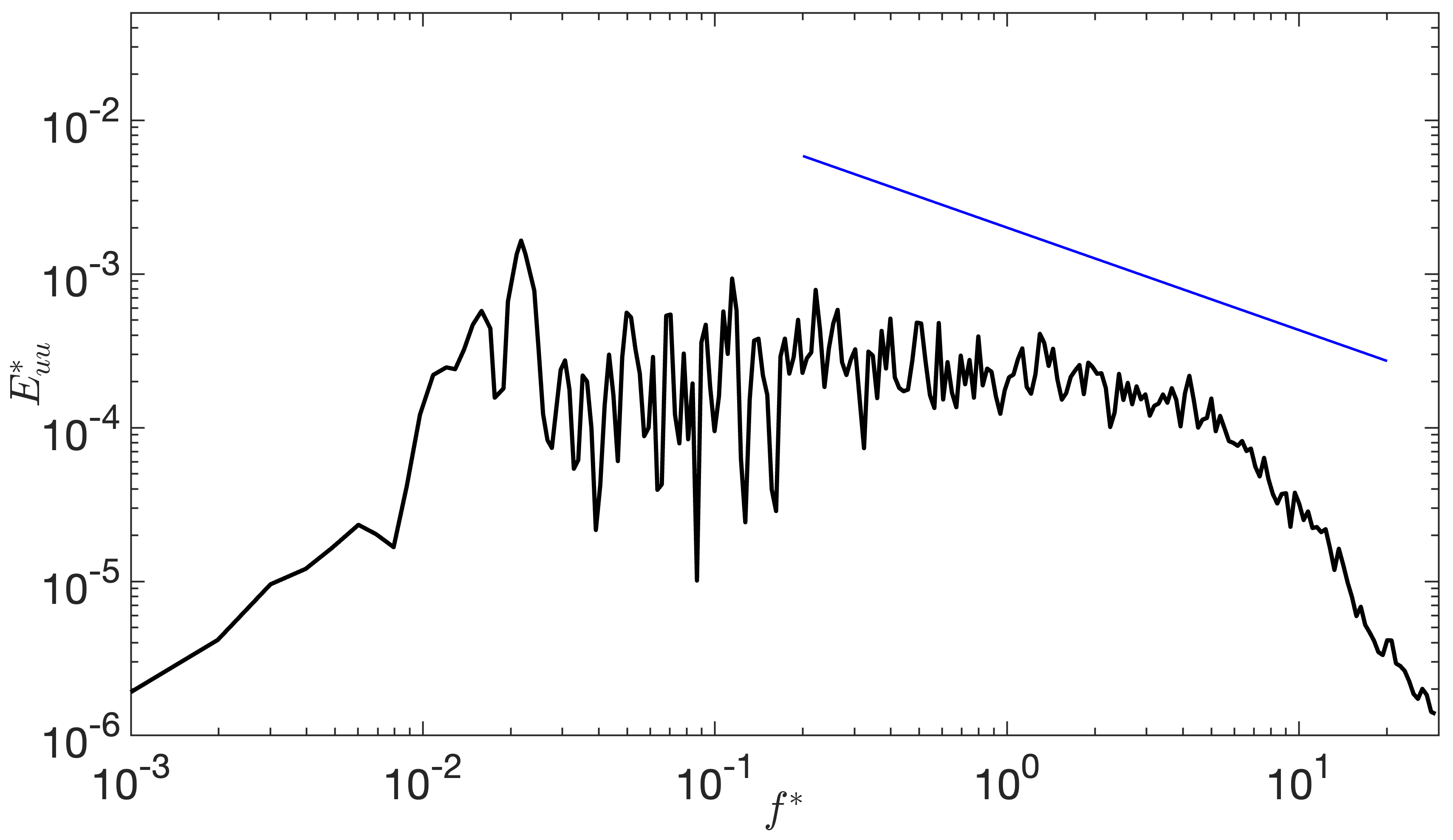}
    }
    \caption{
    Pre-multiplied energy spectra of the streamwise fluctuating velocity component, $E_{uu}^* =\:E_{uu}/\sigma_u$; $f^* = f\:H/U_{ref}$. Velocities were measured at $z/H = 0.083~(2~\text{m})$ at locations as shown in the top figure (see also Table \ref{tab:spectra} for the locations of the numerical probes). The blue line denotes the ~$-2/3$ power law.}
    \label{fig:spectra_1}     
\end{figure}

\begin{figure}
    \centering
    \includegraphics[width=0.60\linewidth]{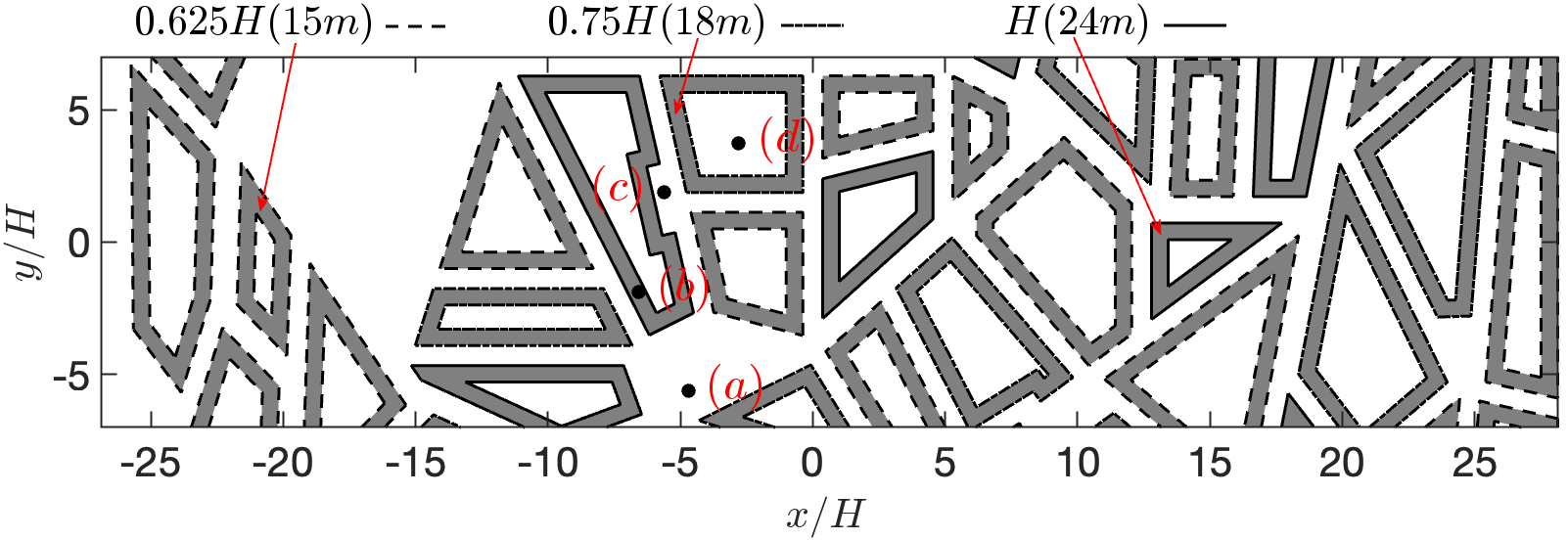}\\
    \subfigure[]{
    \includegraphics[width=0.48\linewidth]{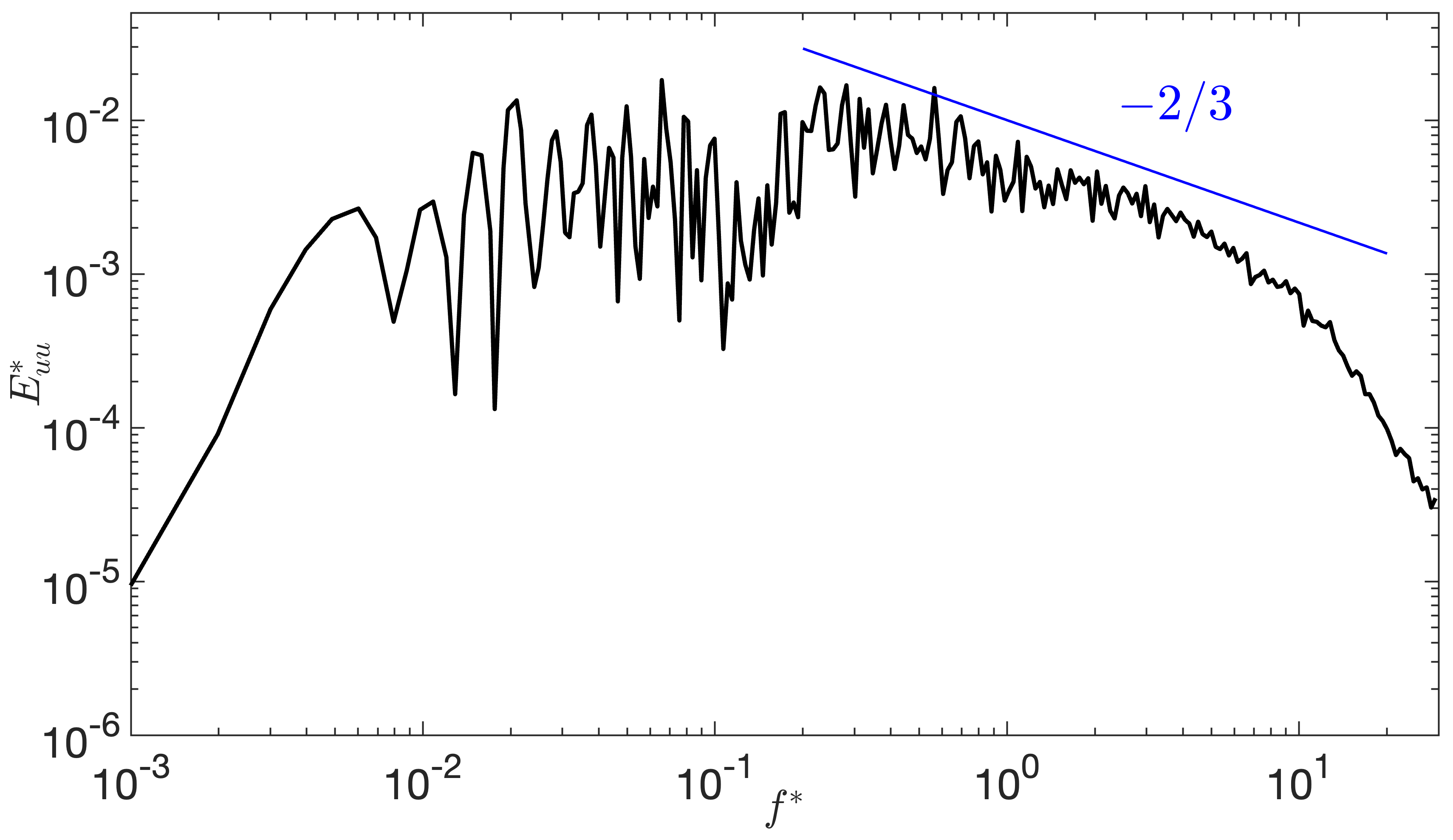}
    }
    \subfigure[]{
    \includegraphics[width=0.48\linewidth]{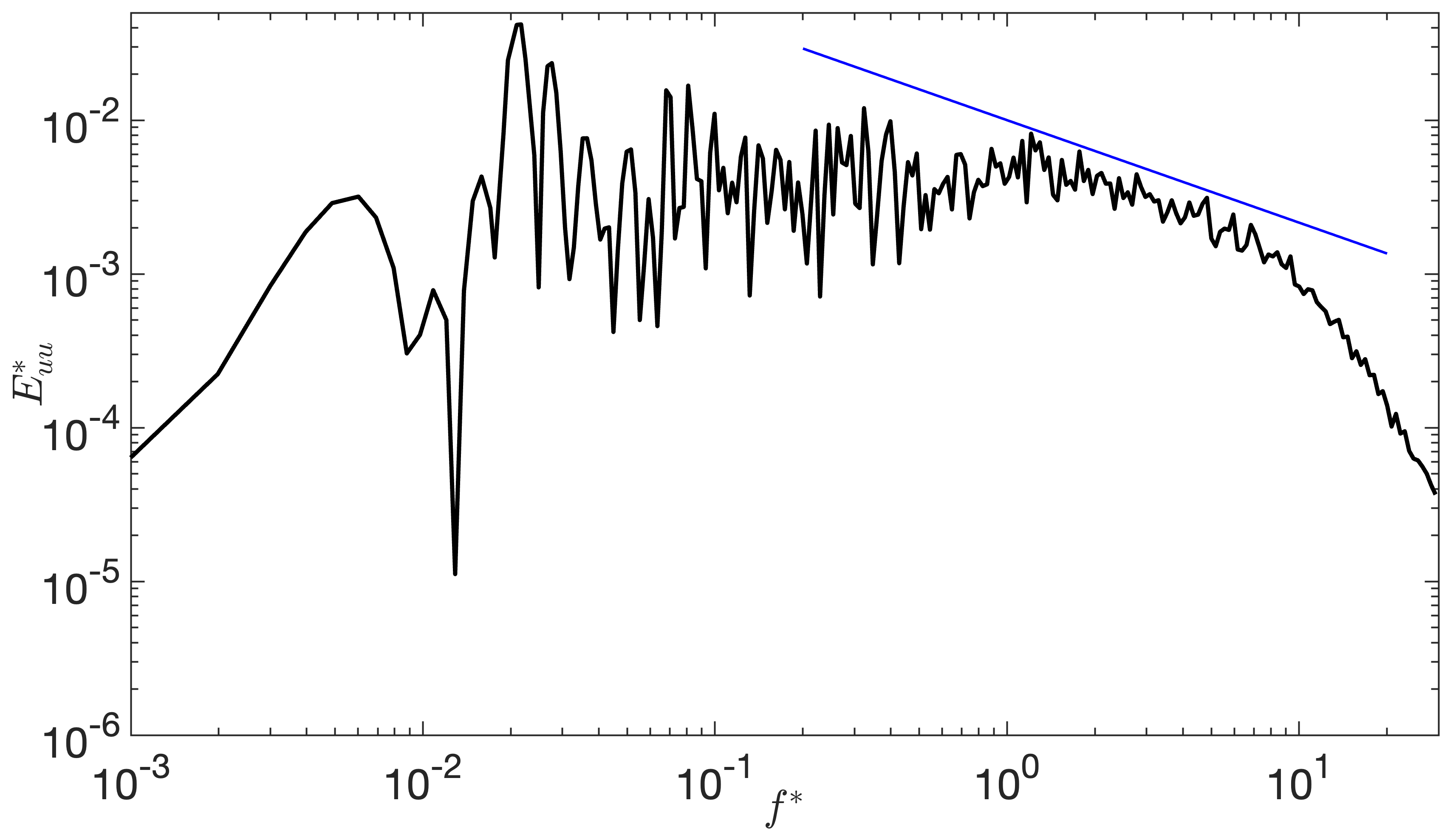}
    }\\
    \subfigure[]{
    \includegraphics[width=0.48\linewidth]{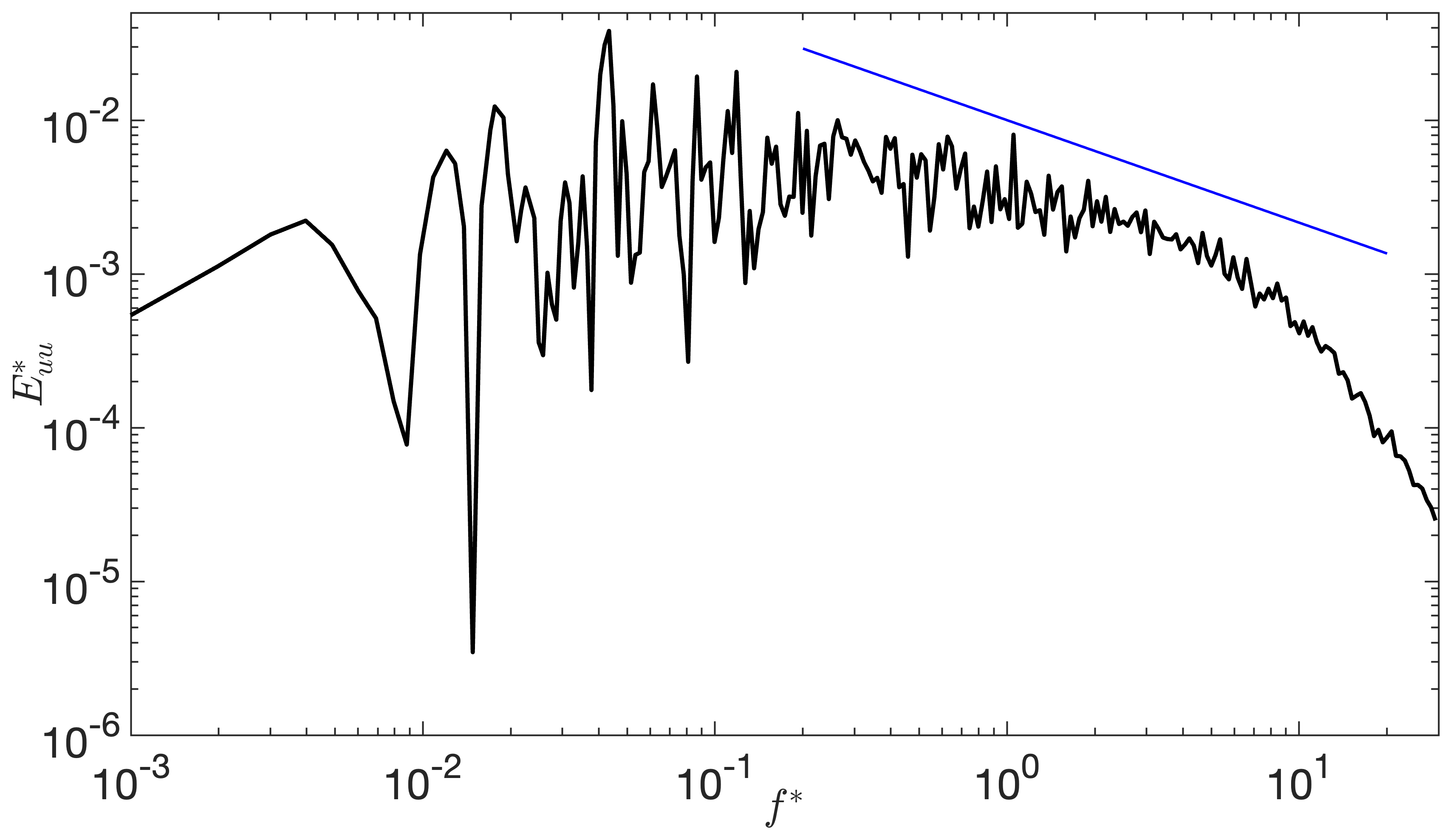}
    }
    \subfigure[]{
    \includegraphics[width=0.48\linewidth]{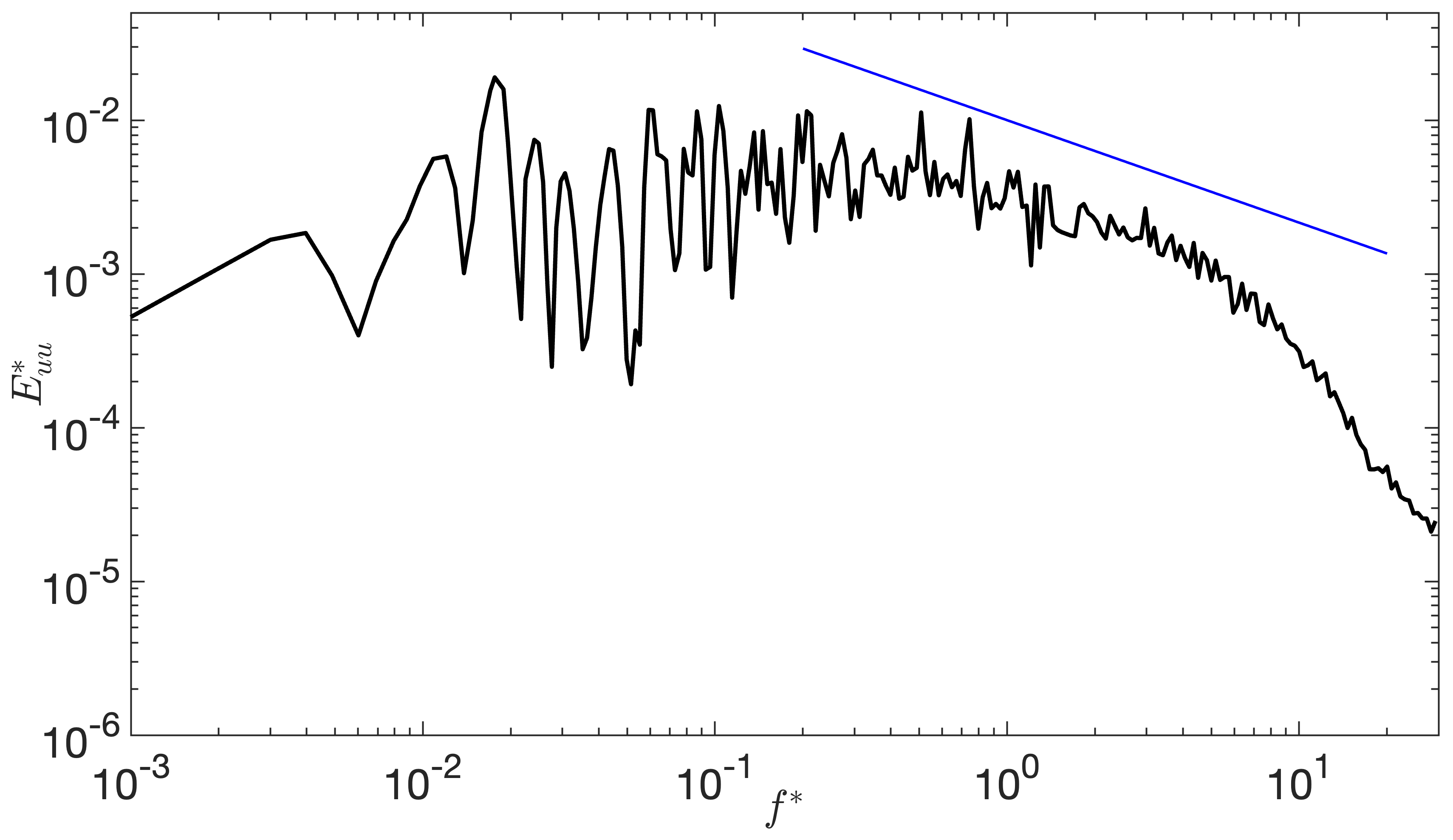}
    }
    \caption{Pre-multiplied energy spectra of the streamwise fluctuating velocity component, $E_{uu}^*=\:E_{uu}/\sigma_u$; $f^* = f\:H/U_{ref}$. Velocities were measured at $z/H = 1.25~(30~\text{m})$ at locations as shown in the top figure (see also Table \ref{tab:spectra} for the exact location of the numerical probes). The blue line denotes the ~$-2/3$ power law.}
    \label{fig:spectra_2}              
\end{figure}

Figures \ref{fig:spectra_1} and \ref{fig:spectra_2} present the pre-multiplied energy spectra of the streamwise fluctuating velocity component  for Case 2. \vet{The energy spectra is non-dimensionalised by the standard deviation of the velocity fluctuations $\sigma_u=\sqrt{u'^2}$, i.e. $E^*_{uu}=E_{uu}/\sigma_u^2$, and the frequency is non-dimensionalised as $f^*=f\:H/U_{ref}$.}
The selected locations are representative of the flow field in different areas, including open spaces enclosed by buildings, rooftops, streets, and road intersections. The exact location of each probe is given in Table \ref{tab:spectra}. 

\vet{The energy spectra exhibit a well-defined inertial subrange with the expected $-2/3$ power-law decay when the energy density is multiplied by the frequency \cite{roth2000review}, regardless of the numerical probe's location.  For the probes located within the urban canopy layer (Fig.~\ref{fig:spectra_1}), the energy spectra decay spans just over one frequency decade. This behavior is consistent with the findings of \citet{Poggi2010}, who noted that within the urban canopy layer, the isotropic assumption becomes less valid due to wake production caused by interactions with buildings. As a result, the inertial subrange only emerges beyond the scales associated with wake production.  

Above the urban canopy layer (Fig.~\ref{fig:spectra_2}), a clearer inertial sub-range extending over approximately two decades is observed. This indicates a better separation between energy-containing scales and smaller dissipative scales, as turbulence becomes more isotropic with distance from the ground. Additionally, the energy spectra in both figures show a distinct filter cutoff around frequencies of $\mathcal{O}(10)$, where a noticeable change in slope occurs.  

Several key conclusions can be drawn from the analysis of these results. First, 
an accurate capture of the inertial sub-range highlights the high-quality of the computations. The high-order methods used in the current implementation do not produce spurious oscillations at high frequencies or excessive numerical dissipation. The latter would result in a faster decay in the inertial sub-range with a slope steeper than that predicted by the Kolmogorov hypothesis. Moreover, there is no energy pile-up at high frequencies, which is typically indicative of poor resolution when using  low-dissipation numerical schemes or insufficient dissipation  of the SGS model\cite{DAVIDSON20091016}.

Second, the resolution used in these simulations is sufficient to distinctly capture the separation between energy-containing scales and dissipative scales. This is particularly important in the context of LES, where the subgrid-scale model requires a computational mesh fine enough for the filter cutoff to occur well within the inertial sub-range. Achieving this ensures that the simulations yield accurate results and effectively represent the underlying physics of turbulent flows.}

\subsection{On the temporal convergence}\label{sec: tmpConv}

\vet{In urban flow simulations, ensuring the convergence of turbulence statistics over time is essential for capturing the complex dynamics of wind interactions with buildings and street canyons. Due to the inherently unsteady and multi-scale nature of urban turbulence, statistical quantities such as mean velocity and turbulent fluctuations might require a relatively long averaging period to provide reliable results. Without proper temporal convergence, assessment of flow structures, shear layers, and recirculation zones may be misleading, affecting the accuracy of numerical studies. Given the sensitivity of second-order statistics to turbulent fluctuations, a thorough evaluation of convergence is necessary to ensure that simulations yield robust and physically meaningful results.
On the other hand,  achieving temporal convergence in turbulence statistics can significantly increase computational costs. Thus, the integration time is a trade-off between statistical accuracy and computational feasibility, as longer simulations demand greater resources, particularly when high-resolution LES are used. 
As discussed in Section \ref{comp_detail}, statistics are collected over a period of 40$FT$s in both cases, which corresponds to 192$T$ and 500$T$, respectively. It is worth noting that although this integration time is similar to those reported in the literature for comparable cases \cite{Castro2009c,GarciaSanchez2018,Duan2019,Stuck2021,Shaukat2024}, a detailed assessment of time convergence is still necessary.  In what follows we provide analysis of the time convergence of the computed statistics.}

\begin{figure}
    \centering
    \subfigure[]{
    \includegraphics[width=0.45\linewidth]{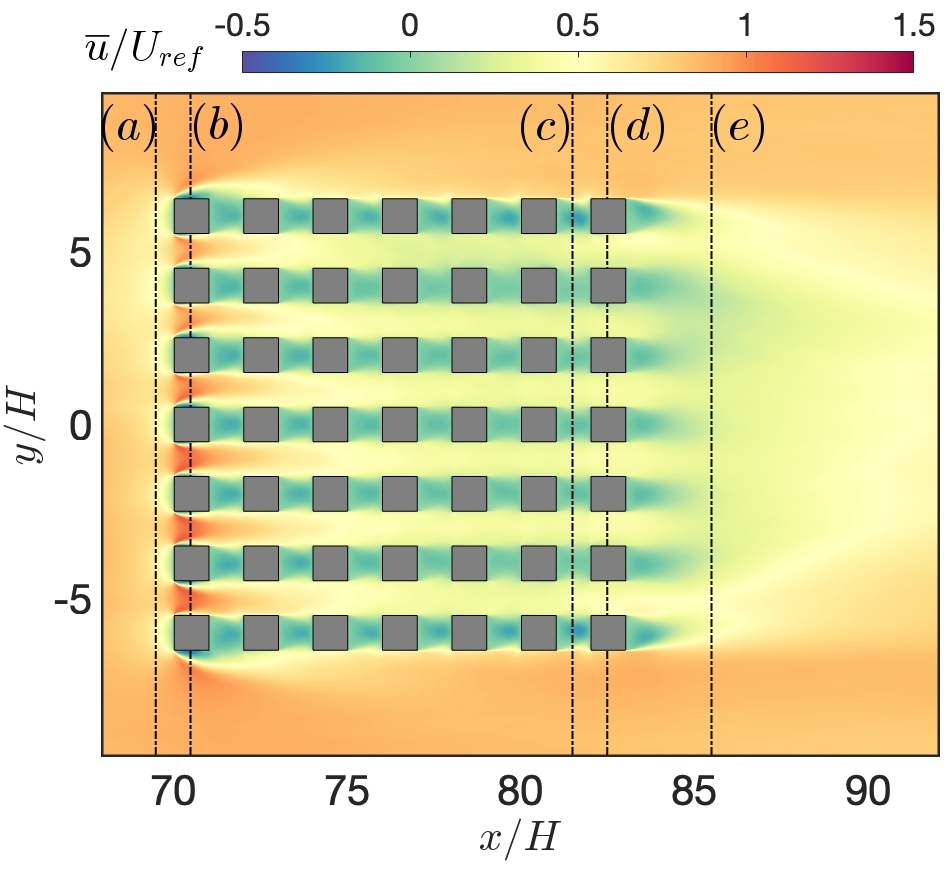}
    }
    \subfigure[]{
    \includegraphics[width=0.45\linewidth]{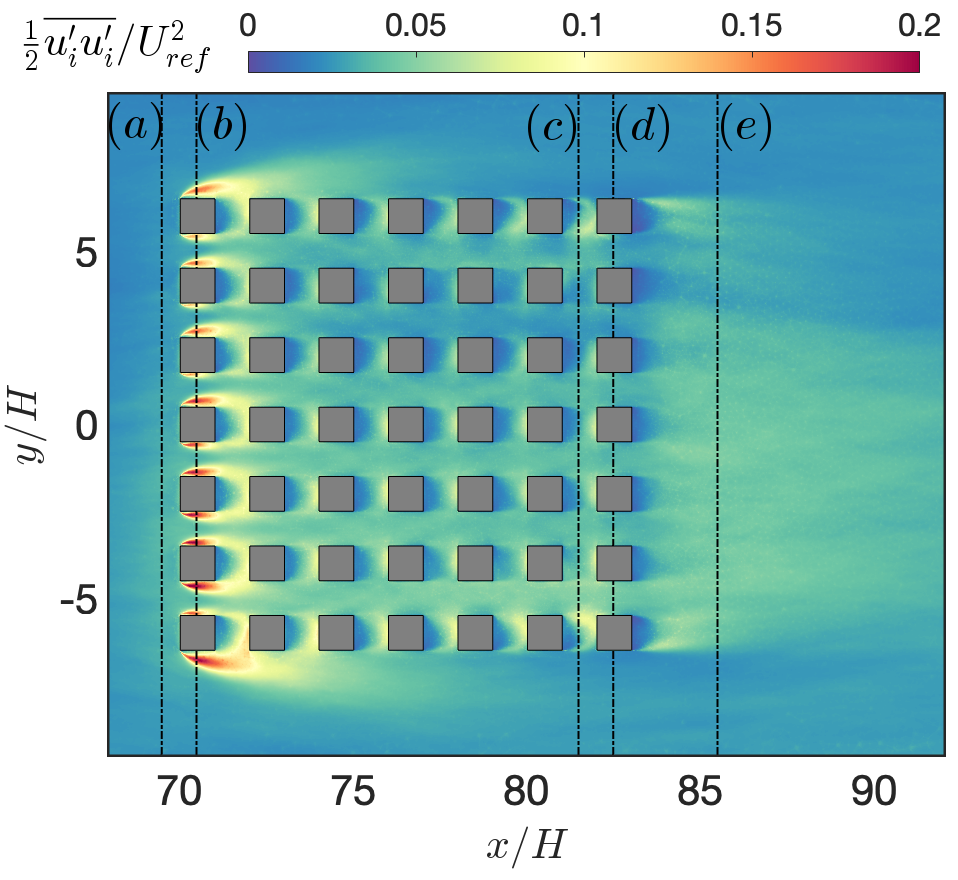}
    }
    \subfigure[]{
    \includegraphics[width=0.48\linewidth]{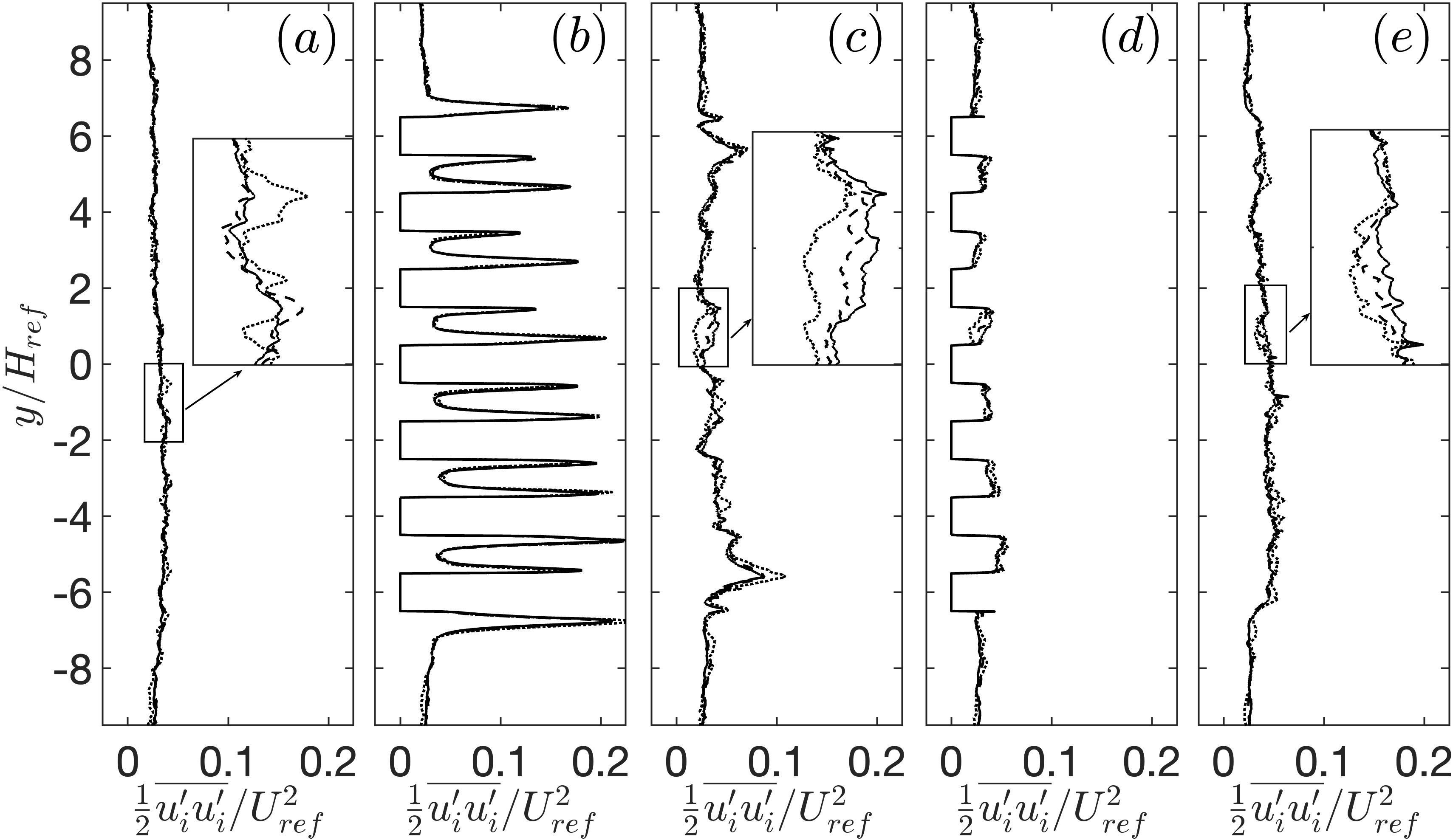}
    }
    \subfigure[]{
    \includegraphics[width=0.48\linewidth]{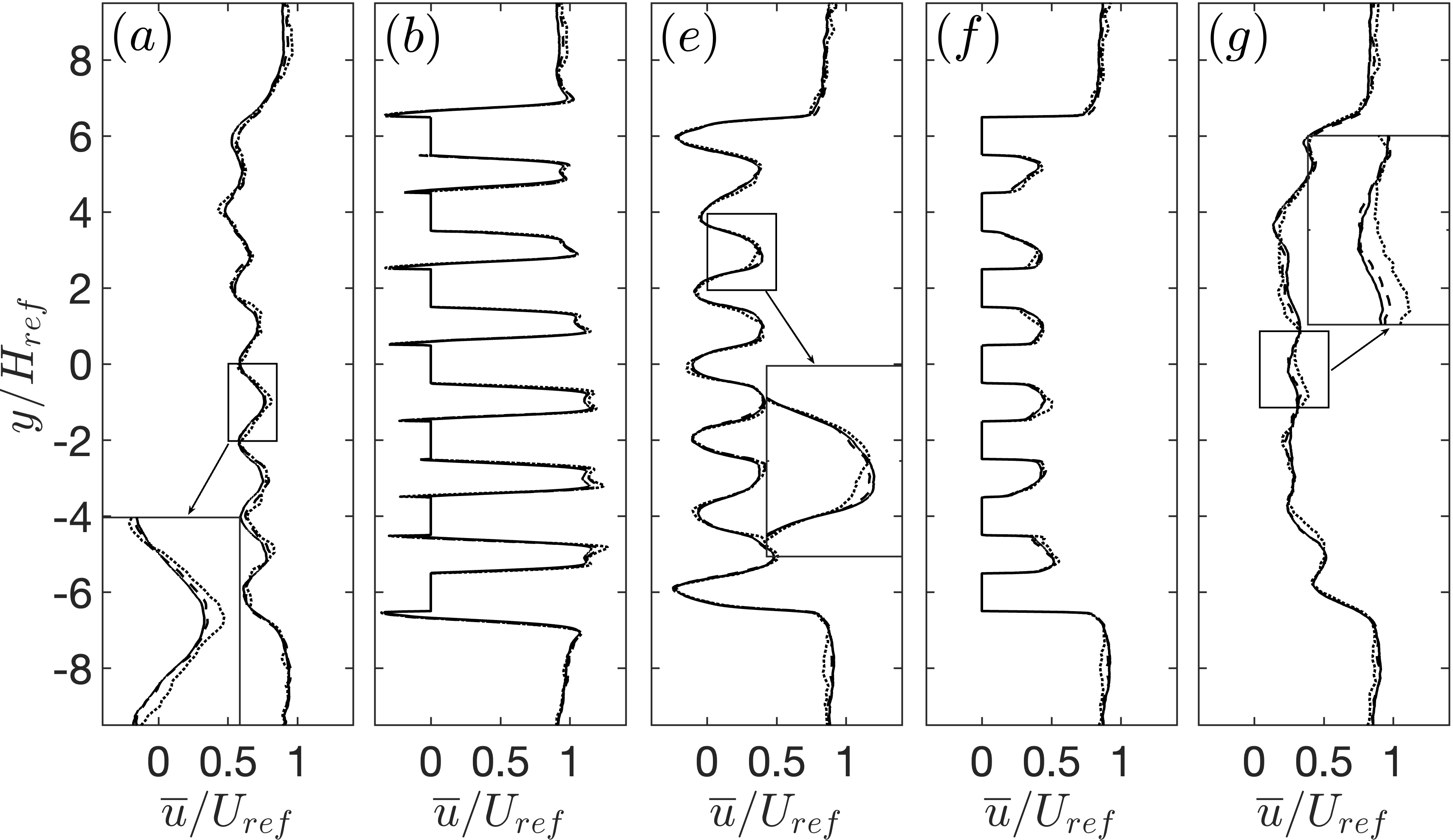}
    }    
    \caption{Case 1: Contour of $(a)$ streamwise mean-velocity, $\overline{u}/U_{ref}$, and $(b)$ $TKE$, $\frac{1}{2}\overline{u'_iu'_i}/U^2_{ref}$, in the $x-y$ plane at a distance of $z/H = 0.5$ away from the ground. Profiles at selected locations shown in $(a)$ and $(b)$ are plotted in $(c)$ and $(d)$. \dashed~20$FT$s; \chndash~40$FT$s; \solid~80$FT$s.}
    \label{fig:tmpConvg}              
\end{figure}

Figure \ref{fig:tmpConvg} presents the temporal convergence study of Case 1, examining both $\overline{u}/U_{ref}$ and $TKE$. Case 2 is not shown as it exhibits a similar convergence behaviour. A comparison of three sampling lengths (see Fig.\ref{fig:tmpConvg}$(c, ~d)$) reveals that while 20 $FT$s exhibit noticeable variations in both streamwise velocity and $TKE$, an averaging period of 40$FT$s is sufficient to achieve a temporal convergence of second-order statistics, i.e, $TKE$ in this case. Profiles of $\overline{u}/U_{ref}$ converge substantially faster than those of $TKE$. This is an expected result, as $TKE$ is affected by small-scales fluctuations which require a much longer integration to statistically capture the full range of fluctuations.
This is specially true in those zones  featuring wake interactions, recirculations, separations and reattachment, where flow complexity drives slower convergence of flow statistics.

\section{Conclusions}\label{sec: conclusions}

The study has presented  wall-modeled large-eddy simulation (LES) of an atmospheric boundary layer (ABL) over idealized urban roughnesses. Two cases are considered: a three-dimensional cubic prism array and the “Michel-stadt” urban model. These cases are selected to investigate the accuracy of LES in predicting urban turbulence and to validate the methodology against wind tunnel experiments. The simulations have been performed using a high-resolution, low-dissipation numerical scheme, with a spatial resolution below 0.75m within the urban canopy, exceeding typical LES studies in the literature. A precursor simulation is used to generate realistic inflow conditions, ensuring that the simulated ABL matches the experimental setup. 

The inflow of the study domain is driven by an online precursor simulation, ensuring a physically consistent ABL that matched the wind tunnel inflow conditions. This approach eliminates the need for synthetic turbulence generation and allows for a realistic turbulent inflow that dynamically adapts to the simulation of study domain. A comparison of the inflow velocity profile with experimental measurements indicate excellent agreement, reinforcing the validity of the ABL development in the numerical model. The precursor simulation approach contributes to the accurate reproduction of mean-velocity profiles and turbulence statistics, further supporting the reliability of the LES setup.

The mean-velocity profiles and velocity fluctuations show excellent agreement between LES and experiments at most heights, particularly for the streamwise component. The LES accurately captures the mean-flow patterns, including wake regions and the acceleration over rooftops, which are key characteristics of urban flows. The turbulence kinetic energy ($TKE$) has been also well predicted at most locations, with only slight underestimations near the surface, which could be attributed to SGS modeling effects or unresolved small-scale turbulence near walls. However, for the spanwise turbulence fluctuations, LES shows a general tendency to slightly underestimate \ming{root-mean-square} values at lower heights, which might be related to measurement constraints in the wind tunnel, including probe misalignment or blockage effects limiting lateral velocity fluctuations together with some insufficient resolution near wall.

The validation against wind tunnel experiments demonstrates a strong agreement in the streamwise velocity component, particularly at higher elevations where discrepancies observed is minimal. The validation metrics, $FAC2$ and hit rate ($HR$), exceeds the commonly accepted thresholds of 0.3 and 0.66, respectively, confirming that LES predictions are well within standard accuracy levels for urban flow modeling. However, larger discrepancies have been observed in the spanwise velocity component, which could be attributed to a combination of numerical and experimental factors, including differences in turbulence generation, lateral boundary effects in the wind tunnel, and Reynolds number mismatches between full-scale and wind tunnel conditions.

While the validation metrics ($FAC2$ and $HR$) confirm the reliability of LES in predicting mean velocities, they do not fully characterize turbulence structure or spectral energy distribution. A detailed spectral analysis has been therefore conducted to evaluate whether LES properly captures the turbulent energy cascade across scales. The energy spectra plots confirmed that LES maintains low numerical dissipation, as there is no excessive damping of small-scale fluctuations, and the inertial sub-range follows the expected turbulence decay slopes. This is a key finding, as it validates that the LES formulation does not introduce excessive numerical diffusion, which could otherwise affect turbulence dynamics in urban flows.

The LES resolution of $0.75m$ within the urban canopy, which is finer than typical LES studies in urban environments, allows for accurate representation of turbulent structures. The spectral resolution analysis demonstrated that LES resolves a significant portion of the inertial sub-range, ensuring that most of the energy-containing turbulence scales are explicitly captured. The Kolmogorov $-2/3$ slope in the pre-multiplied spectra has been well reproduced below and above the urban canopy, indicating that the energy cascade is correctly represented. However, inside the urban canopy, some deviations have been observed due to the influence of building-induced turbulence and wake interactions. These findings reinforce the high quality of the LES predictions and the suitability of the numerical scheme for urban flow modeling.

Overall, this study highlights the effectiveness of the current methodology for accurately capturing the physics of urban flows, particularly when high-order, low-dissipation schemes and fine spatial resolution are used. Despite the limitations of $FAC2$ and $HR$ as validation metrics, the findings surpass standard validation criteria, emphasizing the need to complement the analyses with spectral decomposition to fully assess turbulence dynamics. Furthermore, this work sets the foundation for a more comprehensive study that will encompass a more in-depth analysis of turbulent quantities and roughness characterization.

\appendix
\section{Case 2 -- model validations}\label{sec: appedix}

\begin{figure}
    \centering
    \includegraphics[width=0.495\linewidth]{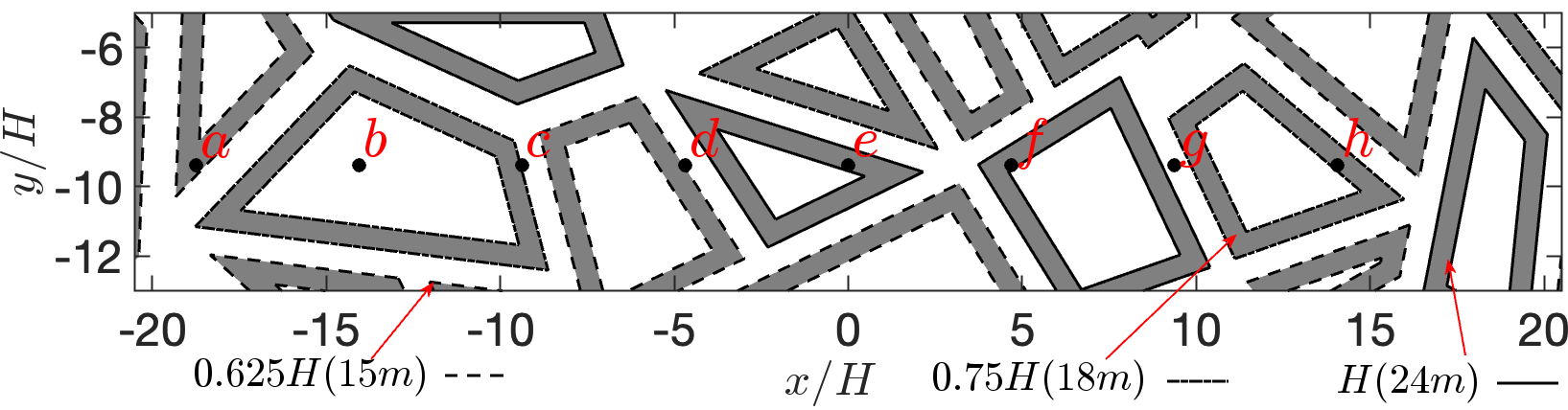}\\
    \vspace{0.1in}
    \includegraphics[width=0.495\linewidth]{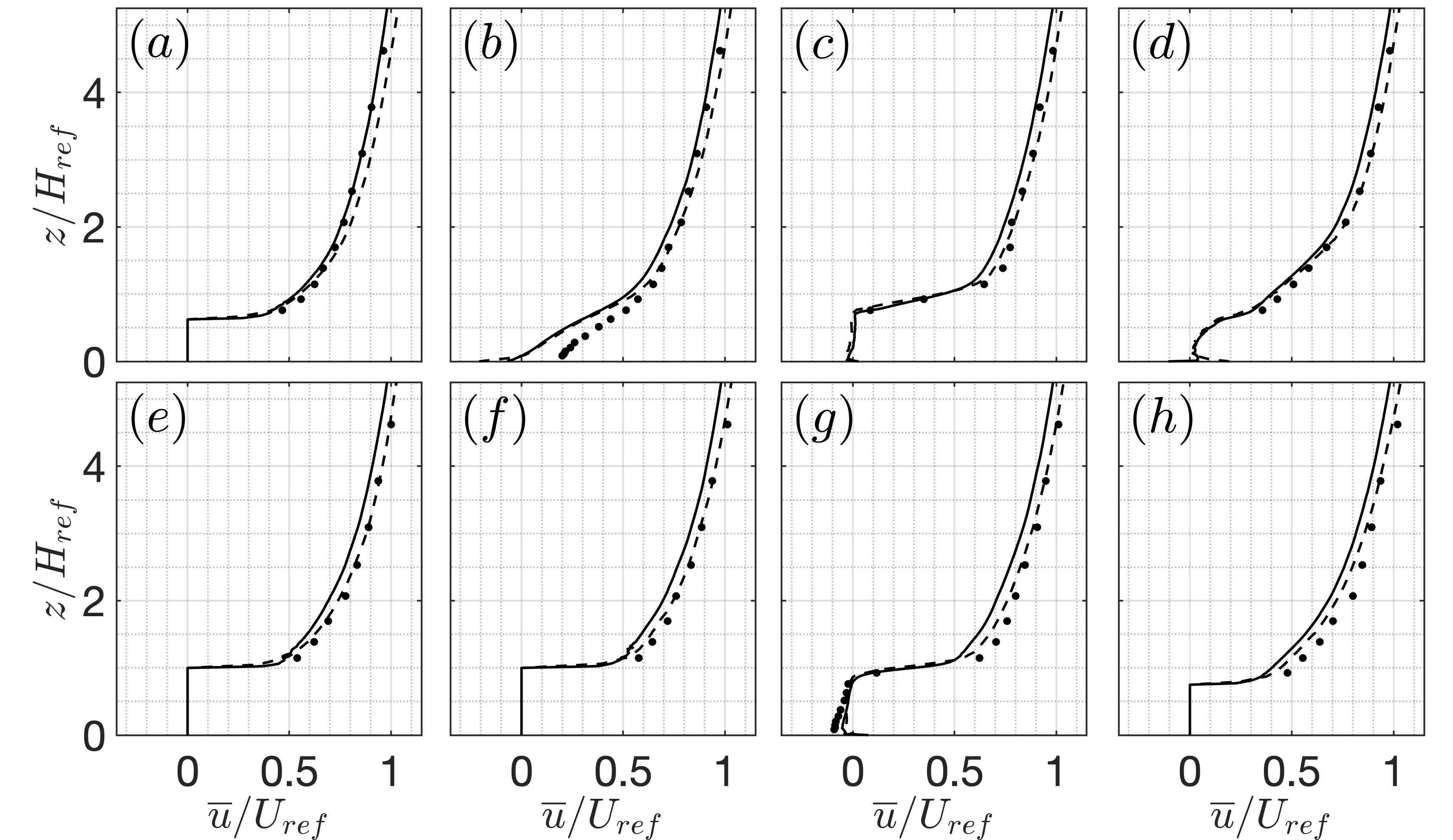}
    \caption{Case 2: comparison of streamwise mean-velocity profiles, $\overline{u}/U_{ref}$,  with those of 
    experimental results at selected locations ($y/H =-9.375$)
    as shown in the top figure: 
    \color{black}\dashed~coarse mesh; 
    \color{black}\solid~fine mesh; $\bullet$~ 
    \citet{leitl2024}.}
    \label{fig:gridCongergence_MichelStadt3_u}
\end{figure}

\begin{figure}
    \centering
    \includegraphics[width=0.495\linewidth]{Figures/Geo2D_MichelStat_3.png}\\
    \vspace{0.1in} 
    \includegraphics[width=0.495\linewidth]{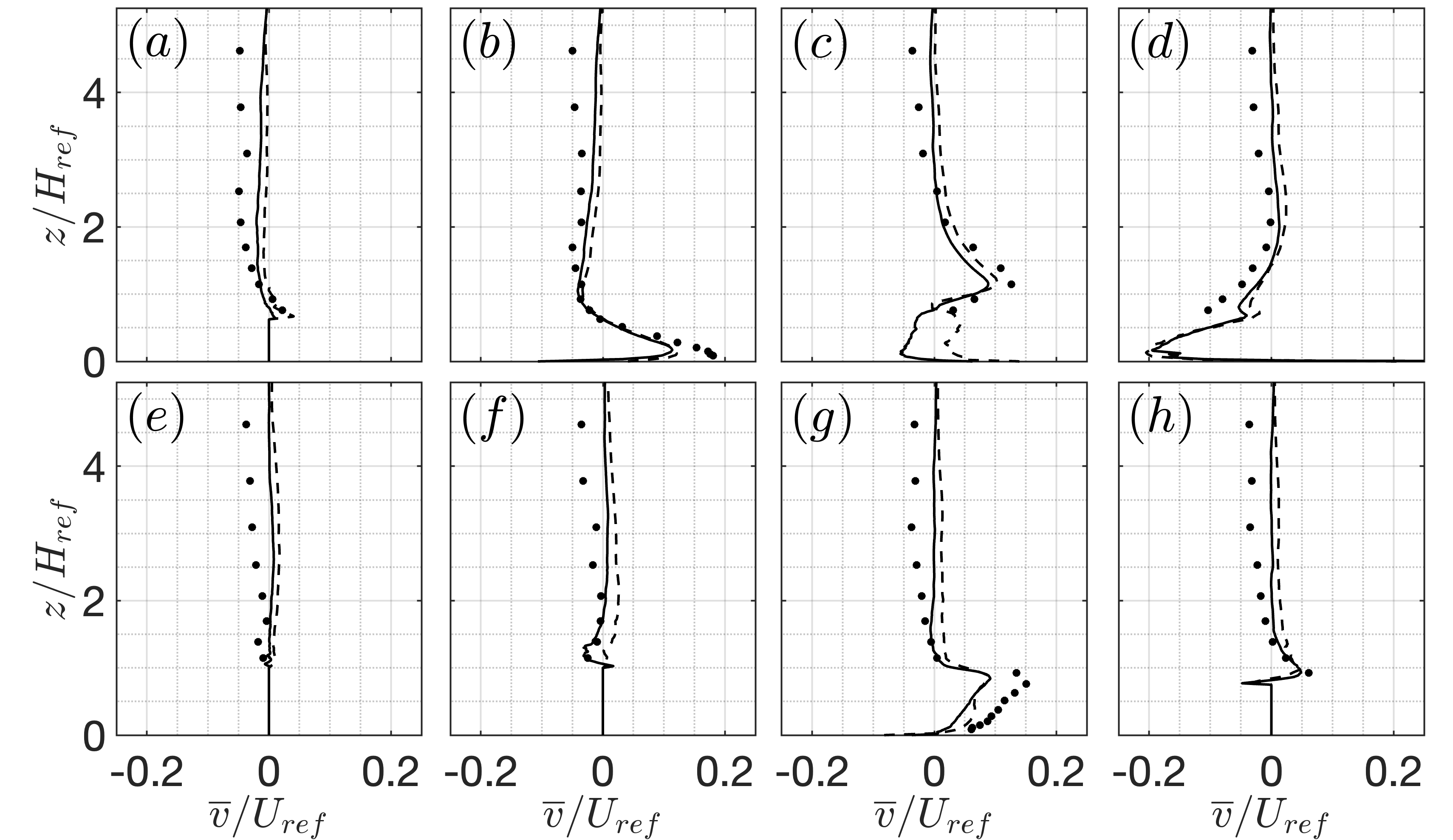} 
    \caption{Case 2: comparison of spanwise mean-velocity profiles, $\overline{v}/U_{ref}$, with those of 
    experimental results at selected locations ($y/H =-9.375$) 
    as shown in the top figure: 
    \color{black}\dashed~coarse mesh; 
    \color{black}\solid~fine mesh; $\bullet$~ 
    \citet{leitl2024}.}
    \label{fig:gridCongergence_MichelStadt3_v}
\end{figure}

\begin{figure}
    \centering
    \includegraphics[width=0.495\linewidth]{Figures/Geo2D_MichelStat_3.png}\\
    \vspace{0.1in} 
    \includegraphics[width=0.495\linewidth]{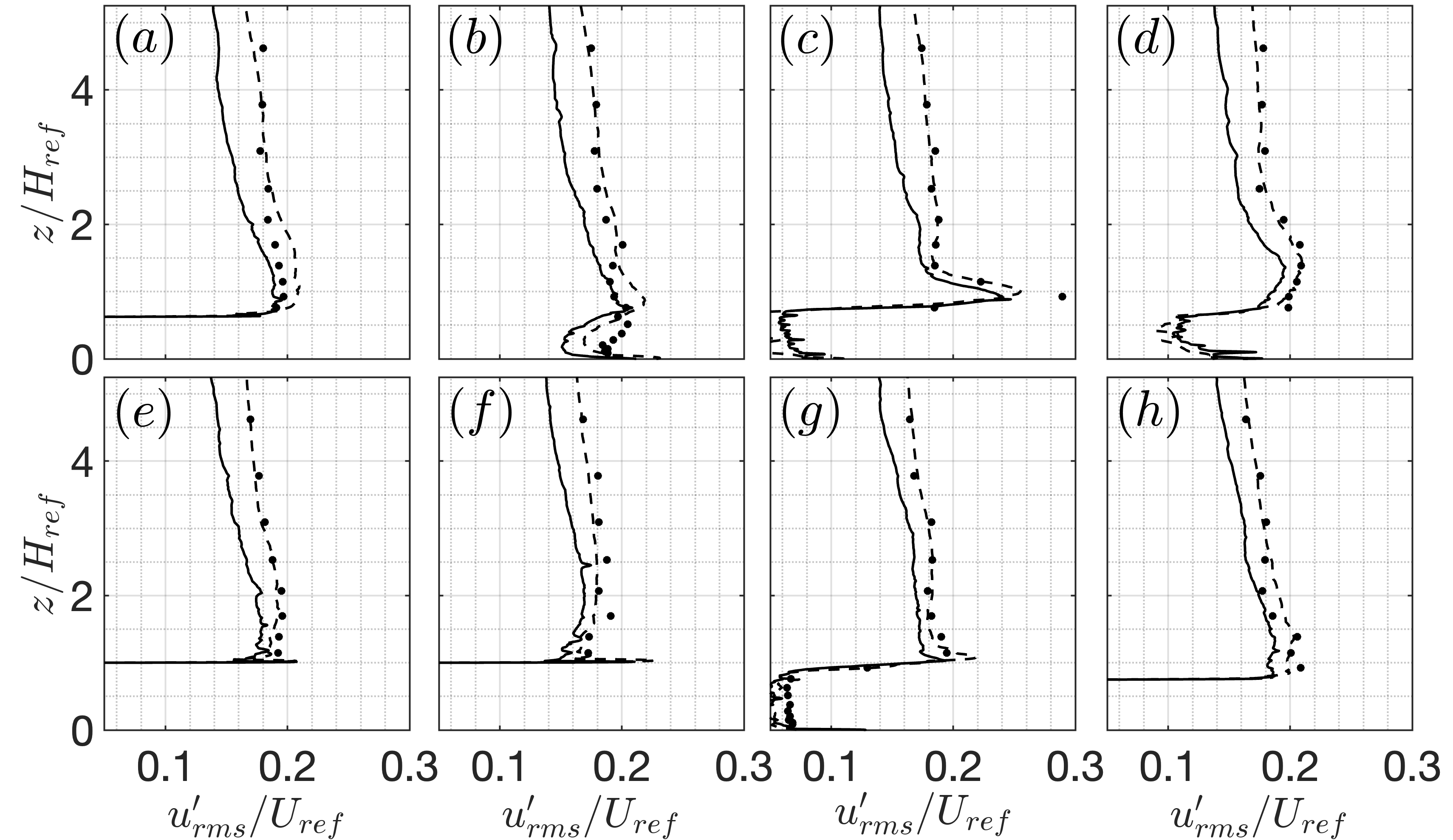}    
    \caption{Case 2: comparison of \textit{rms} streamwise velocity fluctuations profiles, $u_{rms}'/U_{ref}$, with those of 
    experimental results at selected locations ($y/H =-9.375$) 
    as shown in the top figure: 
    \color{black}\dashed~coarse mesh; 
    \color{black}\solid~fine mesh; $\bullet$~ 
    \citet{leitl2024}.}
    \label{fig:gridCongergence_MichelStadt3_RMS}
\end{figure}

\begin{figure}
    \centering
    \includegraphics[width=0.495\linewidth]{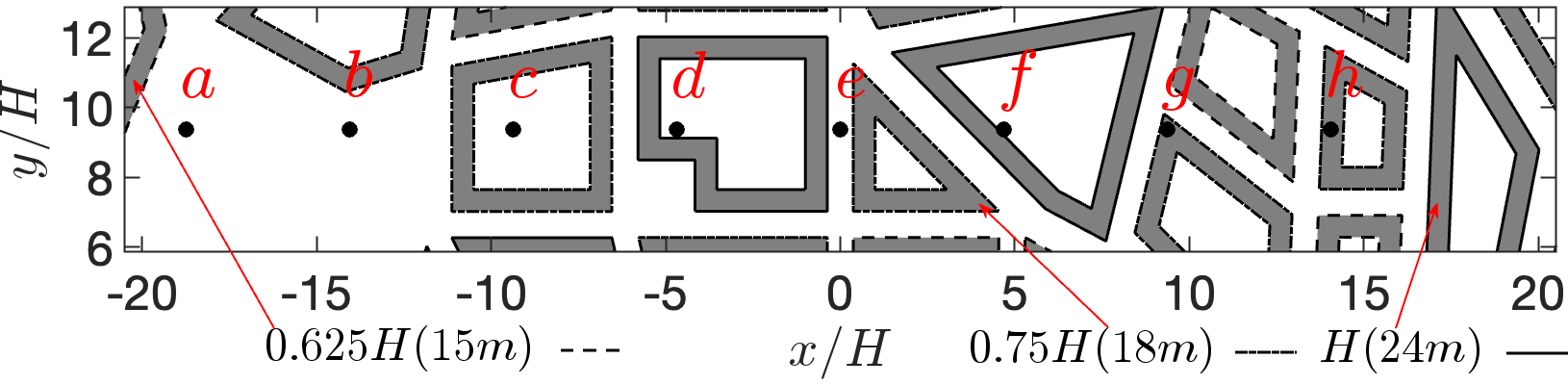}\\
    \vspace{0.1in}
    \includegraphics[width=0.495\linewidth]{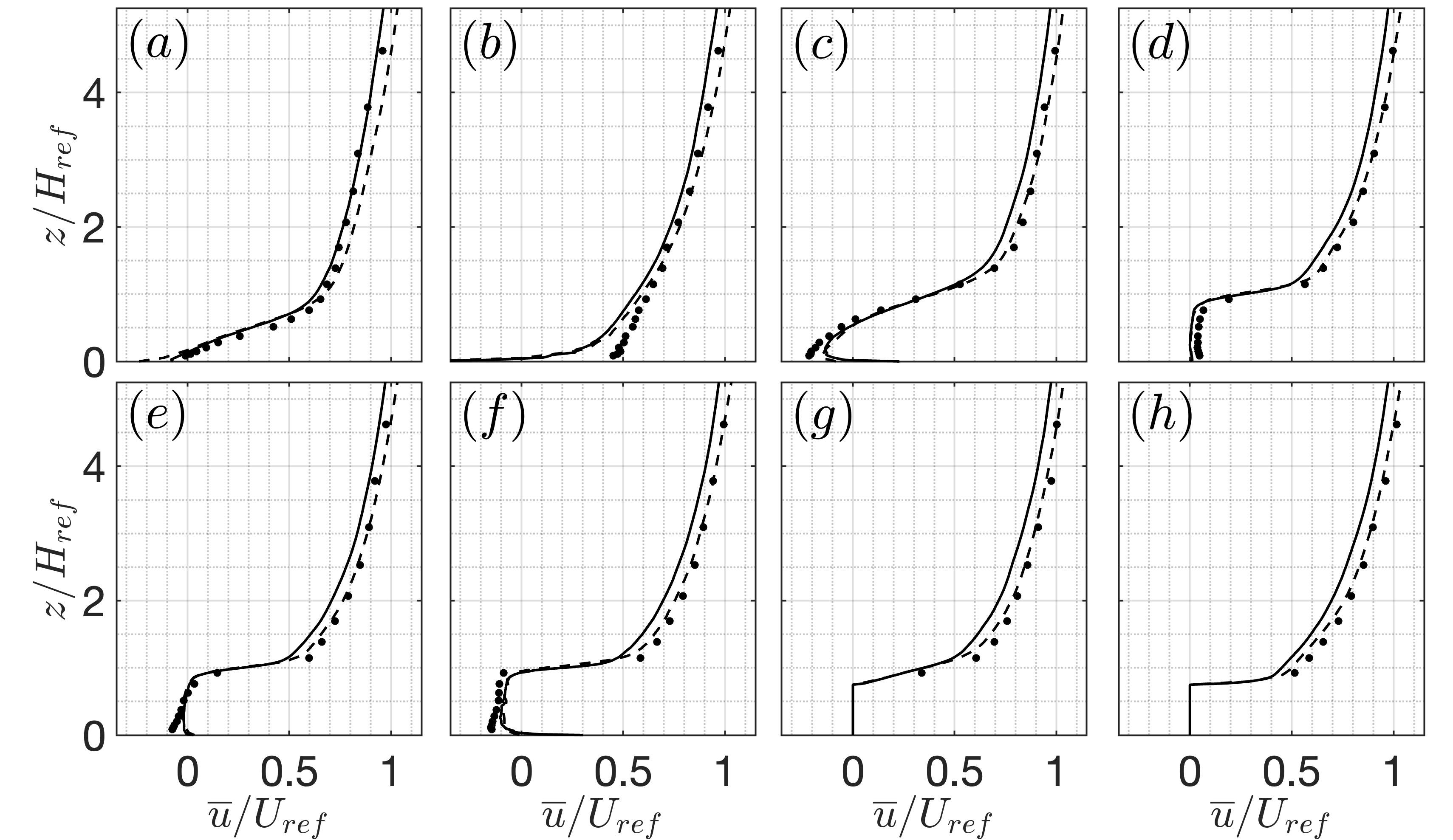}
 
    \caption{Case 2: comparison of streamwise mean-velocity profile, $\overline{u}/U_{ref}$, with those of experimental results at selected locations ($y/H = 9.375$) 
    as shown in the top figure: 
    \color{black}\dashed~coarse mesh; 
    \color{black}\solid~fine mesh; $\bullet$~ 
    \citet{leitl2024}.}
    \label{fig:gridCongergence_MichelStadt2_U}
\end{figure}

\begin{figure}
    \centering
    \includegraphics[width=0.495\linewidth]{Figures/Geo2D_MichelStat_2.png}\\
    \vspace{0.1in}
   \includegraphics[width=0.495\linewidth]{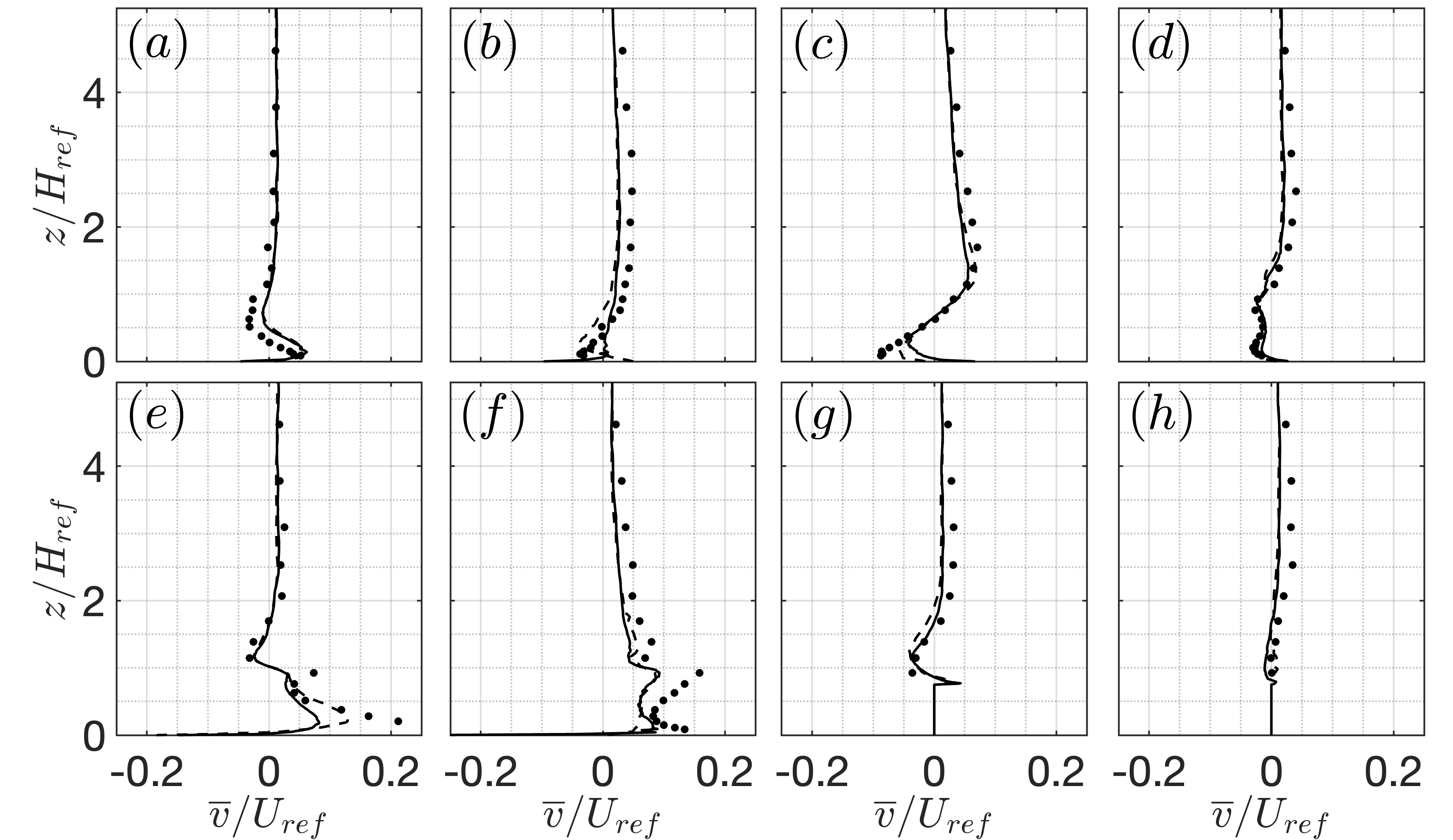}\\  
    \caption{Case 2: comparison of 
    spanwise mean-velocity profile, $\overline{v}/U_{ref}$, with those of 
    experimental results at selected locations ($y/H = 9.375$) 
    as shown in the top figure: 
    \color{black}\dashed~coarse mesh; 
    \color{black}\solid~fine mesh; $\bullet$~ 
    \citet{leitl2024}.}
    \label{fig:gridCongergence_MichelStadt2_V}
\end{figure}

\begin{figure}
    \centering
    \includegraphics[width=0.495\linewidth]{Figures/Geo2D_MichelStat_2.png}\\
    \vspace{0.1in}
   \includegraphics[width=0.495\linewidth]{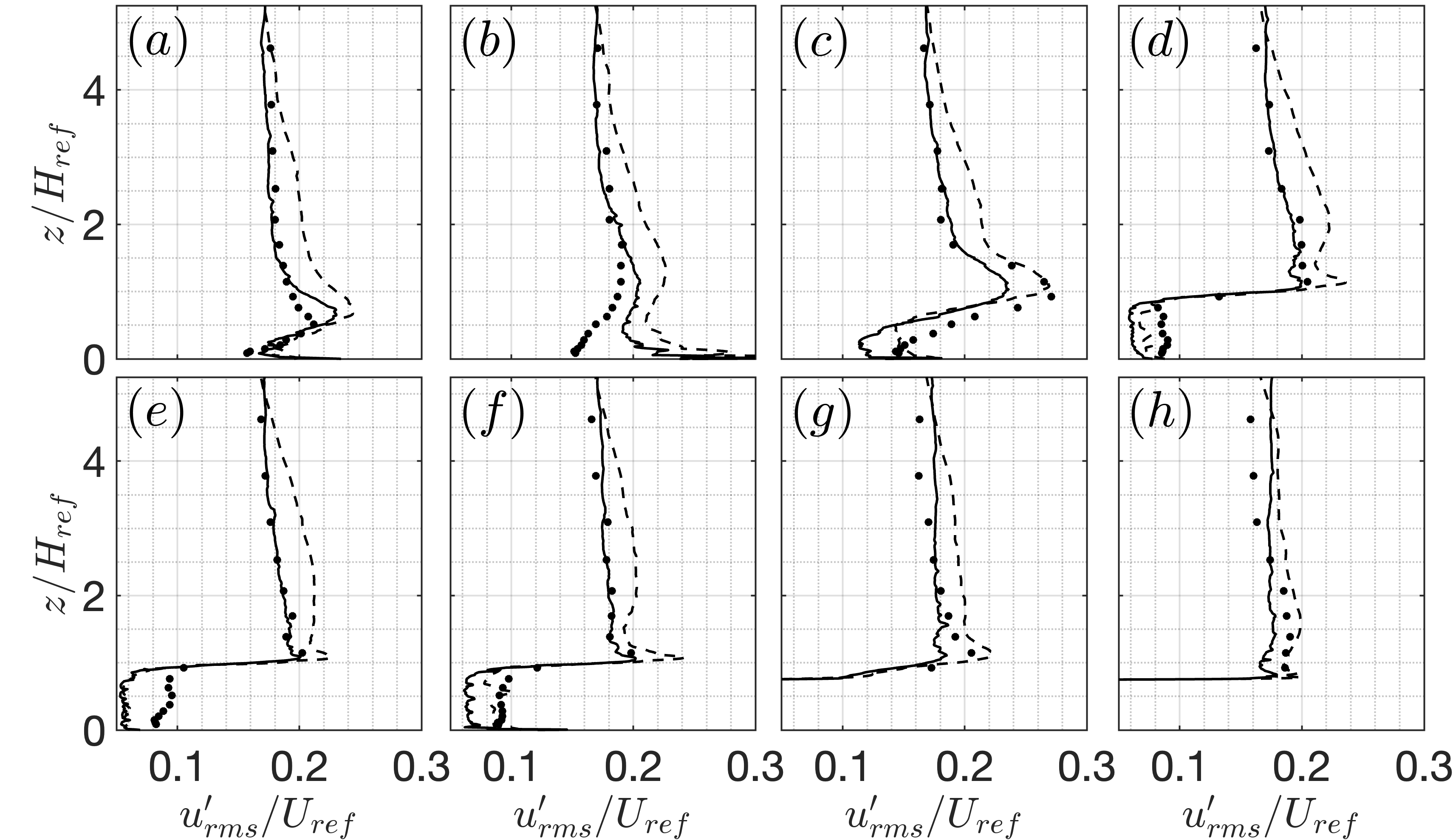}    
    \caption{Case 2: Comparison of \textit{rms} streamwise velocity fluctuations profiles, $u_{rms}'/U_{ref}$, with those of 
    experimental results at selected locations ($y/H = 9.375$) 
    as shown in the top figure: 
    \color{black}\dashed~coarse mesh; 
    \color{black}\solid~fine mesh; $\bullet$~ 
    \citet{leitl2024}.}
    \label{fig:gridCongergence_MichelStadt2_RMS}
\end{figure}

\ming{In addition to the data comparison presented in Section 
\ref{sec: results}, additional comparisons with experimental 
measurements are shown here to further reinforce the 
comprehensive validation of Case 2. Figures \ref{fig:gridCongergence_MichelStadt3_u} to 
\ref{fig:gridCongergence_MichelStadt3_RMS} present the 
comparisons of $\overline{u}/U_{ref}$, $\overline{v}/U_{ref}$ 
and $u'_{rms}$ in the $x-z$ plane along $y/H_{ref} =-9.375$, 
while those along $y/H_{ref} = 9.375$ in the $x-z$ plane are 
shown in Figs. \ref{fig:gridCongergence_MichelStadt2_U} to 
\ref{fig:gridCongergence_MichelStadt2_RMS}. 

The minor discrepancies observed in all profiles between the two 
grid resolutions indicate the grid independence of the numerical 
results. A satisfactory agreement with experimental measurements 
is obtained in $\overline{u}/U_{ref}$ profiles across all 
locations (see Figs \ref{fig:gridCongergence_MichelStadt3_u} and 
\ref{fig:gridCongergence_MichelStadt2_U}). In comparison of 
$\overline{v}/U_{ref}$, discrepancies are largely pronounced at 
pedestrian level around building corners and road intersections
(e.g., Figs. \ref{fig:gridCongergence_MichelStadt3_v}$(c,~g)$ 
and Fig. \ref{fig:gridCongergence_MichelStadt2_V}$(e)$). 
Although the simulations accurately predict the peak values of 
$u'_{rms}/U_{ref}$ above the roofs, an under-prediction within 
the reverse-flow regions between buildings is observed (see 
Fig. \ref{fig:gridCongergence_MichelStadt3_RMS}$(b)$ and Figs. 
\ref{fig:gridCongergence_MichelStadt2_RMS}$(b, ~e)$). 
Additionally, similar under-predictions are also clearly 
identified at 
certain locations far above the roof-level (e.g. Figs. 
\ref{fig:gridCongergence_MichelStadt3_RMS}$(a-d)$).}

\begin{acknowledgments}
This work has been partially financially supported by `Agència de Gestió d'Ajuts Universitaris i de 
Recerca' under the call CLIMA 2023 (ref. 2023 CLIMA 00097) \ming{and the APPWIND project (ref.: PLEC2021-007943).}  O. Lehmkuhl's work is financed by a Ramón y 
Cajal postdoctoral contract by the Ministerio de Economía y Competitividad, Secretaría de Estado de 
Investigación, Desarrollo e Innovación, Spain (RYC2018-025949-I). The authors acknowledge the support of 
the Departament de Recerca i Universitats de la Generalitat de Catalunya through the research group 
Large-scale Computational Fluid Dynamics (ref.: 2021 SGR 00902) and the Turbulence and Aerodynamics 
Research Group (ref.: 2021 SGR 01051).
\end{acknowledgments}

\section*{Data Availability Statement}
Data supporting the findings of this study is available
from the author upon request.

% \clearpage
% \nocite{*}
% \bibliographystyle{Definition/aipnum4-1}
\bibliography{Reference}% Produces the bibliography via BibTeX.

%merlin.mbs aipnum4-1.bst 2010-07-25 4.21a (PWD, AO, DPC) hacked
%Control: key (0)
%Control: author (8) initials jnrlst
%Control: editor formatted (1) identically to author
%Control: production of article title (0) allowed
%Control: page (1) range
%Control: year (1) truncated
%Control: production of eprint (0) enabled
\providecommand{\noopsort}[1]{}\providecommand{\singleletter}[1]{#1}%
\begin{thebibliography}{54}%
\makeatletter
\providecommand \@ifxundefined [1]{%
 \@ifx{#1\undefined}
}%
\providecommand \@ifnum [1]{%
 \ifnum #1\expandafter \@firstoftwo
 \else \expandafter \@secondoftwo
 \fi
}%
\providecommand \@ifx [1]{%
 \ifx #1\expandafter \@firstoftwo
 \else \expandafter \@secondoftwo
 \fi
}%
\providecommand \natexlab [1]{#1}%
\providecommand \enquote  [1]{``#1''}%
\providecommand \bibnamefont  [1]{#1}%
\providecommand \bibfnamefont [1]{#1}%
\providecommand \citenamefont [1]{#1}%
\providecommand \href@noop [0]{\@secondoftwo}%
\providecommand \href [0]{\begingroup \@sanitize@url \@href}%
\providecommand \@href[1]{\@@startlink{#1}\@@href}%
\providecommand \@@href[1]{\endgroup#1\@@endlink}%
\providecommand \@sanitize@url [0]{\catcode `\\12\catcode `\$12\catcode `\&12\catcode `\#12\catcode `\^12\catcode `\_12\catcode `\%12\relax}%
\providecommand \@@startlink[1]{}%
\providecommand \@@endlink[0]{}%
\providecommand \url  [0]{\begingroup\@sanitize@url \@url }%
\providecommand \@url [1]{\endgroup\@href {#1}{\urlprefix }}%
\providecommand \urlprefix  [0]{URL }%
\providecommand \Eprint [0]{\href }%
\providecommand \doibase [0]{http://dx.doi.org/}%
\providecommand \selectlanguage [0]{\@gobble}%
\providecommand \bibinfo  [0]{\@secondoftwo}%
\providecommand \bibfield  [0]{\@secondoftwo}%
\providecommand \translation [1]{[#1]}%
\providecommand \BibitemOpen [0]{}%
\providecommand \bibitemStop [0]{}%
\providecommand \bibitemNoStop [0]{.\EOS\space}%
\providecommand \EOS [0]{\spacefactor3000\relax}%
\providecommand \BibitemShut  [1]{\csname bibitem#1\endcsname}%
\let\auto@bib@innerbib\@empty
%</preamble>
\bibitem [{\citenamefont {Meinders}\ and\ \citenamefont {Hanjali{\'c}}(1999)}]{meinders1999vortex}%
  \BibitemOpen
  \bibfield  {author} {\bibinfo {author} {\bibfnamefont {E.~R.}\ \bibnamefont {Meinders}}\ and\ \bibinfo {author} {\bibfnamefont {K.}~\bibnamefont {Hanjali{\'c}}},\ }\bibfield  {title} {\enquote {\bibinfo {title} {Vortex structure and heat transfer in turbulent flow over a wall-mounted matrix of cubes},}\ }\href@noop {} {\bibfield  {journal} {\bibinfo  {journal} {International Journal of Heat and Fluid Flow}\ }\textbf {\bibinfo {volume} {20}},\ \bibinfo {pages} {255--267} (\bibinfo {year} {1999})}\BibitemShut {NoStop}%
\bibitem [{\citenamefont {Mathey}, \citenamefont {Fr{\"o}hlich},\ and\ \citenamefont {Rodi}(1999)}]{mathey1999large}%
  \BibitemOpen
  \bibfield  {author} {\bibinfo {author} {\bibfnamefont {F.}~\bibnamefont {Mathey}}, \bibinfo {author} {\bibfnamefont {J.}~\bibnamefont {Fr{\"o}hlich}}, \ and\ \bibinfo {author} {\bibfnamefont {W.}~\bibnamefont {Rodi}},\ }\bibfield  {title} {\enquote {\bibinfo {title} {Large-eddy simulation of the flow over a matrix of surface-mounted cubes},}\ }in\ \href@noop {} {\emph {\bibinfo {booktitle} {Ind. Environ. Appl. of DLES}}}\ (\bibinfo {organization} {Springer},\ \bibinfo {year} {1999})\ pp.\ \bibinfo {pages} {153--163}\BibitemShut {NoStop}%
\bibitem [{\citenamefont {Xie}\ and\ \citenamefont {Castro}(2006)}]{xie2006and}%
  \BibitemOpen
  \bibfield  {author} {\bibinfo {author} {\bibfnamefont {Z.}~\bibnamefont {Xie}}\ and\ \bibinfo {author} {\bibfnamefont {I.~P.}\ \bibnamefont {Castro}},\ }\bibfield  {title} {\enquote {\bibinfo {title} {{LES} and {RANS} for turbulent flow over arrays of wall-mounted obstacles},}\ }\href@noop {} {\bibfield  {journal} {\bibinfo  {journal} {Flow, Turbulence and Combustion}\ }\textbf {\bibinfo {volume} {76}},\ \bibinfo {pages} {291--312} (\bibinfo {year} {2006})}\BibitemShut {NoStop}%
\bibitem [{\citenamefont {Coceal}\ \emph {et~al.}(2006)\citenamefont {Coceal}, \citenamefont {Thomas}, \citenamefont {Castro},\ and\ \citenamefont {Belcher}}]{Coceal2006}%
  \BibitemOpen
  \bibfield  {author} {\bibinfo {author} {\bibfnamefont {O.}~\bibnamefont {Coceal}}, \bibinfo {author} {\bibfnamefont {T.~G.}\ \bibnamefont {Thomas}}, \bibinfo {author} {\bibfnamefont {I.~P.}\ \bibnamefont {Castro}}, \ and\ \bibinfo {author} {\bibfnamefont {S.~E.}\ \bibnamefont {Belcher}},\ }\bibfield  {title} {\enquote {\bibinfo {title} {{Mean flow and turbulence statistics over groups of urban-like cubical obstacles}},}\ }\href@noop {} {\bibfield  {journal} {\bibinfo  {journal} {Boundary-Layer Meteorology}\ }\textbf {\bibinfo {volume} {121}},\ \bibinfo {pages} {491--519} (\bibinfo {year} {2006})}\BibitemShut {NoStop}%
\bibitem [{\citenamefont {Blunn}\ \emph {et~al.}(2022)\citenamefont {Blunn}, \citenamefont {Coceal}, \citenamefont {Nazarian}, \citenamefont {Barlow}, \citenamefont {Plant}, \citenamefont {Bohnenstengel},\ and\ \citenamefont {Lean}}]{Blunn2022}%
  \BibitemOpen
  \bibfield  {author} {\bibinfo {author} {\bibfnamefont {L.~P.}\ \bibnamefont {Blunn}}, \bibinfo {author} {\bibfnamefont {O.}~\bibnamefont {Coceal}}, \bibinfo {author} {\bibfnamefont {N.}~\bibnamefont {Nazarian}}, \bibinfo {author} {\bibfnamefont {J.~F.}\ \bibnamefont {Barlow}}, \bibinfo {author} {\bibfnamefont {R.~S.}\ \bibnamefont {Plant}}, \bibinfo {author} {\bibfnamefont {S.~I.}\ \bibnamefont {Bohnenstengel}}, \ and\ \bibinfo {author} {\bibfnamefont {H.~W.}\ \bibnamefont {Lean}},\ }\bibfield  {title} {\enquote {\bibinfo {title} {{Turbulence characteristics across a range of idealized urban canopy geometries}},}\ }\href@noop {} {\bibfield  {journal} {\bibinfo  {journal} {Boundary-Layer Meteorology}\ }\textbf {\bibinfo {volume} {182}},\ \bibinfo {pages} {275--307} (\bibinfo {year} {2022})}\BibitemShut {NoStop}%
\bibitem [{\citenamefont {Cheng}\ and\ \citenamefont {Castro}(2002)}]{cheng2002near}%
  \BibitemOpen
  \bibfield  {author} {\bibinfo {author} {\bibfnamefont {H.}~\bibnamefont {Cheng}}\ and\ \bibinfo {author} {\bibfnamefont {I.~P.}\ \bibnamefont {Castro}},\ }\bibfield  {title} {\enquote {\bibinfo {title} {Near wall flow over urban-like roughness},}\ }\href@noop {} {\bibfield  {journal} {\bibinfo  {journal} {Boundary-Layer Meteorology}\ }\textbf {\bibinfo {volume} {104}},\ \bibinfo {pages} {229--259} (\bibinfo {year} {2002})}\BibitemShut {NoStop}%
\bibitem [{\citenamefont {Castro}, \citenamefont {Cheng},\ and\ \citenamefont {Reynolds}(2006)}]{castro2006turbulence}%
  \BibitemOpen
  \bibfield  {author} {\bibinfo {author} {\bibfnamefont {I.~P.}\ \bibnamefont {Castro}}, \bibinfo {author} {\bibfnamefont {H.}~\bibnamefont {Cheng}}, \ and\ \bibinfo {author} {\bibfnamefont {R.}~\bibnamefont {Reynolds}},\ }\bibfield  {title} {\enquote {\bibinfo {title} {Turbulence over urban-type roughness: deductions from wind-tunnel measurements},}\ }\href@noop {} {\bibfield  {journal} {\bibinfo  {journal} {Boundary-Layer Meteorology}\ }\textbf {\bibinfo {volume} {118}},\ \bibinfo {pages} {109--131} (\bibinfo {year} {2006})}\BibitemShut {NoStop}%
\bibitem [{\citenamefont {Inagaki}\ and\ \citenamefont {Kanda}(2008)}]{inagaki2008turbulent}%
  \BibitemOpen
  \bibfield  {author} {\bibinfo {author} {\bibfnamefont {A.}~\bibnamefont {Inagaki}}\ and\ \bibinfo {author} {\bibfnamefont {M.}~\bibnamefont {Kanda}},\ }\bibfield  {title} {\enquote {\bibinfo {title} {{T}urbulent flow similarity over an array of cubes in near-neutrally stratified atmospheric flow},}\ }\href@noop {} {\bibfield  {journal} {\bibinfo  {journal} {Journal of Fluid Mechanics}\ }\textbf {\bibinfo {volume} {615}},\ \bibinfo {pages} {101--120} (\bibinfo {year} {2008})}\BibitemShut {NoStop}%
\bibitem [{\citenamefont {Chew}, \citenamefont {Aliabadi},\ and\ \citenamefont {Norford}(2018)}]{Chew2018}%
  \BibitemOpen
  \bibfield  {author} {\bibinfo {author} {\bibfnamefont {L.~W.}\ \bibnamefont {Chew}}, \bibinfo {author} {\bibfnamefont {A.~A.}\ \bibnamefont {Aliabadi}}, \ and\ \bibinfo {author} {\bibfnamefont {L.~K.}\ \bibnamefont {Norford}},\ }\bibfield  {title} {\enquote {\bibinfo {title} {Flows across high aspect ratio street canyons: Reynolds number independence revisited},}\ }\href {\doibase 10.1007/s10652-018-9601-0} {\bibfield  {journal} {\bibinfo  {journal} {Environmental Fluid Mechanics}\ }\textbf {\bibinfo {volume} {18}},\ \bibinfo {pages} {1275--1291} (\bibinfo {year} {2018})}\BibitemShut {NoStop}%
\bibitem [{\citenamefont {Barlow}\ and\ \citenamefont {Coceal}(2009)}]{Barlow2009}%
  \BibitemOpen
  \bibfield  {author} {\bibinfo {author} {\bibfnamefont {J.~F.}\ \bibnamefont {Barlow}}\ and\ \bibinfo {author} {\bibfnamefont {O.}~\bibnamefont {Coceal}},\ }\bibfield  {title} {\enquote {\bibinfo {title} {A review of urban roughness sublayer turbulence},}\ }\href@noop {} {\bibfield  {journal} {\bibinfo  {journal} {Meteorology Research and Development Technical Report}\ }\textbf {\bibinfo {volume} {527}} (\bibinfo {year} {2009})}\BibitemShut {NoStop}%
\bibitem [{\citenamefont {Antoniou}\ \emph {et~al.}(2019)\citenamefont {Antoniou}, \citenamefont {Montazeri}, \citenamefont {Neophytou},\ and\ \citenamefont {Blocken}}]{antoniou2019cfd}%
  \BibitemOpen
  \bibfield  {author} {\bibinfo {author} {\bibfnamefont {N.}~\bibnamefont {Antoniou}}, \bibinfo {author} {\bibfnamefont {H.}~\bibnamefont {Montazeri}}, \bibinfo {author} {\bibfnamefont {M.}~\bibnamefont {Neophytou}}, \ and\ \bibinfo {author} {\bibfnamefont {B.}~\bibnamefont {Blocken}},\ }\bibfield  {title} {\enquote {\bibinfo {title} {{CFD} simulation of urban microclimate: {V}alidation using high-resolution field measurements},}\ }\href@noop {} {\bibfield  {journal} {\bibinfo  {journal} {Science of the Total Environment}\ }\textbf {\bibinfo {volume} {695}},\ \bibinfo {pages} {133743} (\bibinfo {year} {2019})}\BibitemShut {NoStop}%
\bibitem [{\citenamefont {Brozovsky}, \citenamefont {Simonsen},\ and\ \citenamefont {Gaitani}(2021)}]{brozovsky2021validation}%
  \BibitemOpen
  \bibfield  {author} {\bibinfo {author} {\bibfnamefont {J.}~\bibnamefont {Brozovsky}}, \bibinfo {author} {\bibfnamefont {A.}~\bibnamefont {Simonsen}}, \ and\ \bibinfo {author} {\bibfnamefont {N.}~\bibnamefont {Gaitani}},\ }\bibfield  {title} {\enquote {\bibinfo {title} {Validation of a {CFD} model for the evaluation of urban microclimate at high latitudes: A case study in {T}rondheim, {N}orway},}\ }\href@noop {} {\bibfield  {journal} {\bibinfo  {journal} {Building and Environment}\ }\textbf {\bibinfo {volume} {205}},\ \bibinfo {pages} {108175} (\bibinfo {year} {2021})}\BibitemShut {NoStop}%
\bibitem [{\citenamefont {Stuck}\ \emph {et~al.}(2021)\citenamefont {Stuck}, \citenamefont {Vidal}, \citenamefont {Torres}, \citenamefont {Nagib}, \citenamefont {Wark},\ and\ \citenamefont {Vinuesa}}]{Stuck2021}%
  \BibitemOpen
  \bibfield  {author} {\bibinfo {author} {\bibfnamefont {M.}~\bibnamefont {Stuck}}, \bibinfo {author} {\bibfnamefont {A.}~\bibnamefont {Vidal}}, \bibinfo {author} {\bibfnamefont {P.}~\bibnamefont {Torres}}, \bibinfo {author} {\bibfnamefont {H.~M.}\ \bibnamefont {Nagib}}, \bibinfo {author} {\bibfnamefont {C.}~\bibnamefont {Wark}}, \ and\ \bibinfo {author} {\bibfnamefont {R.}~\bibnamefont {Vinuesa}},\ }\bibfield  {title} {\enquote {\bibinfo {title} {Spectral-element simulation of the turbulent flow in an urban environment},}\ }\href {\doibase 10.3390/app11146472} {\bibfield  {journal} {\bibinfo  {journal} {Applied Sciences}\ }\textbf {\bibinfo {volume} {11}},\ \bibinfo {pages} {6472} (\bibinfo {year} {2021})}\BibitemShut {NoStop}%
\bibitem [{\citenamefont {Li}\ \emph {et~al.}(2010)\citenamefont {Li}, \citenamefont {Britter}, \citenamefont {Koh}, \citenamefont {Norford}, \citenamefont {Liu}, \citenamefont {Entekhabi},\ and\ \citenamefont {Leung}}]{li2010large}%
  \BibitemOpen
  \bibfield  {author} {\bibinfo {author} {\bibfnamefont {X.-X.}\ \bibnamefont {Li}}, \bibinfo {author} {\bibfnamefont {R.~E.}\ \bibnamefont {Britter}}, \bibinfo {author} {\bibfnamefont {T.~Y.}\ \bibnamefont {Koh}}, \bibinfo {author} {\bibfnamefont {L.~K.}\ \bibnamefont {Norford}}, \bibinfo {author} {\bibfnamefont {C.-H.}\ \bibnamefont {Liu}}, \bibinfo {author} {\bibfnamefont {D.}~\bibnamefont {Entekhabi}}, \ and\ \bibinfo {author} {\bibfnamefont {D.~Y.~C.}\ \bibnamefont {Leung}},\ }\bibfield  {title} {\enquote {\bibinfo {title} {Large-eddy simulation of flow and pollutant transport in urban street canyons with ground heating},}\ }\href@noop {} {\bibfield  {journal} {\bibinfo  {journal} {Boundary-Layer Meteorology}\ }\textbf {\bibinfo {volume} {137}},\ \bibinfo {pages} {187--204} (\bibinfo {year} {2010})}\BibitemShut {NoStop}%
\bibitem [{\citenamefont {Xie}\ and\ \citenamefont {Castro}(2009)}]{xie2009large}%
  \BibitemOpen
  \bibfield  {author} {\bibinfo {author} {\bibfnamefont {Z.}~\bibnamefont {Xie}}\ and\ \bibinfo {author} {\bibfnamefont {I.~P.}\ \bibnamefont {Castro}},\ }\bibfield  {title} {\enquote {\bibinfo {title} {Large-eddy simulation for flow and dispersion in urban streets},}\ }\href@noop {} {\bibfield  {journal} {\bibinfo  {journal} {Atmospheric Environment}\ }\textbf {\bibinfo {volume} {43}},\ \bibinfo {pages} {2174--2185} (\bibinfo {year} {2009})}\BibitemShut {NoStop}%
\bibitem [{\citenamefont {Gousseau}\ \emph {et~al.}(2011)\citenamefont {Gousseau}, \citenamefont {Blocken}, \citenamefont {Stathopoulos},\ and\ \citenamefont {Van~Heijst}}]{gousseau2011cfd}%
  \BibitemOpen
  \bibfield  {author} {\bibinfo {author} {\bibfnamefont {P.}~\bibnamefont {Gousseau}}, \bibinfo {author} {\bibfnamefont {B.}~\bibnamefont {Blocken}}, \bibinfo {author} {\bibfnamefont {T.}~\bibnamefont {Stathopoulos}}, \ and\ \bibinfo {author} {\bibfnamefont {G.~J.~F.}\ \bibnamefont {Van~Heijst}},\ }\bibfield  {title} {\enquote {\bibinfo {title} {{CFD} simulation of near-field pollutant dispersion on a high-resolution grid: a case study by {LES} and {RANS} for a building group in downtown {M}ontreal},}\ }\href@noop {} {\bibfield  {journal} {\bibinfo  {journal} {Atmospheric Environment}\ }\textbf {\bibinfo {volume} {45}},\ \bibinfo {pages} {428--438} (\bibinfo {year} {2011})}\BibitemShut {NoStop}%
\bibitem [{\citenamefont {Hassan}\ \emph {et~al.}(2022)\citenamefont {Hassan}, \citenamefont {Akter}, \citenamefont {Nag}, \citenamefont {Molla}, \citenamefont {Khan},\ and\ \citenamefont {Hasan}}]{hassan2022large}%
  \BibitemOpen
  \bibfield  {author} {\bibinfo {author} {\bibfnamefont {S.}~\bibnamefont {Hassan}}, \bibinfo {author} {\bibfnamefont {U.~H.}\ \bibnamefont {Akter}}, \bibinfo {author} {\bibfnamefont {P.}~\bibnamefont {Nag}}, \bibinfo {author} {\bibfnamefont {M.~M.}\ \bibnamefont {Molla}}, \bibinfo {author} {\bibfnamefont {A.}~\bibnamefont {Khan}}, \ and\ \bibinfo {author} {\bibfnamefont {M.~F.}\ \bibnamefont {Hasan}},\ }\bibfield  {title} {\enquote {\bibinfo {title} {{L}arge-eddy simulation of airflow and pollutant dispersion in a model street canyon intersection of {D}haka city},}\ }\href@noop {} {\bibfield  {journal} {\bibinfo  {journal} {Atmosphere}\ }\textbf {\bibinfo {volume} {13}},\ \bibinfo {pages} {1028} (\bibinfo {year} {2022})}\BibitemShut {NoStop}%
\bibitem [{\citenamefont {Tolias}\ \emph {et~al.}(2018)\citenamefont {Tolias}, \citenamefont {Koutsourakis}, \citenamefont {Hertwig}, \citenamefont {Efthimiou}, \citenamefont {Venetsanos},\ and\ \citenamefont {Bartzis}}]{tolias2018large}%
  \BibitemOpen
  \bibfield  {author} {\bibinfo {author} {\bibfnamefont {I.~C.}\ \bibnamefont {Tolias}}, \bibinfo {author} {\bibfnamefont {N.}~\bibnamefont {Koutsourakis}}, \bibinfo {author} {\bibfnamefont {D.}~\bibnamefont {Hertwig}}, \bibinfo {author} {\bibfnamefont {G.~C.}\ \bibnamefont {Efthimiou}}, \bibinfo {author} {\bibfnamefont {A.~G.}\ \bibnamefont {Venetsanos}}, \ and\ \bibinfo {author} {\bibfnamefont {J.~G.}\ \bibnamefont {Bartzis}},\ }\bibfield  {title} {\enquote {\bibinfo {title} {Large eddy simulation study on the structure of turbulent flow in a complex city},}\ }\href@noop {} {\bibfield  {journal} {\bibinfo  {journal} {Journal of Wind Engineering and Industrial Aerodynamics}\ }\textbf {\bibinfo {volume} {177}},\ \bibinfo {pages} {101--116} (\bibinfo {year} {2018})}\BibitemShut {NoStop}%
\bibitem [{\citenamefont {Kirkil}\ and\ \citenamefont {Lin}(2020)}]{kirkil2020large}%
  \BibitemOpen
  \bibfield  {author} {\bibinfo {author} {\bibfnamefont {G.}~\bibnamefont {Kirkil}}\ and\ \bibinfo {author} {\bibfnamefont {C.}~\bibnamefont {Lin}},\ }\bibfield  {title} {\enquote {\bibinfo {title} {{L}arge eddy simulation of wind flow over a realistic urban area},}\ }\href@noop {} {\bibfield  {journal} {\bibinfo  {journal} {Computation}\ }\textbf {\bibinfo {volume} {8}},\ \bibinfo {pages} {47} (\bibinfo {year} {2020})}\BibitemShut {NoStop}%
\bibitem [{\citenamefont {Auvinen}\ \emph {et~al.}(2020)\citenamefont {Auvinen}, \citenamefont {Boi}, \citenamefont {Hellsten}, \citenamefont {Tanhuanp{\"a}{\"a}},\ and\ \citenamefont {J{\"a}rvi}}]{auvinen2020study}%
  \BibitemOpen
  \bibfield  {author} {\bibinfo {author} {\bibfnamefont {M.}~\bibnamefont {Auvinen}}, \bibinfo {author} {\bibfnamefont {S.}~\bibnamefont {Boi}}, \bibinfo {author} {\bibfnamefont {A.}~\bibnamefont {Hellsten}}, \bibinfo {author} {\bibfnamefont {T.}~\bibnamefont {Tanhuanp{\"a}{\"a}}}, \ and\ \bibinfo {author} {\bibfnamefont {L.}~\bibnamefont {J{\"a}rvi}},\ }\bibfield  {title} {\enquote {\bibinfo {title} {Study of realistic urban boundary layer turbulence with high-resolution large-eddy simulation},}\ }\href@noop {} {\bibfield  {journal} {\bibinfo  {journal} {Atmosphere}\ }\textbf {\bibinfo {volume} {11}},\ \bibinfo {pages} {201} (\bibinfo {year} {2020})}\BibitemShut {NoStop}%
\bibitem [{\citenamefont {Cheng}\ and\ \citenamefont {Yang}(2023)}]{cheng2023scaling}%
  \BibitemOpen
  \bibfield  {author} {\bibinfo {author} {\bibfnamefont {W.}~\bibnamefont {Cheng}}\ and\ \bibinfo {author} {\bibfnamefont {Y.}~\bibnamefont {Yang}},\ }\bibfield  {title} {\enquote {\bibinfo {title} {{S}caling of flows over realistic urban geometries: {A} large-eddy simulation study},}\ }\href@noop {} {\bibfield  {journal} {\bibinfo  {journal} {Boundary-Layer Meteorology}\ }\textbf {\bibinfo {volume} {186}},\ \bibinfo {pages} {125--144} (\bibinfo {year} {2023})}\BibitemShut {NoStop}%
\bibitem [{\citenamefont {Oh}, \citenamefont {Yang},\ and\ \citenamefont {Choi}(2024)}]{OH2024105682}%
  \BibitemOpen
  \bibfield  {author} {\bibinfo {author} {\bibfnamefont {G.}~\bibnamefont {Oh}}, \bibinfo {author} {\bibfnamefont {M.}~\bibnamefont {Yang}}, \ and\ \bibinfo {author} {\bibfnamefont {J.}~\bibnamefont {Choi}},\ }\bibfield  {title} {\enquote {\bibinfo {title} {Large-eddy simulation-based wind and thermal comfort assessment in urban environments},}\ }\href {\doibase https://doi.org/10.1016/j.jweia.2024.105682} {\bibfield  {journal} {\bibinfo  {journal} {Journal of Wind Engineering and Industrial Aerodynamics}\ }\textbf {\bibinfo {volume} {246}},\ \bibinfo {pages} {105682} (\bibinfo {year} {2024})}\BibitemShut {NoStop}%
\bibitem [{\citenamefont {Shaukat}\ and\ \citenamefont {Giljarhus}(2024)}]{Shaukat2024}%
  \BibitemOpen
  \bibfield  {author} {\bibinfo {author} {\bibfnamefont {U.}~\bibnamefont {Shaukat}}\ and\ \bibinfo {author} {\bibfnamefont {K.~E.~T.}\ \bibnamefont {Giljarhus}},\ }\bibfield  {title} {\enquote {\bibinfo {title} {Precursor turbulent inflow dataset for large eddy simulation of a semi-idealized european generic city},}\ }\href@noop {} {\bibfield  {journal} {\bibinfo  {journal} {Data in Brief}\ }\textbf {\bibinfo {volume} {54}},\ \bibinfo {pages} {110467} (\bibinfo {year} {2024})}\BibitemShut {NoStop}%
\bibitem [{\citenamefont {Tian}\ \emph {et~al.}(2024)\citenamefont {Tian}, \citenamefont {Ma}, \citenamefont {Chen}, \citenamefont {Wan},\ and\ \citenamefont {Chen}}]{Tian2024}%
  \BibitemOpen
  \bibfield  {author} {\bibinfo {author} {\bibfnamefont {G.}~\bibnamefont {Tian}}, \bibinfo {author} {\bibfnamefont {Y.}~\bibnamefont {Ma}}, \bibinfo {author} {\bibfnamefont {Y.}~\bibnamefont {Chen}}, \bibinfo {author} {\bibfnamefont {M.}~\bibnamefont {Wan}}, \ and\ \bibinfo {author} {\bibfnamefont {S.}~\bibnamefont {Chen}},\ }\bibfield  {title} {\enquote {\bibinfo {title} {Impact of urban canopy characteristics on turbulence dynamics},}\ }\href {\doibase 10.1016/j.buildenv.2024.111183} {\bibfield  {journal} {\bibinfo  {journal} {Building and Environment}\ }\textbf {\bibinfo {volume} {250}},\ \bibinfo {pages} {111183} (\bibinfo {year} {2024})}\BibitemShut {NoStop}%
\bibitem [{\citenamefont {Toparlar}\ \emph {et~al.}(2017)\citenamefont {Toparlar}, \citenamefont {Blocken}, \citenamefont {Maiheu},\ and\ \citenamefont {van Heijst}}]{toparlar2017review}%
  \BibitemOpen
  \bibfield  {author} {\bibinfo {author} {\bibfnamefont {Y.}~\bibnamefont {Toparlar}}, \bibinfo {author} {\bibfnamefont {B.}~\bibnamefont {Blocken}}, \bibinfo {author} {\bibfnamefont {B.}~\bibnamefont {Maiheu}}, \ and\ \bibinfo {author} {\bibfnamefont {G.~J.~F.}\ \bibnamefont {van Heijst}},\ }\bibfield  {title} {\enquote {\bibinfo {title} {A review on the {CFD} analysis of urban microclimate},}\ }\href@noop {} {\bibfield  {journal} {\bibinfo  {journal} {Renewable and Sustainable Energy Reviews}\ }\textbf {\bibinfo {volume} {80}},\ \bibinfo {pages} {1613--1640} (\bibinfo {year} {2017})}\BibitemShut {NoStop}%
\bibitem [{\citenamefont {Moon}\ \emph {et~al.}(2014)\citenamefont {Moon}, \citenamefont {Hwang}, \citenamefont {Kim}, \citenamefont {Lee},\ and\ \citenamefont {Choi}}]{moon2014large}%
  \BibitemOpen
  \bibfield  {author} {\bibinfo {author} {\bibfnamefont {K.}~\bibnamefont {Moon}}, \bibinfo {author} {\bibfnamefont {J.}~\bibnamefont {Hwang}}, \bibinfo {author} {\bibfnamefont {B.}~\bibnamefont {Kim}}, \bibinfo {author} {\bibfnamefont {C.}~\bibnamefont {Lee}}, \ and\ \bibinfo {author} {\bibfnamefont {J.}~\bibnamefont {Choi}},\ }\bibfield  {title} {\enquote {\bibinfo {title} {{L}arge-eddy simulation of turbulent flow and dispersion over a complex urban street canyon},}\ }\href@noop {} {\bibfield  {journal} {\bibinfo  {journal} {Environmental Fluid Mechanics}\ }\textbf {\bibinfo {volume} {14}},\ \bibinfo {pages} {1381--1403} (\bibinfo {year} {2014})}\BibitemShut {NoStop}%
\bibitem [{\citenamefont {Chew}, \citenamefont {Glicksman},\ and\ \citenamefont {Norford}(2018)}]{chew2018buoyant}%
  \BibitemOpen
  \bibfield  {author} {\bibinfo {author} {\bibfnamefont {L.~W.}\ \bibnamefont {Chew}}, \bibinfo {author} {\bibfnamefont {L.~R.}\ \bibnamefont {Glicksman}}, \ and\ \bibinfo {author} {\bibfnamefont {L.~K.}\ \bibnamefont {Norford}},\ }\bibfield  {title} {\enquote {\bibinfo {title} {{B}uoyant flows in street canyons: Comparison of {RANS} and {LES} at reduced and full scales},}\ }\href@noop {} {\bibfield  {journal} {\bibinfo  {journal} {Building and Environment}\ }\textbf {\bibinfo {volume} {146}},\ \bibinfo {pages} {77--87} (\bibinfo {year} {2018})}\BibitemShut {NoStop}%
\bibitem [{\citenamefont {Brown}\ \emph {et~al.}(2001)\citenamefont {Brown}, \citenamefont {Lawson}, \citenamefont {DeCroix},\ and\ \citenamefont {Lee}}]{BRO01}%
  \BibitemOpen
  \bibfield  {author} {\bibinfo {author} {\bibfnamefont {M.~J.}\ \bibnamefont {Brown}}, \bibinfo {author} {\bibfnamefont {R.~E.}\ \bibnamefont {Lawson}}, \bibinfo {author} {\bibfnamefont {D.~S.}\ \bibnamefont {DeCroix}}, \ and\ \bibinfo {author} {\bibfnamefont {R.~L.}\ \bibnamefont {Lee}},\ }\href@noop {} {\enquote {\bibinfo {title} {{Comparison of Centreline Velocity Measurements Obtained around 2D and 3D Building Arrays in a Wind Tunnel}},}\ }\bibinfo {type} {Tech. Rep.}\ \bibinfo {number} {LA-UR-01-4138}\ (\bibinfo  {institution} {Los Alamos National Laboratory},\ \bibinfo {year} {2001})\BibitemShut {NoStop}%
\bibitem [{\citenamefont {Brown}\ and\ \citenamefont {Lawson}(2001)}]{brown2001comparison}%
  \BibitemOpen
  \bibfield  {author} {\bibinfo {author} {\bibfnamefont {M.~J.}\ \bibnamefont {Brown}}\ and\ \bibinfo {author} {\bibfnamefont {R.~E.}\ \bibnamefont {Lawson}},\ }\href@noop {} {\enquote {\bibinfo {title} {Comparison of centerline velocity measurements obtained around 2{D} and 3{D} building arrays in a wind tunnel},}\ }\bibinfo {type} {Tech. Rep.}\ \bibinfo {number} {LA-UR-01-4131}\ (\bibinfo  {institution} {Los Alamos National Lab},\ \bibinfo {year} {2001})\BibitemShut {NoStop}%
\bibitem [{\citenamefont {Leitl}\ and\ \citenamefont {Harms}(2017)}]{leitl2024}%
  \BibitemOpen
  \bibfield  {author} {\bibinfo {author} {\bibfnamefont {B.}~\bibnamefont {Leitl}}\ and\ \bibinfo {author} {\bibfnamefont {F.}~\bibnamefont {Harms}},\ }\href@noop {} {\enquote {\bibinfo {title} {{CEDVAL-LES}: {COMPLEXITY} 3},}\ }\bibinfo {howpublished} {\url{https://www.mi.uni-hamburg.de/en/arbeitsgruppen/windkanallabor/data-sets.html}} (\bibinfo {year} {2017}),\ \bibinfo {note} {{U}niv. of Hamburg, Environmental Wind Tunnel Laboratory. Accessed: 2024-11-20}\BibitemShut {NoStop}%
\bibitem [{\citenamefont {Hertwig}\ \emph {et~al.}(2012)\citenamefont {Hertwig}, \citenamefont {Efthimiou}, \citenamefont {Bartzis},\ and\ \citenamefont {Leitl}}]{Hertwig2012}%
  \BibitemOpen
  \bibfield  {author} {\bibinfo {author} {\bibfnamefont {D.}~\bibnamefont {Hertwig}}, \bibinfo {author} {\bibfnamefont {G.~C.}\ \bibnamefont {Efthimiou}}, \bibinfo {author} {\bibfnamefont {J.~G.}\ \bibnamefont {Bartzis}}, \ and\ \bibinfo {author} {\bibfnamefont {B.}~\bibnamefont {Leitl}},\ }\bibfield  {title} {\enquote {\bibinfo {title} {{CFD-RANS model validation of turbulent flow in a semi-idealized urban canopy}},}\ }\href@noop {} {\bibfield  {journal} {\bibinfo  {journal} {Journal of Wind Engineering and Industrial Aerodynamics}\ }\textbf {\bibinfo {volume} {111}},\ \bibinfo {pages} {61--72} (\bibinfo {year} {2012})}\BibitemShut {NoStop}%
\bibitem [{\citenamefont {Vreman}(2004)}]{vreman2004}%
  \BibitemOpen
  \bibfield  {author} {\bibinfo {author} {\bibfnamefont {A.~W.}\ \bibnamefont {Vreman}},\ }\bibfield  {title} {\enquote {\bibinfo {title} {An eddy-viscosity subgrid-scale model for turbulent shear flow: {A}lgebraic theory and applications},}\ }\href@noop {} {\bibfield  {journal} {\bibinfo  {journal} {Physics of Fluids}\ }\textbf {\bibinfo {volume} {16}},\ \bibinfo {pages} {3670--3681} (\bibinfo {year} {2004})}\BibitemShut {NoStop}%
\bibitem [{\citenamefont {Gasparino}, \citenamefont {Spiga},\ and\ \citenamefont {Lehmkuhl}(2024{\natexlab{a}})}]{gasparino2024}%
  \BibitemOpen
  \bibfield  {author} {\bibinfo {author} {\bibfnamefont {L.}~\bibnamefont {Gasparino}}, \bibinfo {author} {\bibfnamefont {F.}~\bibnamefont {Spiga}}, \ and\ \bibinfo {author} {\bibfnamefont {O.}~\bibnamefont {Lehmkuhl}},\ }\bibfield  {title} {\enquote {\bibinfo {title} {{SOD2D}: A {G}{P}{U}-enabled {S}pectral {F}inite {E}lements {M}ethod for compressible scale-resolving simulations},}\ }\href@noop {} {\bibfield  {journal} {\bibinfo  {journal} {Computer Physics Communications}\ }\textbf {\bibinfo {volume} {297}},\ \bibinfo {pages} {109067} (\bibinfo {year} {2024}{\natexlab{a}})}\BibitemShut {NoStop}%
\bibitem [{\citenamefont {Gasparino}, \citenamefont {Spiga},\ and\ \citenamefont {Lehmkuhl}(2024{\natexlab{b}})}]{sod2drepo}%
  \BibitemOpen
  \bibfield  {author} {\bibinfo {author} {\bibfnamefont {L.}~\bibnamefont {Gasparino}}, \bibinfo {author} {\bibfnamefont {F.}~\bibnamefont {Spiga}}, \ and\ \bibinfo {author} {\bibfnamefont {O.}~\bibnamefont {Lehmkuhl}},\ }\href@noop {} {\enquote {\bibinfo {title} {{SOD2D} gitlab repository},}\ }\bibinfo {howpublished} {\url{https://gitlab.com/bsc_sod2d/sod2d_gitlab}} (\bibinfo {year} {2024}{\natexlab{b}})\BibitemShut {NoStop}%
\bibitem [{\citenamefont {Kennedy}\ and\ \citenamefont {Gruber}(2008)}]{kennedy2008}%
  \BibitemOpen
  \bibfield  {author} {\bibinfo {author} {\bibfnamefont {C.~A.}\ \bibnamefont {Kennedy}}\ and\ \bibinfo {author} {\bibfnamefont {A.}~\bibnamefont {Gruber}},\ }\bibfield  {title} {\enquote {\bibinfo {title} {{R}educed aliasing formulations of the convective terms within the {N}avier--{S}tokes equations for a compressible fluid},}\ }\href@noop {} {\bibfield  {journal} {\bibinfo  {journal} {Journal of Computational Physics}\ }\textbf {\bibinfo {volume} {227}},\ \bibinfo {pages} {1676--1700} (\bibinfo {year} {2008})}\BibitemShut {NoStop}%
\bibitem [{\citenamefont {Karniadakis}, \citenamefont {Israeli},\ and\ \citenamefont {Orszag}(1991)}]{Karniadakis1991}%
  \BibitemOpen
  \bibfield  {author} {\bibinfo {author} {\bibfnamefont {G.~E.}\ \bibnamefont {Karniadakis}}, \bibinfo {author} {\bibfnamefont {M.}~\bibnamefont {Israeli}}, \ and\ \bibinfo {author} {\bibfnamefont {S.~A.}\ \bibnamefont {Orszag}},\ }\bibfield  {title} {\enquote {\bibinfo {title} {{H}igh-order splitting methods for the incompressible {N}avier-{S}tokes equations},}\ }\href@noop {} {\bibfield  {journal} {\bibinfo  {journal} {Journal of Computational Physics}\ }\textbf {\bibinfo {volume} {97}},\ \bibinfo {pages} {414--443} (\bibinfo {year} {1991})}\BibitemShut {NoStop}%
\bibitem [{\citenamefont {Folk}, \citenamefont {Cheng},\ and\ \citenamefont {Yates}(1999)}]{folk1999hdf5}%
  \BibitemOpen
  \bibfield  {author} {\bibinfo {author} {\bibfnamefont {M.}~\bibnamefont {Folk}}, \bibinfo {author} {\bibfnamefont {A.}~\bibnamefont {Cheng}}, \ and\ \bibinfo {author} {\bibfnamefont {K.}~\bibnamefont {Yates}},\ }\bibfield  {title} {\enquote {\bibinfo {title} {{HDF5: A file format and I/O library for high performance computing applications}},}\ }in\ \href@noop {} {\emph {\bibinfo {booktitle} {Proceedings of Supercomputing}}},\ Vol.~\bibinfo {volume} {99}\ (\bibinfo {year} {1999})\ pp.\ \bibinfo {pages} {5--33}\BibitemShut {NoStop}%
\bibitem [{\citenamefont {Franke}\ \emph {et~al.}(2007)\citenamefont {Franke}, \citenamefont {Hellsten}, \citenamefont {Schl{\"u}nzen},\ and\ \citenamefont {Carissimo}}]{franke2007best}%
  \BibitemOpen
  \bibfield  {author} {\bibinfo {author} {\bibfnamefont {J.}~\bibnamefont {Franke}}, \bibinfo {author} {\bibfnamefont {A.}~\bibnamefont {Hellsten}}, \bibinfo {author} {\bibfnamefont {H.}~\bibnamefont {Schl{\"u}nzen}}, \ and\ \bibinfo {author} {\bibfnamefont {B.}~\bibnamefont {Carissimo}},\ }\href {https://hal.science/hal-04181390} {\enquote {\bibinfo {title} {{Best Practice Guideline for the {CFD} Simulation of Flows in The Urban Environment}},}\ }\bibinfo {type} {Tech. Rep.}\ \bibinfo {number} {hal-04181390}\ (\bibinfo  {institution} {{COST European Cooperation in Science and Technology}},\ \bibinfo {year} {2007})\BibitemShut {NoStop}%
\bibitem [{\citenamefont {Xie}\ and\ \citenamefont {Castro}(2008)}]{Xie2008}%
  \BibitemOpen
  \bibfield  {author} {\bibinfo {author} {\bibfnamefont {Z.~T.}\ \bibnamefont {Xie}}\ and\ \bibinfo {author} {\bibfnamefont {I.~P.}\ \bibnamefont {Castro}},\ }\bibfield  {title} {\enquote {\bibinfo {title} {{Efficient generation of inflow conditions for large eddy simulation of street-scale flows}},}\ }\href@noop {} {\bibfield  {journal} {\bibinfo  {journal} {Flow, Turbulence and Combustion}\ }\textbf {\bibinfo {volume} {81}},\ \bibinfo {pages} {449--470} (\bibinfo {year} {2008})}\BibitemShut {NoStop}%
\bibitem [{\citenamefont {Oke}(1988)}]{oke1988street}%
  \BibitemOpen
  \bibfield  {author} {\bibinfo {author} {\bibfnamefont {T.~R.}\ \bibnamefont {Oke}},\ }\bibfield  {title} {\enquote {\bibinfo {title} {Street design and urban canopy layer climate},}\ }\href@noop {} {\bibfield  {journal} {\bibinfo  {journal} {Energy and Buildings}\ }\textbf {\bibinfo {volume} {11}},\ \bibinfo {pages} {103--113} (\bibinfo {year} {1988})}\BibitemShut {NoStop}%
\bibitem [{\citenamefont {Vranckx}\ \emph {et~al.}(2015)\citenamefont {Vranckx}, \citenamefont {Vos}, \citenamefont {Maiheu},\ and\ \citenamefont {Janssen}}]{Vranckx2015}%
  \BibitemOpen
  \bibfield  {author} {\bibinfo {author} {\bibfnamefont {S.}~\bibnamefont {Vranckx}}, \bibinfo {author} {\bibfnamefont {P.}~\bibnamefont {Vos}}, \bibinfo {author} {\bibfnamefont {B.}~\bibnamefont {Maiheu}}, \ and\ \bibinfo {author} {\bibfnamefont {S.}~\bibnamefont {Janssen}},\ }\bibfield  {title} {\enquote {\bibinfo {title} {{Impact of trees on pollutant dispersion in street canyons: A numerical study of the annual average effects in Antwerp, Belgium}},}\ }\href {\doibase 10.1016/j.scitotenv.2015.06.032} {\bibfield  {journal} {\bibinfo  {journal} {Science of the Total Environment}\ }\textbf {\bibinfo {volume} {532}},\ \bibinfo {pages} {474--483} (\bibinfo {year} {2015})}\BibitemShut {NoStop}%
\bibitem [{\citenamefont {Yu}\ and\ \citenamefont {Th{\'{e}}}(2017)}]{Yu2017}%
  \BibitemOpen
  \bibfield  {author} {\bibinfo {author} {\bibfnamefont {H.}~\bibnamefont {Yu}}\ and\ \bibinfo {author} {\bibfnamefont {J.}~\bibnamefont {Th{\'{e}}}},\ }\href {\doibase 10.1080/10962247.2016.1232667} {\bibfield  {journal} {\bibinfo  {journal} {Journal of the Air and Waste Management Association}\ }\textbf {\bibinfo {volume} {67}},\ \bibinfo {pages} {517--536} (\bibinfo {year} {2017})}\BibitemShut {NoStop}%
\bibitem [{\citenamefont {Ding}\ \emph {et~al.}(2022)\citenamefont {Ding}, \citenamefont {Zhou}, \citenamefont {Wu},\ and\ \citenamefont {Chen}}]{Ding2022}%
  \BibitemOpen
  \bibfield  {author} {\bibinfo {author} {\bibfnamefont {P.}~\bibnamefont {Ding}}, \bibinfo {author} {\bibfnamefont {X.}~\bibnamefont {Zhou}}, \bibinfo {author} {\bibfnamefont {H.}~\bibnamefont {Wu}}, \ and\ \bibinfo {author} {\bibfnamefont {Q.}~\bibnamefont {Chen}},\ }\bibfield  {title} {\enquote {\bibinfo {title} {{An efficient numerical approach for simulating airflows around an isolated building}},}\ }\href@noop {} {\bibfield  {journal} {\bibinfo  {journal} {Building and Environment}\ }\textbf {\bibinfo {volume} {210}} (\bibinfo {year} {2022})}\BibitemShut {NoStop}%
\bibitem [{\citenamefont {Schatzmann}, \citenamefont {Olesen},\ and\ \citenamefont {Franke}(2010)}]{Schatzmann2010}%
  \BibitemOpen
  \bibfield  {author} {\bibinfo {author} {\bibfnamefont {M.}~\bibnamefont {Schatzmann}}, \bibinfo {author} {\bibfnamefont {H.}~\bibnamefont {Olesen}}, \ and\ \bibinfo {author} {\bibfnamefont {J.}~\bibnamefont {Franke}},\ }\href@noop {} {\emph {\bibinfo {title} {{COST} 732 {M}odel {E}valuation {C}ase {S}tudies: Approach and Results}}}\ (\bibinfo  {publisher} {University of Hamburg},\ \bibinfo {year} {2010})\ p.~\bibinfo {pages} {44}\BibitemShut {NoStop}%
\bibitem [{\citenamefont {Franke}, \citenamefont {Sturm},\ and\ \citenamefont {Kalmbach}(2012)}]{Franke2012}%
  \BibitemOpen
  \bibfield  {author} {\bibinfo {author} {\bibfnamefont {J.}~\bibnamefont {Franke}}, \bibinfo {author} {\bibfnamefont {M.}~\bibnamefont {Sturm}}, \ and\ \bibinfo {author} {\bibfnamefont {C.}~\bibnamefont {Kalmbach}},\ }\bibfield  {title} {\enquote {\bibinfo {title} {{Validation of OpenFOAM 1.6.x with the German VDI guideline for obstacle resolving micro-scale models}},}\ }\href {\doibase 10.1016/j.jweia.2012.02.021} {\bibfield  {journal} {\bibinfo  {journal} {Journal of Wind Engineering and Industrial Aerodynamics}\ }\textbf {\bibinfo {volume} {104-106}},\ \bibinfo {pages} {350--359} (\bibinfo {year} {2012})}\BibitemShut {NoStop}%
\bibitem [{\citenamefont {Kakosimos}\ and\ \citenamefont {Assael}(2013)}]{Kakosimos2013}%
  \BibitemOpen
  \bibfield  {author} {\bibinfo {author} {\bibfnamefont {K.~E.}\ \bibnamefont {Kakosimos}}\ and\ \bibinfo {author} {\bibfnamefont {M.~J.}\ \bibnamefont {Assael}},\ }\bibfield  {title} {\enquote {\bibinfo {title} {{Application of Detached Eddy Simulation to neighbourhood scale gases atmospheric dispersion modelling}},}\ }\href {\doibase 10.1016/j.jhazmat.2013.08.018} {\bibfield  {journal} {\bibinfo  {journal} {Journal of Hazardous Materials}\ }\textbf {\bibinfo {volume} {261}},\ \bibinfo {pages} {653--668} (\bibinfo {year} {2013})}\BibitemShut {NoStop}%
\bibitem [{\citenamefont {Hanna}\ and\ \citenamefont {Chang}(2012)}]{Hanna2012}%
  \BibitemOpen
  \bibfield  {author} {\bibinfo {author} {\bibfnamefont {S.}~\bibnamefont {Hanna}}\ and\ \bibinfo {author} {\bibfnamefont {J.}~\bibnamefont {Chang}},\ }\bibfield  {title} {\enquote {\bibinfo {title} {{Acceptance criteria for urban dispersion model evaluation}},}\ }\href {\doibase 10.1007/s00703-011-0177-1} {\bibfield  {journal} {\bibinfo  {journal} {Meteorology and Atmospheric Physics}\ }\textbf {\bibinfo {volume} {116}},\ \bibinfo {pages} {133--146} (\bibinfo {year} {2012})}\BibitemShut {NoStop}%
\bibitem [{\citenamefont {{Verein Deutscher Ingenieure (VDI)}}(2000)}]{VDI2000}%
  \BibitemOpen
  \bibfield  {author} {\bibinfo {author} {\bibnamefont {{Verein Deutscher Ingenieure (VDI)}}},\ }\bibfield  {title} {\enquote {\bibinfo {title} {Environmental meteorology—physical modelling of flow and dispersion processes in the atmospheric boundary layer—application of wind tunnels},}\ }\href@noop {} {\bibfield  {journal} {\bibinfo  {journal} {VDI-Richtlinien}\ }\bibinfo {series} {VDI Guideline 3783, Part 12} (\bibinfo {year} {2000})}\BibitemShut {NoStop}%
\bibitem [{\citenamefont {Roth}(2000)}]{roth2000review}%
  \BibitemOpen
  \bibfield  {author} {\bibinfo {author} {\bibfnamefont {M.}~\bibnamefont {Roth}},\ }\bibfield  {title} {\enquote {\bibinfo {title} {Review of atmospheric turbulence over cities},}\ }\href@noop {} {\bibfield  {journal} {\bibinfo  {journal} {Quarterly Journal of the Royal Meteorological Society}\ }\textbf {\bibinfo {volume} {126}},\ \bibinfo {pages} {941--990} (\bibinfo {year} {2000})}\BibitemShut {NoStop}%
\bibitem [{\citenamefont {Poggi}\ and\ \citenamefont {Katul}(2010)}]{Poggi2010}%
  \BibitemOpen
  \bibfield  {author} {\bibinfo {author} {\bibfnamefont {D.}~\bibnamefont {Poggi}}\ and\ \bibinfo {author} {\bibfnamefont {G.~G.}\ \bibnamefont {Katul}},\ }\bibfield  {title} {\enquote {\bibinfo {title} {{Evaluation of the turbulent kinetic energy dissipation rate inside canopies by zero- and level-crossing density methods}},}\ }\href@noop {} {\bibfield  {journal} {\bibinfo  {journal} {Boundary-Layer Meteorology}\ }\textbf {\bibinfo {volume} {136}},\ \bibinfo {pages} {219--233} (\bibinfo {year} {2010})}\BibitemShut {NoStop}%
\bibitem [{\citenamefont {Davidson}(2009)}]{DAVIDSON20091016}%
  \BibitemOpen
  \bibfield  {author} {\bibinfo {author} {\bibfnamefont {L.}~\bibnamefont {Davidson}},\ }\bibfield  {title} {\enquote {\bibinfo {title} {Large eddy simulations: How to evaluate resolution},}\ }\href {https://www.sciencedirect.com/science/article/pii/S0142727X09001039} {\bibfield  {journal} {\bibinfo  {journal} {International Journal of Heat and Fluid Flow}\ }\textbf {\bibinfo {volume} {30}},\ \bibinfo {pages} {1016--1025} (\bibinfo {year} {2009})}\BibitemShut {NoStop}%
\bibitem [{\citenamefont {Castro}\ and\ \citenamefont {Xie}(2009)}]{Castro2009c}%
  \BibitemOpen
  \bibfield  {author} {\bibinfo {author} {\bibfnamefont {I.~P.}\ \bibnamefont {Castro}}\ and\ \bibinfo {author} {\bibfnamefont {Z.-T.}\ \bibnamefont {Xie}},\ }\bibfield  {title} {\enquote {\bibinfo {title} {{LES for modelling the urban environment}},}\ }\href@noop {} {\bibfield  {journal} {\bibinfo  {journal} {ERCOFTAC Bulletin}\ }\textbf {\bibinfo {volume} {78}},\ \bibinfo {pages} {4--10} (\bibinfo {year} {2009})}\BibitemShut {NoStop}%
\bibitem [{\citenamefont {García-Sánchez}, \citenamefont {van Beeck},\ and\ \citenamefont {Gorlé}(2018)}]{GarciaSanchez2018}%
  \BibitemOpen
  \bibfield  {author} {\bibinfo {author} {\bibfnamefont {C.}~\bibnamefont {García-Sánchez}}, \bibinfo {author} {\bibfnamefont {J.}~\bibnamefont {van Beeck}}, \ and\ \bibinfo {author} {\bibfnamefont {C.}~\bibnamefont {Gorlé}},\ }\bibfield  {title} {\enquote {\bibinfo {title} {Predictive large eddy simulations for urban flows: Challenges and opportunities},}\ }\href {\doibase 10.1016/j.buildenv.2018.05.007} {\bibfield  {journal} {\bibinfo  {journal} {Building and Environment}\ }\textbf {\bibinfo {volume} {139}},\ \bibinfo {pages} {146--156} (\bibinfo {year} {2018})}\BibitemShut {NoStop}%
\bibitem [{\citenamefont {Duan}\ and\ \citenamefont {Ngan}(2019)}]{Duan2019}%
  \BibitemOpen
  \bibfield  {author} {\bibinfo {author} {\bibfnamefont {G.}~\bibnamefont {Duan}}\ and\ \bibinfo {author} {\bibfnamefont {K.}~\bibnamefont {Ngan}},\ }\bibfield  {title} {\enquote {\bibinfo {title} {{Sensitivity of turbulent flow around a 3-D building array to urban boundary-layer stability}},}\ }\href {\doibase 10.1016/j.jweia.2019.103958} {\bibfield  {journal} {\bibinfo  {journal} {Journal of Wind Engineering and Industrial Aerodynamics}\ }\textbf {\bibinfo {volume} {193}},\ \bibinfo {pages} {103958} (\bibinfo {year} {2019})}\BibitemShut {NoStop}%
\end{thebibliography}%

\end{document}